\documentclass[aps,twocolumn,amsmath,amssymb,reprint,floatfix,superscriptaddress]{revtex4-2}
\usepackage{amsmath,amssymb,bm,graphicx}
\usepackage{times}
\usepackage{dsfont}
\usepackage{xcolor}

\usepackage{hyperref}
\hypersetup{
colorlinks=true,
linkcolor=blue,           
citecolor=blue,           
filecolor=magenta,        
urlcolor=cyan
}
\usepackage{cleveref}

\newcommand{\tr}[1]{\mathrm{Tr}\left[ {#1} \right]} 

\newcommand{\ket}[1]{\vert #1\rangle}
\newcommand{\bra}[1]{\langle #1\vert}

\newcommand{\papertitle}{Quantum theory of a three-photon Kerr parametric oscillator: from exact solution to the Kerr-cat qutrit}

\begin{document}

\title{\papertitle}

\author{Alessandro Bruno}
\affiliation{PSI Center for Photon Science, 5232 Villigen PSI, Switzerland}
\affiliation{Swiss Nanoscience Institute, University of Basel, Klingelbergstrasse 82, 4056 Basel, Switzerland}
\author{Patrick P. Potts}
\affiliation{Department of Physics, University of Basel, Klingelbergstrasse 82, CH-4056 Basel, Switzerland}
\author{Alexander Grimm}
\affiliation{PSI Center for Photon Science, 5232 Villigen PSI, Switzerland}
\author{Matteo Brunelli}
\email{matteo.brunelli@college-de-france.fr}
\affiliation{JEIP, UAR 3573 CNRS, Coll\`ege de France, PSL Research University, 11 Place Marcelin Berthelot, 75321 Paris Cedex 05, France}


\begin{abstract}
We investigate the quantum properties of a nonlinear Kerr oscillator driven by a three-photon pump. We identify an exact spectral degeneracy at which the interacting ground state admits a nonperturbative analytical solution. In the large-amplitude regime, the resulting manifold becomes threefold quasi-degenerate and forms delocalized three-headed quantum superpositions. 
Away from the exact degeneracy, we derive an approximate description of the threefold quasi-degenerate manifold in terms of  
$\mathbb{Z}_3$-symmetric superpositions of squeezed coherent states. These states differ qualitatively from conventional three-component cat states due to the presence of squeezing with a distinctive parametric dependence. By varying the pump-oscillator detuning, 
the squeezing can be enhanced, suppressed, or even reversed, leading to a squeezing-to-anti-squeezing transition. In the large-amplitude regime, the threefold manifold defines a noise-biased bosonic qutrit, which we term a Kerr-cat qutrit. 
We show that photon loss predominantly induces cyclic shift errors, while clock errors are suppressed, and quantitatively characterize the resulting noise bias. We further show that the detuning 
controls the noise bias through the squeezing. 
Going beyond stabilization, we construct arbitrary operations on a single Kerr-cat qutrit together with an entangling two-qutrit gate, providing a complete set of logical operations for universal qutrit quantum computation, and validate their action through full Lindblad simulations. Our results establish the three-photon Kerr oscillator both as a rich nonlinear quantum system with analytically tractable macroscopic states and as a platform for noise-biased bosonic qutrit quantum computation.
\end{abstract}

\maketitle

\section{introduction}

A driven and interacting single bosonic mode provides a simple and yet formidably rich arena for the study of classical and quantum phenomena. A special role is played by nonlinear Kerr oscillators subject to parametric driving and dissipation, which offer insight into the role of symmetries in open quantum systems~\cite{Bartolo2016,Minganti2016SciRep,Roberts2020}, and into the phenomenology of quantum~\cite{Grigoriou2023} as well as dissipative~\cite{Minganti2018,Zhang2021,Minganti2023nphoton, Soriente2021_phasetrans, Ameye2025} phase transitions. 
For standard parametric driving, one obtains the two-photon Kerr parametric oscillator (2KPO)~\cite{Wielinga1993,wustmann_parametric_2019,dykman_fluctuating_2012,Eichler2023_parametricphenomena}. 
In the quantum regime, the 2KPO hosts a ground-state manifold spanned by quantum superpositions of two macroscopically distinct classical states, known as Schrödinger's cat states~\cite{Puri2017Kerr,Goto2016, cochrane_macroscopically_1999}. The stabilization of the cat-state manifold can be exploited for encoding and manipulating quantum information in a hardware-efficient manner and offers a promising route toward quantum error correction~\cite{mirrahimi_dynamically_2014, leghtas_confining_2015, touzard_coherent_2018, Lescanne2020, grimm2020, Frattini2024Arrhenius, hajr_high-coherence_2024, qing_benchmarking_2024, ding_quantum_2025, adinolfi_enhancing_2025, Puri2020,darmawan_practical_2021, chono_two-qubit_2022, Gravina2023, Rousseau25,Putterman2022, Ruiz2023,Masuda2025, Su2024, BlaisRescueKCQ, Marquet2024, guillaud_repetition_2019, guillaud_error_2021, chamberland_building_2022, gautier_combined_2022, xu_engineering_2022, Venkatraman2024,putterman_hardware-efficient_2025,albornoz_oscillatory_2024,ruiz_ldpc-cat_2025,reglade_quantum_2024,kanao_quantum_gate_2022}. 

A natural higher-order extension of this paradigm is obtained by replacing the two-photon parametric drive with a three-photon process. 
In circuit quantum electrodynamics (cQED) a three-photon parametric pump is readily available, e.g. by modulating the external flux of a SNAIL superconducting circuit in proximity of three times its natural frequency. Three-photon interactions in the form of a cubic phase gate  have been recently demonstrated in a SNAIL resonator~\cite{Eriksson_2024_cubic}. Three-photon spontaneous parametric down-conversion has also been demonstrated in driven superconducting resonators~\cite{Svensson2017_periodtripling, Chang_2020_3spdc}, including the generation of states with negative Wigner functions~\cite{kwon_realisation_2026}. From a theory standpoint, period-tripling dynamics induced by three-photon pump has been discussed in Kerr oscillators~\cite{Zhang2017, Zhang2019, Lorch2019, Gosner2020} as well as 
Josephson junctions coupled to a microwave resonator~\cite{Arndt2022, Arndt2022_beyondRWA, Lang_2021_multiphoton}. Excited-state quantum phase transitions and dissipative phase transitions in Kerr oscillators driven by three- and four-photon parametric pumps have been investigated~\cite{Reynoso_2025, Minganti2023nphoton}.

The quantum regime of three-photon Kerr parametric oscillator (3KPO), 
however, remains considerably less understood than that of the 2KPO.
While Ref.~\cite{kwon_realisation_2026} shows the initialization and Wigner tomography of quantum states in such a system, theoretical insight into the structure of its low-energy states, the effect of perturbations and the benefits of encoding and manipulating quantum information is still lacking. 
Moreover, unlike for the 2KPO~\cite{Ruiz2023}, the role of control parameters such as cavity-pump detuning, has not been systematically investigated.

In this work we study the quantum properties of the 3KPO through both exact and approximate analytical treatments. We provide analytical insight into the structure of its eigenstates, especially on the ground-state manifold,  yielding the same simple conceptual understanding  that has proven invaluable for the quantum description of the 2KPO~\cite{ puri2019, Puri2017Kerr}.
We first identify an exact degeneracy in the spectrum, defined by a quadratic dependence of the detuning on the three-photon pump, for which we provide an exact, nonperturbative analytical solution for the interacting ground state. The ground state is asymptotically threefold degenerate and consists of quantum superpositions with threefold rotational symmetry, 
see Fig.~\ref{f:CerberusExact}(a)-(c), distinct from three-legged Schr\"odinger's cat states. 
This exact quantum solution corresponds to the unique four-stable configuration of the  semiclassical potential, in which the zero-amplitude and the three finite-amplitude stationary points have the same energy. 

Away from this special point, we identify a broad regime of threefold quasi-degeneracy associated with tristability at the semiclassical level~\cite{Zhang2017, Zhang2019}. 
We derive an approximate description of the ground-state manifold in terms of three-legged squeezed cat states, given by $\mathbb{Z}_3$-symmetric superpositions of three squeezed coherent states localized around the semiclassical stationary points.
These analytical expressions reveal a distinctive feature of the 3KPO: 
by varying the detuning relative to the pump strength, quantum fluctuations are displaced and concomitantly squeezed along the direction of displacement, and can  be continuously tuned from squeezing to anti-squeezing.
We further analyze the preparation of three-legged squeezed cat states in the presence of single-photon loss, the resulting steady state of the 3KPO and the leakage from the low-energy manifold.

In the large-amplitude regime, the threefold ground-state manifold defines a three-dimensional encoding of quantum information, which we refer to as a \emph{Kerr-cat qutrit}, in analogy with the Kerr-cat qubit~\cite{grimm2020, Puri2017Kerr, cochrane_macroscopically_1999}. Within this encoding, single-photon loss predominantly produces cyclic shift errors, whereas clock errors are suppressed. The resulting noise bias can be tuned through the oscillator parameters and further enhanced by state-selective dissipation. We finally investigate coherent control of the Kerr-cat qutrit. We construct single-qutrit operations from physical controls of the 3KPO, including clock, shift, and Kerr gates, and show that their generators span the full $\mathfrak{su}(3)$ Lie algebra. We further consider two 3KPO modes coupled by photon exchange and obtain an entangling two-qutrit gate. We validate the logical action of these operations  through simulations of the full oscillator dynamics, including leakage to the excited states.

The paper is organized as follows. In Sec.~\ref{s:System-Semiclassical} we introduce the 3KPO Hamiltonian and discuss its semiclassical structure. In Sec.~\ref{s:Spectrum} we study the energy spectrum and  its  degeneracies. In Sec.~\ref{s:Exact} we derive the exact ground-state solution at the exact spectral degeneracy. In Sec.~\ref{s:Approximate} we develop an approximate description of the 3KPO entire spectrum in the quasi-degenerate regime and introduce the three-legged squeezed cat states. We then study their dissipative dynamics and stabilization and discuss the effect of single-photon loss in Sec.~\ref{s:SinglePhoton} and Sec.~\ref{s:PhotonLoss}. In Sec.~\ref{s:NoiseBias} we develop the corresponding noise-biased qutrit encoding and analyzes its error channels. In Sec.~\ref{s:Gates} we construct universal single-qutrit operations and an entangling two-qutrit gate. We conclude with Sec.~\ref{s:Conclusions}, which contains the conclusions and outlook of our work.

\section{The system and its semiclassical limit}\label{s:System-Semiclassical}

\begin{figure*}[t!]
\centering
\includegraphics[width=1.\textwidth]{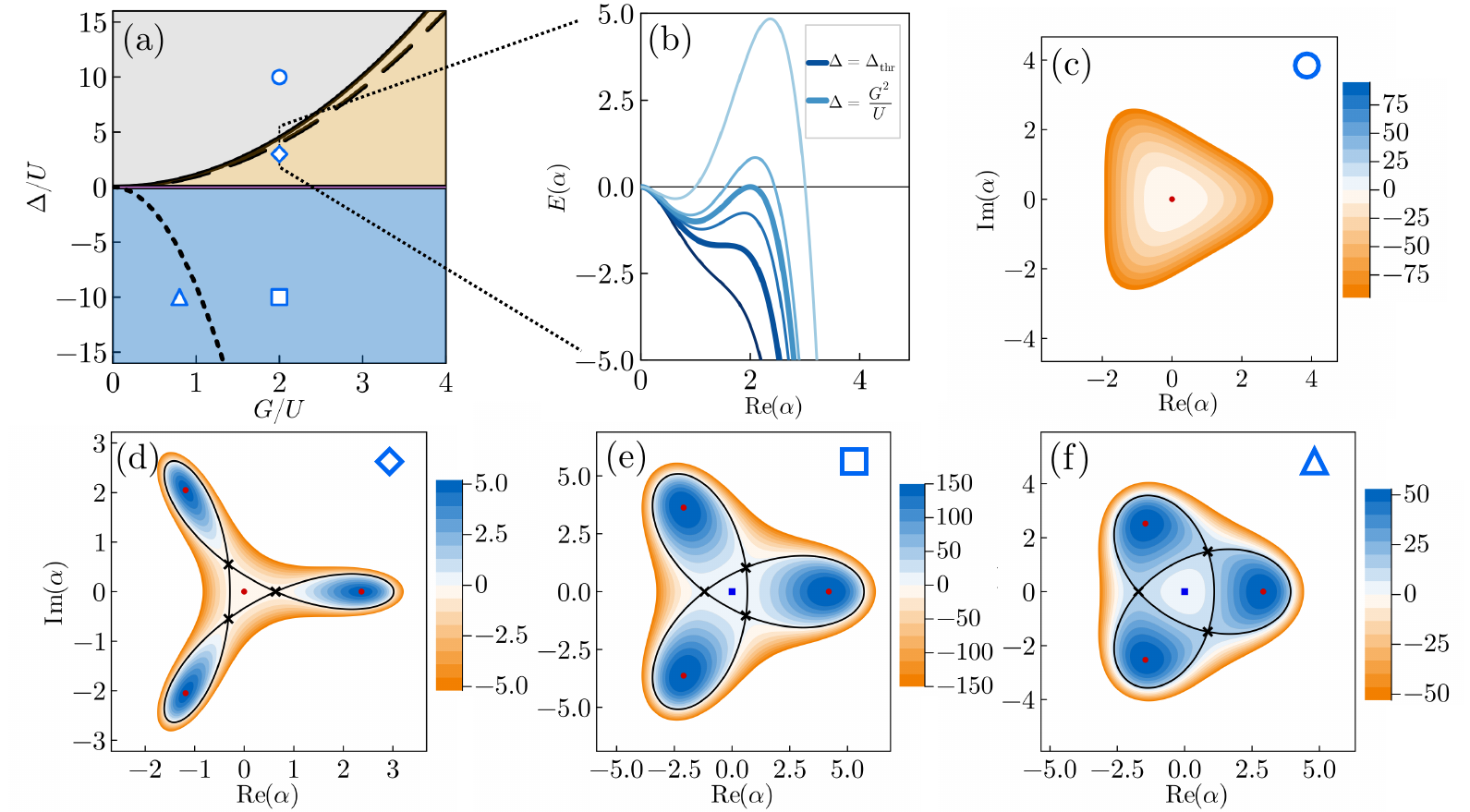}
\caption{\label{f:Model}  
(a) Phase diagram of the three-photon Kerr parametric oscillator (3KPO), obtained from the meta-potential $E(\alpha, \alpha^*)$. Different phases are color coded according to the number of local maxima of the meta-potential: single maximum (gray), four maxima (yellow) and three maxima (blue).  The solid black curve corresponds to $\Delta=\Delta_\mathrm{thr}$, the dashed curve to $\Delta=G^2/U$ and the dotted one to $\Delta=-9G^2/U$. The yellow region is characterized by metastability, with either a global maximum and three local maxima $(G^2/U\le \Delta<\Delta_\mathrm{thr})$ or a local maximum and three global maxima $(0\le \Delta <G^2/U)$. (b) Cuts of the meta-potential along $\mathrm{Im}(\alpha)=0$ for $G/U=2$ and different values of detuning, illustrating metastability. The curves correspond, from darker to lighter, to the values of detuning $\Delta=(5U, \, \Delta_\mathrm{thr},\, 0.95 G^2/U,\, G^2/U,\, 1.05 G^2/U,\, 3U)$; the second and fourth value are marked by a thick line. (c)-(f) plot of the meta-potential for the four points highlighted in panel (a). Maxima of the meta-potential are marked by red dots, minima by blue squares, while saddle points by black crosses. 
}
\label{f:Potential}
\end{figure*}

\begin{figure*}[t]
\centering
\includegraphics[width=1.\textwidth]{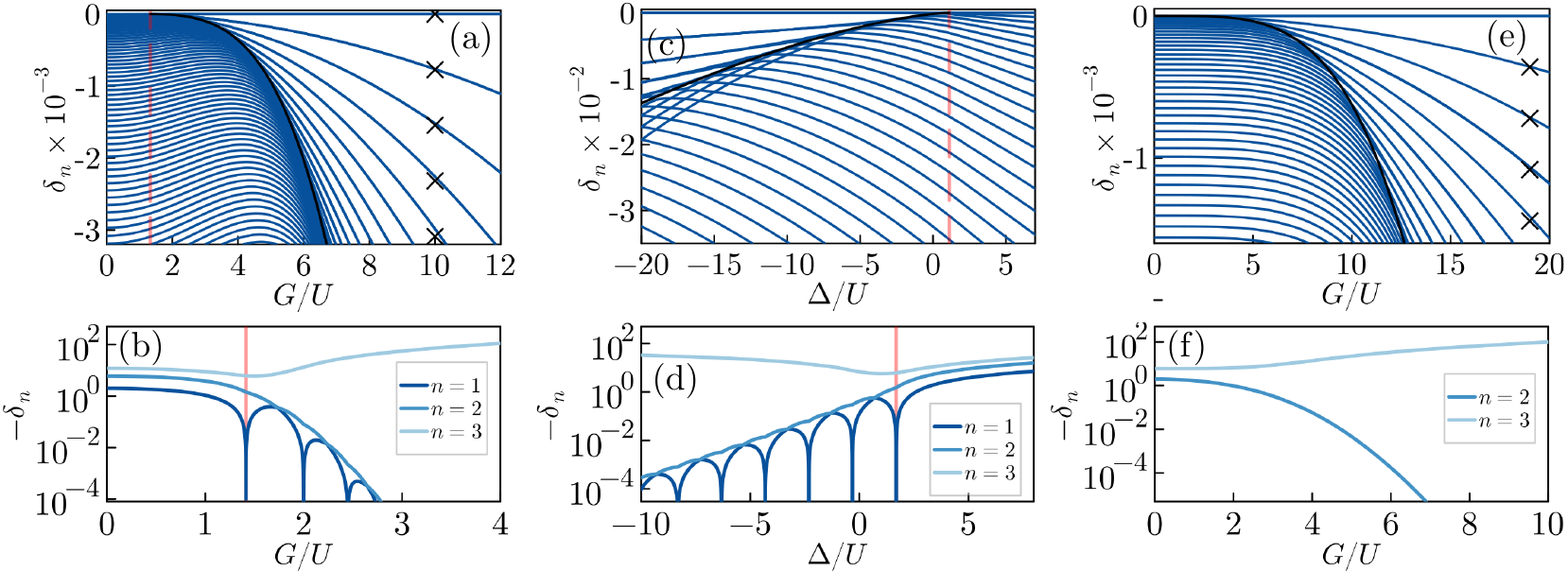}
\caption{\label{f:Spectrum}   
(a) Excitation spectrum $\delta_n=E_n-E_0$ of the 3KPO Hamiltonian Eq.~\eqref{eq:H3KPR} as a function of the three-photon pump for $\Delta=2 U$. For sufficiently large pump amplitudes, the spectrum organizes in quasi-degenerate triplets. The dashed red line marks the threshold value $G_\mathrm{thr}=\frac{2\sqrt2}{3}\sqrt{\Delta U}$, the crosses are the approximate values of the energy given by Eq.~\eqref{eq:Omega}, while the black curve is obtained from the semiclassical potential. (b) Energy gaps $\delta_n$ of the first three excited states, showing exponential convergence to a threefold degenerate ground state. The red solid line marks the first level crossings, corresponding to $G=\sqrt{\Delta U}$, where the energy gap to the first excited state closes. (c), (d) Same as panels (a), (b) but plotted as a function of the detuning, for $G=1.3U$. (e), (f) Same as panels (a), (b) but for the choice of pump-dependent detuning $\Delta=G^2/U$, where $\delta_{n=1}$ is identically zero.}
\end{figure*}

The 3KPO consists of a Kerr oscillator described by the annihilation operator $\hat a$ $([\hat a,\hat a^\dagger]=1)$ which is driven by a three-photon pump. The Hamiltonian is given by  $(\hbar=1)$
\begin{equation}\label{eq:H3KPR}
\hat H=-\Delta \hat a^{\dagger}\hat a -U\hat a^{\dagger\,2}\hat a^{2}+G \left(\hat a^{\dagger\,3}+\hat a^{3}\right)\,,
\end{equation}
where $\Delta=\omega_p/3-\omega_c$ is the pump-cavity detuning,  $\omega_c$ and $\omega_p$ being respectively the oscillator frequency and the modulation frequency, $U$ is the strength of the Kerr non-linearity and $G$ the strength of the drive; we choose $G,\,U>0$ without loss of generality. 
The three-photon pump term can be obtained by parametric driving of a third order nonlinearity $2G\cos (\omega_p t) (\hat a^{\dagger}+\hat a )^3$ at the frequency $\omega_p=3(\omega_c +\Delta)$ and retaining the resonant contributions.

We start with a semiclassical analysis of the 3KPO by inspecting the semiclassical potential associated to Eq.~\eqref{eq:H3KPR}. We obtain it 
by taking the expectation value $E(\alpha,\alpha^*)=\langle \alpha\vert \hat H \vert \alpha\rangle$ over the coherent state $\hat a \vert \alpha\rangle=\alpha \vert \alpha\rangle$, also referred to as meta-potential~\cite{ZhangDykman2017}. In Fig.~\ref{f:Potential}(a) we show a phase diagram as a function of pump strength and detuning, with different phases corresponding to different number of local maxima of the meta-potential. Figs.~\ref{f:Potential} (c)-(f) show instances of $E(\alpha,\alpha^*)$, revealing a characteristic threefold symmetry. Following this symmetry,
we can further set $\arg\alpha=\frac{2\pi}{3}k$, $k\in\mathbb{Z}$, to get the expression
\begin{equation}\label{eq:SemiclassPotential}
E(\vert\alpha\vert)=-\vert\alpha\vert^2\left(U \vert\alpha\vert^2 - 2 G \vert\alpha\vert +\Delta \right).
\end{equation}
This energy function, shown in Fig.~\ref{f:Potential}(b), 
describes a cut through the meta-potential. 

Starting from positive and large values of the detuning, the meta-potential has a single global maximum at $\alpha=0$ with $E(0)=0$, see Fig.~\ref{f:Potential}(c). By decreasing the detuning past the threshold value  $\Delta_\mathrm{thr}=\frac{9}{8}\frac{G^2}{U}$, $\alpha=0$ still remains the global maximum, but new stationary points $\nobreak{\alpha_{\pm,k}=\vert \alpha_{\pm}\vert e^{i\phi_k}}$ appear,
with magnitudes given by
\begin{equation}\label{eq:AlphaPlusMin}
\vert\alpha_{\pm}\vert=\frac{3G\pm\sqrt{9G^2-8U\Delta}}{4U},
\end{equation}
and phases $\phi_k=\frac{2\pi}{3}k$, $k=0,1,2$, for a total of seven stationary points. 
In particular, $\alpha_+$ are local maxima and $\alpha_-$ are saddle points. Notice that $\vert\alpha_-\vert$ corresponds to a minimum of $E(\vert\alpha\vert)$ in the cut of Fig.~\ref{f:Potential}(b). 
The threshold value $\Delta_\mathrm{thr}$ marks the beginning of a coexistence region, where multiple extremal solutions exist. At $\Delta=\Delta_\mathrm{thr}$, the second derivative of Eq.~\eqref{eq:SemiclassPotential} is always negative except for the (spinodal) point $\vert\alpha_+\vert=\vert\alpha_-\vert=3G/4U$, where both the first and the second derivative vanish (akin to the critical isotherm in the $PV$ diagram of van der Waals equation of state). 
When $\frac{G^2}{U}\le \Delta<\Delta_\mathrm{thr}$, the three local maxima with amplitude $\vert\alpha_+\vert$ correspond to metastable solutions. By decreasing the detuning, these local maxima raise in energy until they reach the global maximum at $E=0$, which happens for $\Delta=\frac{G^2}{U}$. 
This special value of the detuning is marked by the dashed line in the phase diagram Fig.~\ref{f:Potential}(a). Along this curve the potential has fourfold stability, i.e., four global maxima with the same energy, see Fig.~\ref{f:Potential}(b). 
In the range of detuning $\frac{G^2}{U}\le \Delta<\Delta_\mathrm{thr}$, the equipotential curves of the meta-potential describe either four closed orbits ($E(\vert\alpha_-\vert)<E<E(\vert\alpha_+\vert)$) or a single closed orbit (for $E<E(\vert\alpha_-\vert)$ or $E(\vert\alpha_+\vert)<E\le 0$). The separatrix is shown as a black curve in Fig.~\ref{f:Potential}(c)-(f).

When the detuning is further decreased, $0\le \Delta<\frac{G^2}{U}$, 
local and global maxima switch role and 
the zero amplitude solution becomes metastable: the points $\alpha_{+,k}$ become global maxima, acquiring positive energy $E(\vert\alpha_+\vert)>0$, while $\alpha=0$ remains at zero energy, thus becoming a local maximum. 
In this range of detuning the meta-potential can have a single closed orbit $(E<E(\vert\alpha_-\vert))$, four closed orbits $(E(\vert\alpha_-\vert)<E<0)$, or three closed orbits $(0<E<E(\vert\alpha_+\vert))$.
This tri-stability has been discussed in~\cite{Zhang2017, Zhang2019}.

For $\Delta=0$ the saddle points approach the origin and merge with the local maximum, see  Eq.~\eqref{eq:AlphaPlusMin}. 
For negative detunings $\Delta<0$ the saddle points reappear but are now shifted by a $\pi$-phase, i.e., 
$\arg\alpha_-=\frac{\pi}{3}k$. As a result, at $\Delta=0$ the zero amplitude solution changes from a local maximum to a minimum. This can be seen from Fig.~\ref{f:Potential}(e), (f) and
from the cut in Fig.~\ref{f:Potential}(b) develops a minimum at the origin. The negative detuning region is then characterized by only three maxima and marked as a distinct phase in Fig.~\ref{f:Potential}(a).
The coexistence region therefore extends through the range of detuning $0<\Delta<\Delta_\mathrm{thr}$. 

Finally, within the negative detuning region, the equipotential curves close to the maxima change from being squashed in the direction of the displacement, see Fig.~\ref{f:Potential}(e), to the orthogonal direction, see Fig.~\ref{f:Potential}(f). This change occurs along the curve $\Delta=-9 G^2/U$, marked by a dotted line in Fig.~\ref{f:Potential}(a). As we will see in Sec.~\ref{s:Approximate}, this line corresponds to a vanishing value of the squeezing of the quantum fluctuations around the stationary solutions. 

For future convenience, we restate the threshold values of the model in terms of the coupling strength. For $\Delta\ge0$, the 3KPO has a threshold for multi-stability at the critical coupling $G_\mathrm{thr}=\frac{2\sqrt2}{3}\sqrt{\Delta U}$. 
When the coupling exceeds this threshold, three metastable solutions with amplitude $\alpha_{+,k}$, $k=0,1,2$ appear alongside the global maximum at the origin. When $G>\sqrt{\Delta U}$ the former become global maxima and the zero-amplitude solution becomes metastable; the special point $G=\sqrt{\Delta U}$ is uniquely characterized by four equal energy solutions. For $\Delta<0$, there is no threshold for multi-stability, as $E(\alpha,\alpha^*)$ has always three global maxima at $\alpha_{+,k}$ and a minimum at the origin.

\section{Energy spectrum}
\label{s:Spectrum}

We proceed by investigating the energy spectrum of the 3KPO Hamiltonian Eq.~\eqref{eq:H3KPR}. The 3KPO Hamiltonian has $\mathbb{Z}_3$ symmetry, as it commutes with the discrete rotation $\hat Z=e^{i \frac{2\pi}{3}\hat a^\dagger \hat a}$, i.e., $[\hat H, \hat Z]=0$.
For simplicity, we will refer to quasi-energy levels simply as energy levels, to the highest quasi-energy state as the ground state, to the second-highest as first excited state and so on.
In Fig.~\ref{f:Spectrum}(a) we show the excitation spectrum $\delta_n=E_n-E_0$, i.e., the energy spectrum shifted by the vacuum energy $E_0$, as a function of the three-photon pump for $\Delta>0$. A close-up of the first three energy gaps $\delta_{n=1,2,3}$ is shown in Fig.~\ref{f:Spectrum}(b) on a logarithmic scale. In the limit of vanishing pump, the Hamiltonian reduces to that of an undriven Kerr oscillator, i.e.,   diagonal in the Fock number basis with eigenenergies $E_n^{(G=0)}=-(\Delta + U(n-1))n$.
For sufficiently large values of the pump, on the other hand,  neighboring energy levels belonging to different $\mathbb{Z}_3$ sectors merge into triplets, resulting in asymptotic triple degeneracies of the whole spectrum. This behavior, analogous to the  `spectral kissings' observed in the 2KPO~\cite{Chavez-Carlos2023, Frattini2024Arrhenius}, was first reported in Ref.~\cite{Zhang2019}.
From Fig.~\ref{f:Spectrum}(a) we also see that higher excited states coalesce at increasingly larger values of the pump, forming an envelope marked by the black curve. The envelop curve corresponds to the energy barrier $\vert E(\alpha_+)-E(\alpha_-)\vert$ obtained from the meta-potential. In the semiclassical model, states above this curve correspond to three distinct  orbits with the same energy, each localized in a separate well. 
Quantum mechanically, the system can be found in a superposition of all three symmetry-broken solutions.
In the quantum model Eq.~\eqref{eq:H3KPR}, we obtain approximate analytical expressions of the degenerate states, see discussion in Sec.~\ref{s:Approximate} and Eq.~\eqref{eq:Omega}. Their energies are marked by the black crosses, showing good agreement with the numerically obtained spectrum.
For any finite value of the pump, quantum tunneling between these states lifts the degeneracy, resulting in an exponentially small correction~\cite{Zhang2017, Zhang2019}.   One can verify that the energy splitting of first and second excited states decreases exponentially in $G$, as shown in Fig.~\ref{f:Spectrum}(b) for the first triplet. At the same time, we see that the gap to the third excited state increases with pump amplitude. This confirms that the model has a nearly triply degenerate ground state manifold, gapped from the rest of the spectrum. 

Fig.~\ref{f:Spectrum}(c) and (d) display the energy spectrum as a function of detuning, showing that a large negative detuning also leads to triple degeneracy of the spectrum. We notice that, for negative detuning, the spectrum in the limit of vanishing pump is no longer decreasingly ordered with respect to $n$ 
(some of the $E_{n\ge1}^{(G=0)}$ become positive). This can lead to extra degeneracies for the choice of specific integer values of the detuning, as recently pointed out in~\cite{Roberts2020, Ruiz2023}. For even values $\Delta/U=-2m$, there are $m+1$ pairs of degenerate levels. For instance, for $m=2$, we have the following degenerate pairs: $\vert 0\rangle$ and $\vert 5\rangle$, $\vert 1\rangle$ and $\vert 4\rangle$, $\vert 2\rangle$ and $\vert 3\rangle$. For odd values $\Delta/U=-(2m+1)$, the highest quasi-energy state  is unique but is followed by $m+1$ degenerate pairs. When $G\neq 0$ these degeneracies are quickly removed by the three-photon pump, which leads to multiple level crossings, as shown in Fig.~\ref{f:Spectrum}(c).

A distinctive feature of the energy spectrum that sets the 3KPO apart from the 2KPO~\cite{Frattini2024Arrhenius, Ruiz2023, Venkatraman2024}, is the presence of level crossings even for positive values of the detuning, which occur at values $G_{m\ge1}=\sqrt{m \Delta U}$ with integer $m$. The first instance of such crossing, which  happens for $G=\sqrt{\Delta U}$, is marked by the solid red line in Figs.~\ref{f:Spectrum}(b), (d). At this fine tuned value the ground state is \emph{exactly} doubly degenerate. 
Put differently, if we choose the detuning to vary quadratically with the three-photon pump $\Delta=\frac{G^2}{U}$, we expect a doubly degenerate  ground state for \emph{any} value of $G$, which becomes triply degenerate for a large enough values of the three-photon pump. This is confirmed by inspecting Figs.~\ref{f:Spectrum}(e), (f), which show the spectrum for $\Delta=\frac{G^2}{U}$, characterized by  $\delta_{n=1}\equiv 0$ identically zero. This observation is directly connected to the exact solution of the ground state manifold that will be discussed in next Section.  

\section{Ground state manifold: Exact solution}\label{s:Exact}

\begin{figure*}[t]
\centering
\includegraphics[width=1.\textwidth]{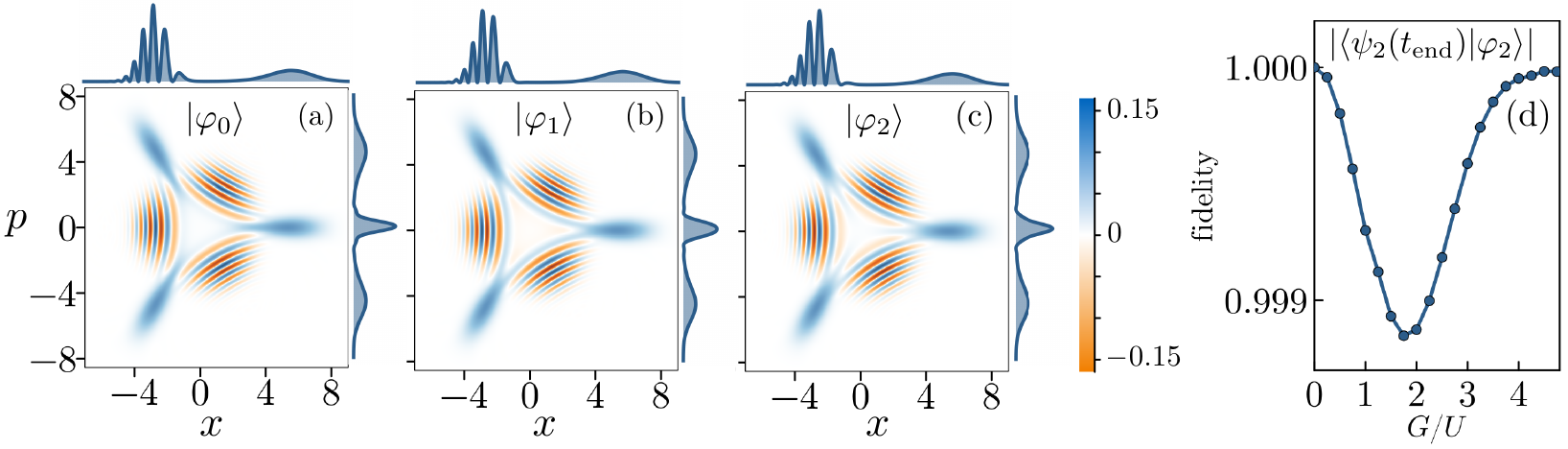}
\caption{\label{f:CerberusExact} (a)-(c) Wigner function of the ground states $\ket{\varphi_0}$ (a), $\ket{\varphi_1}$ (b) and $\ket{\varphi_2}$ (c), respectively given in Eqs.~\eqref{eq:Phi0}-\eqref{eq:Phi2}, for $G/U=4$ and $\Delta=G^2/U$. The marginal probability distributions for position $x=\sqrt2 \mathrm{Re}(\alpha)$ and momentum $p=\sqrt2 \mathrm{Im}(\alpha)$ quadratures are also shown. (d) Fidelity $\vert \langle \psi_2(t_\mathrm{end})\ket{\varphi_2} \vert$ between the adiabatically evolved state $\ket{\psi_k(t_\mathrm{end})}$ at the final time, i.e., the quantum state obtained from the initial state $\ket{n=2}$ by switching on adiabatically the three-photon pump, and the target state $\ket{\varphi_2}$ for different values of the three-photon pump; the lines joining the dots are for visualization purposes only. For $k=0,1$, the  state $\ket{\varphi_k}$ attains unit fidelity with the state adiabatically evolved from $\ket{n=k}$ for any value of $G$.}
\end{figure*}

We want to characterize the ground state manifold of the 3KPO. We first consider the special value  of the detuning $\Delta=G^2/U$, for which it is possible to provide an exact analytical solution. 
As shown in the previous section, when $\Delta=G^2/U$, the ground state has an exact double degeneracy for \emph{any} value of the three-photon pump [cf. Fig.~\ref{f:Spectrum}(f)]. This is analogous to the case of the 2KPO at zero detuning, whose ground state is also exactly doubly degenerate for any value of the two-photon pump~\cite{cochrane_macroscopically_1999, Marthaler2007, Puri2017Kerr}. 

In analogy with the  2KPO at zero detuning, for $\Delta=G^2/U$ the Hamiltonian Eq.~\eqref{eq:H3KPR} can be  factorized as
\begin{align}\label{eq:H3KPRSpecial}
\hat H&= -U\left(\hat a^{\dagger\,2}-g \,\hat a\right)\left(\hat a^{2}-g\,\hat a^\dagger\right)+U g^2 \,,
\end{align}
where for convenience we introduced the dimensionless parameter $g=G/U$. From this form it is clear that a state $\ket{\varphi}$ that satisfies 
\begin{equation}\label{eq:Dark}
(\hat a^2-g\,\hat a^\dagger)\ket{\varphi}=0 \,,
\end{equation}
is an eigenstate  $\hat H\ket{\varphi}=E\ket{\varphi}$ with $E=G^2/U$. Different from the case of the 2KPO, however, Eq.~\eqref{eq:Dark} cannot be further factorized in terms of coherent states with equal and opposite amplitudes, due to the presence of the creation operator.
Nevertheless, we can still project Eq.~\eqref{eq:Dark} onto the Fock basis to get the  recurrence relation $\sqrt{n(n-1)}c^\varphi_n-g\sqrt{n-2}c^\varphi_{n-3}=0$, for the coefficients $c^\varphi_n=\bra{n}\varphi\rangle$. We solve the recurrence relation by choosing as boundary term either the vacuum state $\ket{n=0}$ or the single-photon state $\ket{n=1}$, i.e., the two degenerate eigenstates of the undriven Kerr oscillator.  
By doing so we get the following states
\begin{align}
\ket{\varphi_0}&=\mathcal{M}_0  \sum_{n=0}^{+\infty}g^n\sqrt{\frac{\Gamma(n+\frac13)}{3^n n!\Gamma(n+\frac23)}}\ket{3n} \,, \label{eq:Phi0} \\
\ket{\varphi_1}&=\mathcal{M}_1 \sum_{n=0}^{+\infty}g^n\sqrt{\frac{\Gamma(n+\frac23)}{3^n n!\Gamma(n+\frac43)}}\ket{3n+1} \,, \label{eq:Phi1}
\end{align}
where the normalization factors are given by
$\mathcal{M}_\ell= \bigl(3^{\frac{(-1)^{\ell+1}}{6}}\sqrt{\pi} g^{(-1)^{\ell}/3} e^{g^2/6} I_{(-1)^{\ell+1}/6}(g^2/6)\bigr)^{-\frac12}$, with $\ell=0,1$,
and where $I_\nu(x)$ indicates the modified Bessel function of the first kind. 
We provide additional details on the states $\ket{\varphi_{0}},\,\ket{\varphi_{1}}$ in Appendix~\ref{app:WaveFunction}.
Furthermore, 
from our previous analysis of the energy spectrum, we know that the ground state becomes nearly triply degenerate in the large pump limit, see Fig.~\ref{f:Spectrum}(f). This suggests that
in this limit one could try to solve the recurrence relation also for the choice of boundary term $\ket{n=2}$, which yields
\begin{align}\label{eq:Phi2}
\ket{\varphi_2}&=\mathcal{M}_2 \sum_{n=0}^{+\infty}g^n\frac{3^n n!}{\sqrt{\Gamma(3n+3)}}\ket{3n+2} \,,
\end{align}
with normalization $\mathcal{M}_2= \bigl(\frac12 {}_1F_1 (\frac43; \frac 53;\frac{g^2}{3}) \bigr)^{-\frac12}$ and
where ${}_1 F_1(a;b;x)$ is the confluent hypergeometric function. We notice that the state \eqref{eq:Phi2} is not an eigenstate of Eq.~\eqref{eq:Dark} and the validity of this expression must be checked \textit{a posteriori}, as we do below.

The states in Eqs.~\eqref{eq:Phi0},~\eqref{eq:Phi1},~\eqref{eq:Phi2} span the ground state manifold of the 3KPO in the large pump limit. Their Wigner functions, illustrated in Fig.~\ref{f:CerberusExact}(a)-(c), reveal the non-local character of these states in phase space. They display a distinctive threefold symmetry, with maxima centered around the semiclassical ones Eq.~\eqref{eq:AlphaPlusMin} (see discussion in Section~\ref{s:Diagonalization}) and the characteristic interference fringes due to quantum superposition. 
These states contain only photons in multiples of three, i.e., they are eigenstates of the projectors $\hat \Pi_k = \sum_{n=0}^{+\infty} \ket{3n+k}\bra{3n+k}$, with $k=0,1,2 \mod 3$. They are orthogonal for any value of $G$ and thus form an orthonormal basis. 
By increasing $G$, the heads grow further apart and the interference region increases. Notice the relative shift in the interference fringes between different states, also visible from the marginal probability distributions.  

As already anticipated, we stress an important difference between the expressions of $\ket{\varphi_0}$, $\ket{\varphi_1}$ on the one hand, and $\ket{\varphi_2}$ on the other. While Eqs.~\eqref{eq:Phi0} and~\eqref{eq:Phi1} are exact expressions for the ground state of the interacting model Eq.~\eqref{eq:H3KPR}, valid for any value of the pump, this is not the case for Eq.~\eqref{eq:Phi2}. 
This is because the state in the symmetry sector $k=2$ 
is gapped from the ground state manifold and only approaches it in the large pump limit (although exponentially fast), see Fig.~\ref{f:Spectrum}(f). Eq.~\eqref{eq:Phi2} becomes exact in the infinite pump limit. 

For a vanishing value of the pump, the state $\ket{\varphi_k}$ reduces to the Fock state $\ket{n=k}$, with $k=0,1,2$. This, together with the fact that the Hamiltonian conserves the number of photons modulo three, provides a convenient way to prepare the ground states by initializing the 3KPO in the  state $\nobreak{\ket{\psi_k(0)}= \ket{n=k}}$ and adiabatically ramping up the three-photon pump.
The adiabatic theorem guarantees that the system remains in its instantaneous eigenstate if a given perturbation is acting on it slowly enough and as long as the eigenvalue is gapped from the rest of the Hamiltonian  spectrum. 
We consider  a time-dependent three-photon pump of the form $G(t)=G(1-e^{-(t/\tau)^4})$, which ensures that the drive can be turned on adiabatically by controlling the characteristic time $\tau$~\cite{Puri2017Kerr}.
In Fig~\ref{f:CerberusExact}(d) we show the fidelity between the analytic expression Eq.~\eqref{eq:Phi2} and the  states adiabatically evolved from $\ket{n=2}$, confirming that the fidelity is generally below one and that unit fidelity is approached for large  pump values. The same simulation for $\ket{\varphi_0},\,\ket{\varphi_1}$ (not shown) yields unit fidelity for any value of the pump, due to the exact nature of expressions Eqs.~\eqref{eq:Phi0}, ~\eqref{eq:Phi1}. At the same time, the plot shows that Eq.~\eqref{eq:Phi2}  still provides an excellent approximation of the actual state even in the intermediate regime.

\section{3KPO spectrum: Approximate solution}\label{s:Approximate}

\subsection{Diagonalization}\label{s:Diagonalization}
Away from the value $\Delta=G^2/U$, the factorization of Eq.~\eqref{eq:H3KPRSpecial} no longer applies. However, from Fig.~\ref{f:Spectrum}(a)-(d) we know that for both large negative detunings and positive detunings far above threshold $G\gg G_\mathrm{thr}$, the spectrum of the 3KPO is organized in nearly triply degenerate manifolds. We now look for an approximate description of this regime. In order to do that, we consider the Hamiltonian Eq.~\eqref{eq:H3KPR} and move to a displaced reference frame $\hat H'=\hat D^\dagger(\alpha)\hat H \hat D(\alpha)$ defined by $D(\alpha)=e^{\alpha \hat a^\dagger-\alpha^* \hat a}$, which gives
\begin{align}\label{eq:H3KPRDisp}
\hat H'&=-\Delta'\hat a^{\dagger}\hat a-U \hat a^{\dagger\,2}\hat a^{2}+G\left(\hat a^{\dagger\,3}+\hat a^{3}\right) \nonumber\\
&-\left[f_1(\alpha,\alpha^*)\hat a^{\dagger}+f_2(\alpha,\alpha^*)\hat a^{\dagger\,2} +\mathrm{H.c.} \right] \nonumber \\
&- 2U\alpha a^{\dagger\,2}\hat a - 2U\alpha^*\hat a^{\dagger}\hat a^2 \,.
\end{align}
In the above expression, the first line describes a 3KPO with renormalized detuning $\Delta'=\left(4U\vert\alpha\vert^2 +\Delta \right)$, the second line contains single- and two-photon drive terms, with amplitude $f_1(\alpha,\alpha^*)=2U\vert\alpha\vert^2\alpha - 3G\alpha^{*2}+\Delta\alpha$ and $f_2(\alpha,\alpha^*)=U\alpha^2-3G\alpha^*$, respectively, and the third line describes a three-photon process 
that can be thought of as a nonlinear (number dependent) single-photon drive.
In Eq.~\eqref{eq:H3KPRDisp} we omitted constant terms.
We now choose the value of $\alpha=\vert\alpha\vert e^{i \phi}$ such that the amplitude of the linear term $f_1(\alpha,\alpha^*)$ vanishes.
By enforcing this condition we get, besides the trivial solution $\alpha=0$, two families of solutions $\nobreak{\alpha_{\pm,k}=\vert \alpha_{\pm}\vert e^{i\phi_k}}$,
with magnitude as given in Eq.~\eqref{eq:AlphaPlusMin}
and with phase 
$\phi_k=\frac{2\pi }{3}k$, $k=0,1,2 \mod 3$. These solutions coincide with the extremal points from the semiclassical analysis in Sec.~\ref{s:System-Semiclassical}. 
Next, we consider small fluctuations around these amplitudes, i.e., we retain only terms in Eq.~\eqref{eq:H3KPRDisp} that are up to second order in the creation and annihilation
operators. The Hamiltonian can then be approximated as 
\begin{equation}\label{eq:H3KPRSq}
\hat H'_\pm\approx -\omega_\pm \hat a^\dagger \hat a -\lambda_\pm e^{-i\frac{2\pi}{3}k} \hat a^{\dagger\,2} -\lambda_\pm e^{i\frac{2\pi}{3}k} \hat a^{2}\,,
\end{equation}
with $k=0,1,2 \mod 3$. Apart from an overall minus sign, we recognize the Hamiltonian of a degenerate parametric amplifier with frequency 
\begin{equation}\label{eq:OmegaPM}
\omega_\pm=4 U\vert\alpha_\pm\vert^2 +\Delta\,,
\end{equation}
and two-photon pump strength
\begin{equation}\label{eq:LambdaPM}
\lambda_\pm=U\vert\alpha_\pm\vert \left(\vert\alpha_\pm\vert -3\frac{G}{U}\right)\,.
\end{equation}
The Hamiltonian~\eqref{eq:H3KPRSq} can be brought into diagonal form by means of a squeezing transformation, provided that the stability requirement $\omega_\pm>2 \vert \lambda_\pm\vert$ is met, otherwise the system is parametrically unstable. Enforcing this stability requirement selects only $\hat H'_+$, which is always diagonalizable, 
while $\hat H'_-$ is unstable for all values of the parameters, see Appendix~\ref{app:Explicit}. In the following, we therefore drop the subscript and make the identifications 
$\hat H'_+\equiv\hat H'$, $\omega_+\equiv\omega,\,\lambda_+\equiv\lambda$ and $\alpha_{+,k}\equiv\alpha_k$, with $k=0,1,2 \mod 3$. In particular, from now on we will indicate the magnitude $\vert\alpha_{+}\vert$ from Eq.~\eqref{eq:AlphaPlusMin} simply by $\vert\alpha\vert$, unless stated otherwise. 

We introduce the squeezing operator
$\hat S(\xi)=e^{\xi^*\frac{\hat a^{2}}{2}-\xi\frac{\hat a^{\dagger\,2}}{2}}$,
which acts as $\hat S(\xi)^\dagger \hat a \hat S(\xi) = \cosh r \,\hat a - e^{i\theta}\sinh r \,\hat a^\dagger$. The squeezing parameter $\xi=r e^{i\theta}$ is defined by the relation $\tanh 2r=2\lambda/\omega$, 
which reads 
\begin{equation}\label{eq:SqParameter}
\tanh 2r=\frac{3-\sqrt{1-\Delta/\Delta_{\rm thr}}}{3+\sqrt{1-\Delta/\Delta_{\rm thr}}}=
\frac{9G\vert\alpha \vert}{6G\vert\alpha \vert-\Delta}-1
\end{equation}

Given the multiplicity of the solutions, in our case, we have $\xi_k=r e^{i\theta_k}$, with $\theta_k=-\frac{\pi}{3}\bigl(2(k+1)+1\bigr)$, $k=0,1,2$.  Eq.~\eqref{eq:SqParameter} states that the  squeezing parameter depends on the amount of displacement.
This finding underlies the distinctive properties of the ground state of the 3KPO, as we will see later in this Section. 
By applying the squeezing transformation we get 
\begin{align}\label{H"}
\hat H''&=\hat S^\dagger(\xi_k) \hat D^\dagger(\alpha_k)\hat H \hat D(\alpha_k) \hat S(\xi_k)\nonumber \,,\\
&= -\left[ \Omega \hat a^\dagger \hat a +\left(\frac{\Omega-\omega}{2} \right) \right]  \,,
\end{align}
with frequency  
\begin{align}\label{eq:Omega}
\Omega&=\sqrt{\omega^2-4\lambda^2}\,,\nonumber \\
&=\sqrt{12\vert\alpha\vert^4U^2-3\Delta^2}\,.
\end{align}
Therefore, we conclude that, within the approximation of Eq.~\eqref{eq:H3KPRSq}, the eigenstates of the 3KPO 
\begin{equation}\label{eq:DispSqNum}
\hat H\ket{\zeta_k,n}=E_n  \ket{\zeta_k,n}\,, 
\end{equation}
are displaced squeezed number states $\ket{\zeta_k,n}= \hat D(\alpha_k)\hat S(\xi_k)\ket{n}$, where we grouped the displacement and squeezing coefficients in a single parameter $\zeta_k = (\alpha_{k},\xi_k)$. These states are threefold degenerate, with energies given by $E_n=-\left[ \Omega n +\left(\frac{\Omega-\omega}{2} \right) \right]$. Their values are marked by crosses in Fig.~\ref{f:Spectrum}(a), showing good agreement with numerical results.
These approximate values become
less accurate for higher values of $n$, as the eigenstates are less deeply confined in the potential well, and their validity breaks down for states outside of the well, namely  for $n_*\approx \vert E(\alpha_+)-E(\alpha_-)\vert/\Omega$.
Physically, obtaining the triply degenerate states Eq.~\eqref{eq:DispSqNum} corresponds to neglecting the tunneling between the
three wells in the meta-potential, see Fig.~\ref{f:Potential}~(c)-(e), which induces level splitting within each triplet of
the eigenstates~\cite{Zhang2019}.
Approximate expressions of the wavefunction of the highly excited states based on WKB approximation can be found in \cite{Zhang2019}. In the following, we instead focus on the ground state.

\subsection{Squeezed coherent states}

\begin{figure}[t!]
\centering
\includegraphics[width=1.\columnwidth]{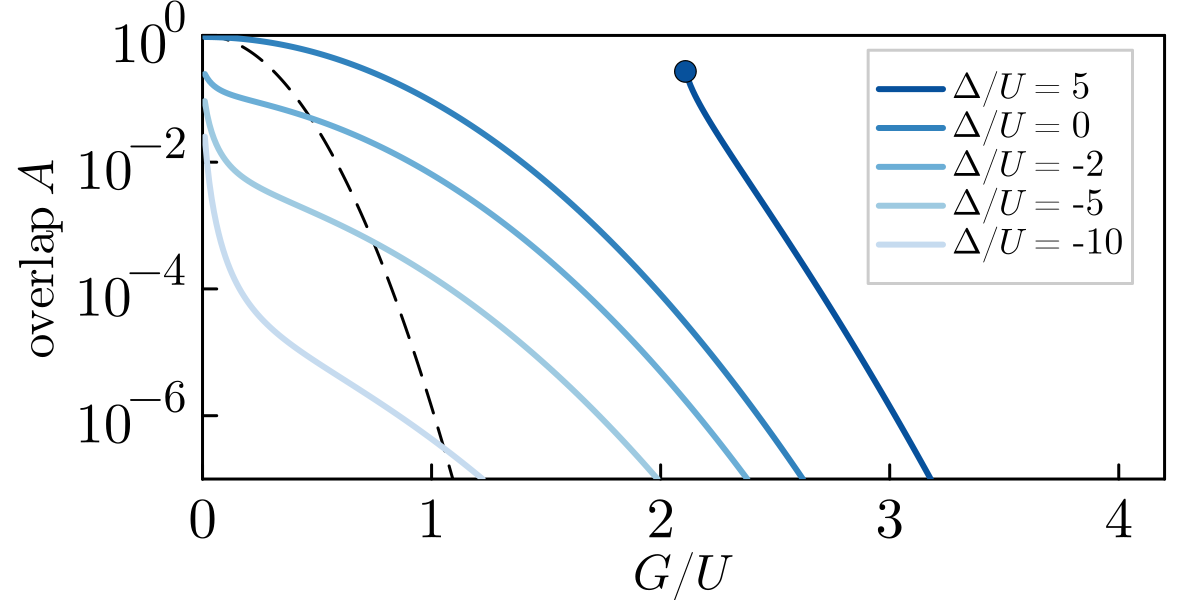}
\caption{Magnitude $A=\vert\langle \zeta_k \ket{\zeta_{k+1}} \vert$ of the overlap between squeezed coherent states $\{ \ket{\zeta_{k}}\}_{k=0,1,2}$, as given by Eq.~\eqref{eq:OverlapA}, for different values of detuning. The blue dot for $\Delta=5U$ marks the threshold value $G=G_\mathrm{thr}$, below which the squeezed coherent states are not defined. The dashed black curve corresponds to magnitude of the overlap evaluated at $\Delta=-9G^2/U$.}
\label{f:Overlap}
\end{figure}

\begin{figure*}[t!]
\centering
\includegraphics[width=1.\linewidth]{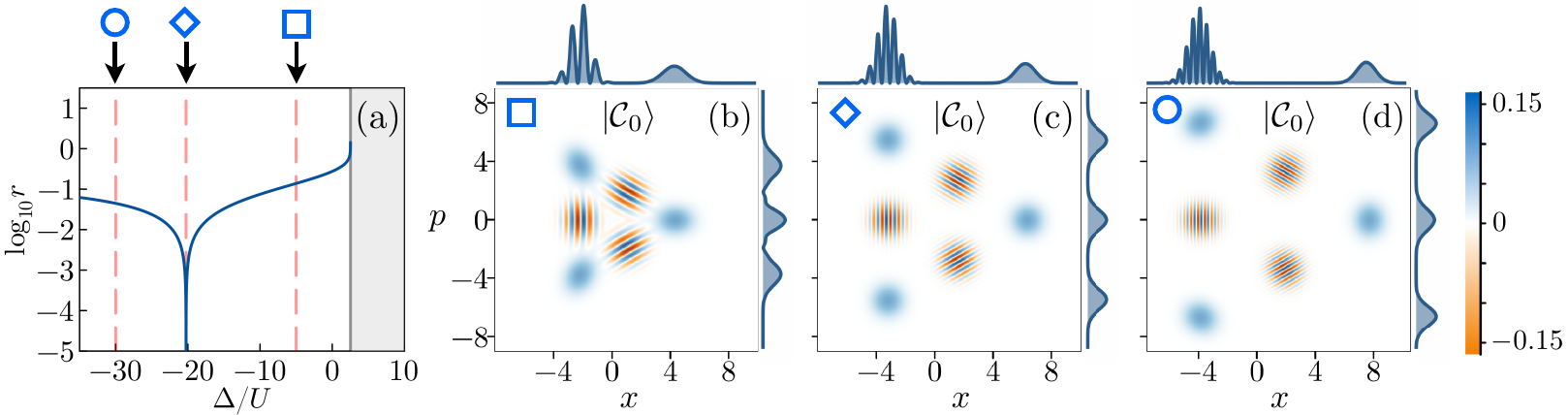}
\caption{
(a) Squeezing parameter Eq.~\eqref{eq:SqParameter} as a function of detuning for $G=1.5U$. The squeezing vanishes for $\Delta=-9 G^2/U$ and diverges for $\Delta=\Delta_\mathrm{thr}$; the gray area marks the region $\Delta>\Delta_\mathrm{thr}$ with no multi-stability. (b)-(d) Wigner function of the $k=0$ squeezed cat state $\ket{\mathcal{C}_0}$ at the values of detuning specified by the dashed lines in panel (a). For vanishing squeezing, the Wigner function of a three-legged cat state (c) is recovered.}
\label{f:WignerDelta}
\end{figure*}

The vacuum manifold is  
spanned by the states $\ket{\zeta_k,0}\equiv\ket{\zeta_k}$ and is separated from the first excited manifold by an energy gap $\Omega$. The  states $\ket{\zeta_k}$ are an instance of squeezed coherent states. We explicitly write them below, together with the corresponding value of the displacement and squeezing phase 
\begin{align}
\ket{\zeta_0}&=\hat D(\alpha_0) \hat S(\xi_0) \ket{0}\,, &\phi_0=0\,, \theta_0=\pi\,, \label{eq:DS0} \\
\ket{\zeta_1}&=\hat D(\alpha_1) \hat S(\xi_1) \ket{0}\,,  &\phi_1=\frac{2\pi}{3}\,, \theta_1=\frac{\pi}{3}\,, \label{eq:DS1} \\
\ket{\zeta_2}&=\hat D(\alpha_2) \hat S(\xi_2) \ket{0}\,, &\phi_2=-\frac{2\pi}{3}\,, \theta_2=-\frac{\pi}{3}\,. \label{eq:DS2}
\end{align}

The states have the same average number of photons $\langle \zeta_k\vert \hat a^\dagger \hat a\ket{\zeta_k}=\vert \alpha\vert^2+\sinh^2 r$ and are related to one another via the discrete rotations $\hat Z=e^{i \frac{2\pi}{3}\hat a^\dagger \hat a}$ and $\hat Z^2$. 
In fact, one has $\hat Z \ket{\zeta_k} = \ket{\zeta_{k+1}}$, which follows from the relation $\hat Z^\dagger \hat D(\alpha_k)\hat S(\xi_k) \hat Z = \hat D(\alpha_{k-1})\hat S(\xi_{k-1})$. 
It follows that different $\ket{\zeta_k}$ are characterized by the same values of $\vert \alpha\vert$ and $r$, since the dependence on $k=0,1,2,$ only appears in the phases. 

The states in Eqs.~\eqref{eq:DS0}-\eqref{eq:DS2} are not orthogonal and their overlap is given by $\langle \zeta_k \ket{\zeta_{k+1}}=A e^{i\Theta}$  with magnitude and phase given by (see Appendix~\ref{app:SqCohOverlap} for the details of the calculation)
\begin{align}
A&=\frac{2^{\frac34}}{(5+3\cosh 4r)^{\frac14}}\exp\left(-\frac{3 \vert \alpha\vert^2}{2\cosh 2r +\sinh 2r}\right)\,,\label{eq:OverlapA} \\
\Theta&= \frac{2\sqrt3 \vert \alpha\vert^2}{1+3 e^{4r}}-\frac{1}{2}\mathrm{Arg}(\cosh^2 r+(-1)^\frac13\sinh^2 r)\,. \label{eq:OverlapTheta}
\end{align}

From Eq.~\eqref{eq:OverlapA} we see that the magnitude of the overlap is exponentially suppressed in $\vert\alpha\vert^2$, while the dependence is non-monotonic in $r$. We should keep in mind, however, that the amplitude of the displacement and the squeezing parameter are not allowed to vary independently; Eq.~\eqref{eq:SqParameter} imposes a constraint between the two quantities. In Fig.~\ref{f:Overlap}, we show the behaviour of $A$ as a function of the pump strength for different values of the detuning. 
For fixed pump strength, the magnitude of the overlap  decreases as the detuning is reduced.
This is consistent with the behavior of the  displacement amplitudes from Eq.~\eqref{eq:AlphaPlusMin}, which shows that the separation between the squeezed coherent states increases accordingly.

Remarkably, Fig.~\ref{f:Overlap} also shows a change in the concavity of the overlap when crossing the dashed black curve, for negative values of the detuning. This behaviour cannot be accounted for by a change in the displacement amplitudes alone.
To understand its origin, we then turn our attention to the squeezing parameter Eq.~\eqref{eq:SqParameter}, which is plotted in Fig.~\ref{f:WignerDelta}(a) as a function of the detuning.
Starting from positive values of detuning $\Delta>0$, the squeezing phase $\theta_k$ is half the value of the displacement phase $\phi_k$, see Eqs.~\eqref{eq:DS0}-\eqref{eq:DS2}. This means that the squeezing direction is orthogonal to the direction of the displacement or, in other words, that the states $\ket{\zeta_k}$, $k=0,1,2$  are stretched (anti-squeezed) in phase space along the direction of the displacement; see the lobes of the Wigner functions in Fig.~\ref{f:CerberusExact}(a)-(c). The anti-squeezing contributes to an increase in $A=\vert \langle \zeta_k \ket{\zeta_{k+1}}\vert$ relative to a reference overlap between non-squeezed states, i.e., compared to the overlap $\vert\langle \alpha_k \ket{\alpha_{k+1}}\vert$ between two coherent states with the same amplitudes. 
This contribution becomes increasingly weaker in relative terms for larger values of the pump, as the squeezing parameter saturates to the constant value $r\rightarrow \frac{\log3}{4}\approx 0.275$ in the limit $G/U \rightarrow +\infty$.

Upon decreasing the detuning and crossing into the negatively detuned regime, the squeezing parameter is progressively reduced until it exactly vanishes. This happens for  $\Delta=-9 G^2/U$, which corresponds to the dotted curve in the phase diagram of Fig.~\ref{f:Potential}(a). The value of the overlap $A$ along this curve is marked by the dashed line in Fig.~\ref{f:Overlap} and the squeezed coherent states in Eqs.~\eqref{eq:DS0}-\eqref{eq:DS2} reduce to three coherent states, $\ket{\zeta_k}\rightarrow \ket{\alpha_k}$. 

By further decreasing the detuning past $\Delta=-9 G^2/U$, the squeezing parameter changes sign or, equivalently, the squeezing phase acquires a $\pi$-shift. In phase space, the states $\ket{\zeta_k}$ are now squeezed along the direction of the displacement. This change in orientation of the squeezing reduces the overlap 
at a faster rate compared to two coherent states of the same amplitude
and hence explains the observed change in the concavity of $A$.
For $\Delta/U \rightarrow -\infty$ (and finite $G$) the squeezing parameter diverges logarithmically as $r\approx \frac{1}{8}\log(\frac{9\Delta_\mathrm{thr}}{\vert\Delta\vert})$.

This characteristic transition between squeezing and anti-squeezing of quantum fluctuations when varying  the detuning is directly reflected in the properties of the corresponding superposition states, as we show next.

\subsection{Three-legged squeezed cat states}

We define the three-legged squeezed cat states as the projected squeezed coherent states
\begin{align}\label{eq:ProjectedState}
\ket{\mathcal{C}_k}&=3\mathcal{N}_k \hat \Pi_k \ket{\zeta_0} \,, 
\end{align}
with $k=0,1,2$, where $\hat \Pi_k = \sum_{n=0}^{+\infty} \ket{3n+k}\bra{3n+k}$ 
is the projector over the $k$-th  symmetry sector  
and the normalization is given by $\mathcal{N}_k=\bigl[3\bigl(1+2A\cos\bigl(\Theta-\frac{2\pi}{3}k\bigr)\bigr)\bigr]^{-1/2}$.
The states $\ket{\mathcal{C}_k}$ are 
exactly orthogonal for any value of $G$ and $\Delta$, thus forming an orthonormal basis of the vacuum manifold.
They do not in general coincide with the true ground state of the 3KPO, since they neglect contributions from the nonlinear terms in Eq.~\eqref{eq:H3KPRDisp}.  
We will verify the accuracy of this approximation later in Section~\ref{subs:Benchmark}.

The explicit expressions of the three-legged squeezed cat states in terms of the squeezed coherent states Eqs.~\eqref{eq:DS0}-\eqref{eq:DS2} read
\begin{align}
\ket{\mathcal{C}_0}&=\mathcal{N}_0 \left(\ket{\zeta_0}+\ket{\zeta_1}+\ket{\zeta_2}\right) \,, \label{eq:Cerb0} \\
\ket{\mathcal{C}_1}&=\mathcal{N}_1 \left(\ket{\zeta_0} + e^{i \frac{4\pi}{3}}\ket{\zeta_1} + e^{i \frac{2\pi}{3}}\ket{\zeta_2}\right)\, , \label{eq:Cerb1}\\
\ket{\mathcal{C}_2}&=\mathcal{N}_2 \left(\ket{\zeta_0} + e^{i \frac{2\pi}{3}}\ket{\zeta_1} + e^{i \frac{4\pi}{3}}\ket{\zeta_2}\right)\, , \label{eq:Cerb2}
\end{align}
or more compactly
\begin{equation}\label{eq:Cerb}
\ket{\mathcal{C}_k}=\mathcal{N}_k \sum_{j=0}^{2} \omega^{-kj}\,\ket{\zeta_j}\,,
\end{equation}
with $\omega=e^{i 2\pi/3}$ the cube root of unity, a notation which we adopt from here on.
Conversely, the discrete Fourier transform of three-legged squeezed cats 
\begin{equation}\label{eq:ZetakApprox}
\frac{1}{\sqrt3}\sum_{k=0}^{2}\omega^{jk}\ket{\mathcal{C}_k} \approx \ket{\zeta_j}\,,
\end{equation}
with $j=0,1,2$, coincides with squeezed coherent states $\ket{\zeta_j}$, up to an additive correction proportional to $A$.  

In Fig.~\ref{f:WignerDelta}(b)-(d) we show the Wigner function of the state $\ket{\mathcal{C}_0}$ for different choices of  detuning; the Wigner functions of the other two states, $\ket{\mathcal{C}_1}$, $\ket{\mathcal{C}_2}$, (not shown) differ only by a relative shift in the interference fringes. These plots confirm that, by continuously varying the detuning, both the orientation and the magnitude of the squeezing are modified, thereby qualitatively altering the properties of the squeezed cat states. The parametric dependence of the squeezing on the detuning thus enables the preparation of distinct macroscopic quantum superpositions. In particular, Fig.~\ref{f:WignerDelta}(c)
recovers the Wigner function of a three-legged Schr\"odinger's cat state, i.e., a quantum superposition of three coherent states with amplitude $\vert\alpha_+\vert$. 
Three-component Schr\"odinger's cat states are known to be stabilized in the ground state of a parametrically driven nonlinear resonator involving only $3$-photon processes, i.e., $\nobreak{\hat H_3= -U_3 \hat a^{\dagger\,3}\hat a^{3}+G \left(\hat a^{\dagger\,3}+\hat a^{3}\right)}$~\cite{Puri2017Kerr}. This approach however requires a sixth-order photon interaction. Remarkably, here three-legged cat states are instead stabilized in the quasi-degenerate ground state of a Kerr oscillator by judiciously choosing the detuning. 
Single instances of Wigner functions of the ground state of the 3KPO, obtained by numerically diagonalizing  Eq.~\eqref{eq:H3KPR}, appeared in recent works~\cite{Lorch2019, Xiao2024}. 

\begin{figure}[t!]
\centering
\includegraphics[width=1.\columnwidth]{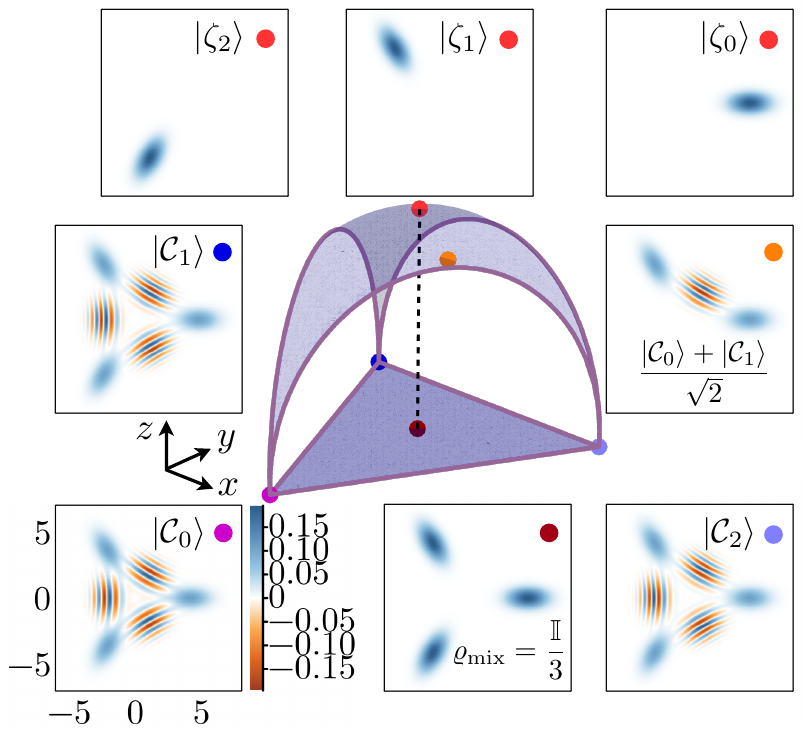}
\caption{The ground state manifold of the 3KPO hosts a qutrit. At the centre is shown the three-dimensional representation of a qutrit, according to Ref.~\cite{Eltschka2021}, an extension of the Bloch sphere representation of a qubit; the Cartesian coordinates indicate expectation values of (combination of) Gell-Mann matrices. The manifold is spanned by the three-legged squeezed cats $\ket{\mathcal{C}_k}$, $k=0,1,2$, Eqs.~\eqref{eq:Cerb0}-\eqref{eq:Cerb2}, and their superpositions. Three-legged squeezed cat states lie at the vertexes of the base triangle, while all other pure states lie on the spherical surface connected to the vertexes. In particular, the squeezed coherent states $\ket{\zeta_k}$, Eqs.~\eqref{eq:DS0}-\eqref{eq:DS2}, are mapped to the same point, on the top of the surface. The  volume underneath the surface as well as all points of the base triangle (except from the vertices) correspond to mixed states, with the maximally mixed state $(\ket{\mathcal{C}_0}\bra{\mathcal{C}_0}+\ket{\mathcal{C}_1}\bra{\mathcal{C}_1}+\ket{\mathcal{C}_2}\bra{\mathcal{C}_2})/3 \approx \mathds{1}/3$ lying at the centre of the triangle. Snapshots of the Wigner function for the states marked by the dots are also shown, for $\Delta=1.5U$ and $G=2.2 U$.
}
\label{f:QutritBall}
\end{figure}

We encode a qutrit in the ground state manifold, taking the three-legged squeezed-cat states as codewords
$\ket{k_L}\equiv\ket{\mathcal{C}_k}$, with $k=0,1,2$. We refer to the 3KPO with this encoding as a Kerr-cat qutrit. We introduce the generalized Pauli operators for a qutrit~\cite{Gottesman1999_higherdim, Wang2020_quditrev}, known as clock $\hat Z_L$ and shift $\hat X_L$ operators, which act as 
\begin{align}
\hat Z_L\ket{k_L}&=\omega^{k}\ket{k_L}\,, \label{eq:Z_L} \\
\hat X_L\ket{k_L}&=\ket{(k-1)_L}\,, \label{eq:X_L}
\end{align}
and satisfy the relations $\hat X_L^3=\hat Z_L^3=\mathds{1}$, $\hat X_L^\dagger=\hat X_L^{2}$ and $\hat Z_L\hat X_L=\omega^{-1}\,\hat X_L \hat Z_L$.
 
Since $\ket{\mathcal{C}_k}$ are eigenstates of the discrete rotation $\hat Z$,
 the physical operator $\hat Z$ is identified with the logical clock, i.e., $\hat Z_L=\sum_k \omega^k \ket{\mathcal{C}_k}\bra{\mathcal{C}_k}$.
This identification is exact and holds for any value of the pump.
Given that the eigenstates of clock and shift operators are  related via discrete
Fourier transform,
the squeezed coherent states $\ket{\zeta_j}$ are, up to exponentially small corrections  in
$|\alpha|^2$, eigenstates of the logical shift operator, i.e.,  $\hat X_L=\sum_{j}\omega^{j}\ket{\zeta_j}\bra{\zeta_j}$.
From now on we drop the subscript and  $\hat Z$, $\hat X$ act as follows:  
\begin{align}
\hat Z\ket{\mathcal{C}_k}&=\omega^k\ket{\mathcal{C}_k}\,, \qquad & \hat Z\ket{\zeta_j}=\ket{\zeta_{j+1}}\,, \label{eq: Z_Lcatzeta} \\
\hat X\ket{\mathcal{C}_k}&=\ket{\mathcal{C}_{k-1}}\,,  \qquad &\hat X\ket{\zeta_j}=\omega^j\ket{\zeta_{j}}\,. \label{eq: X_Lcatzeta}
\end{align}

In the large amplitude limit, the three-legged squeezed cat states are macroscopic quantum superpositions that encode quantum information non-locally in the phase space of the 3KPO, analogously to the cat-qubit encoding in the 2KPO~\cite{Puri2017Kerr, puri2019, grimm2020}. A similar regime has been explored in a recent experimental work~\cite{kwon_realisation_2026}, albeit at small pump amplitudes where a triply degenerate ground state is absent.  
Visualizing high dimensional quantum states is notoriously difficult.
In Fig.~\ref{f:QutritBall} we show an analogue of the Bloch sphere representation for a qutrit. Unlike the actual qubit Bloch sphere, where any point on the surface or in the interior of the sphere corresponds to one and only one state, in the qutrit case the space of states is eight dimensional and therefore cannot be faithfully represented in three dimensions. Here we adopt the three-dimensional representation proposed in Ref.~\cite{Eltschka2021}. We briefly illustrate the structure and refer to Ref.~\cite{Eltschka2021} for all the technical details.
The quantities plotted on the three axes are $x=\tr{\hat \varrho \lambda_3}$, $y=\tr{\hat \varrho \lambda_8}$, where $\lambda_3$ and $\lambda_8$ indicate the diagonal Gell-Mann matrices,  while $z=\sqrt{\sum_{j\neq 3,8}\tr{\hat \varrho \lambda_j}^2}$ is chosen to be a combination of the remaining Gell-Mann matrices~\cite{Eltschka2021}. Three-legged squeezed cat states sit at the vertexes of the equilateral triangle lying in the $xy$ plane. All pure states live on the spherical surface connecting the three vertexes. In particular, all three squeezed coherent states $\ket{\zeta_0}$, $\ket{\zeta_1}$, $\ket{\zeta_2}$ are mapped to the same point, i.e., the head of the dome. The same is true for all pure states $\ket{\psi}$ such that $\vert\langle \mathcal{C}_k \ket{\psi}\vert^2=1/3$ (mutually unbiased basis states). This is due to the above-mentioned impossibility of a faithful representation, by which quasi-orthogonal states are not necessarily mapped to distinct points.  Analogously to the qubit Bloch sphere, mixed states lie inside the volume enclosed within the surface, the triangular basis, and the semicircular cuts, with the maximally mixed state $(\ket{\mathcal{C}_0}\bra{\mathcal{C}_0}+\ket{\mathcal{C}_1}\bra{\mathcal{C}_1}+\ket{\mathcal{C}_2}\bra{\mathcal{C}_2})/3 \approx \mathds{1}/3$ lying at the centre of the triangle. Finally, the semicircular cuts correspond to sections of the qubit Bloch sphere spanned by pairs of three-legged squeezed cat states. Similar to the qubit case, for a given cut the pure states live on the arc. For instance, in the mid-right panel of Fig.~\ref{f:QutritBall} we show the pure state $(\ket{\mathcal{C}_0}+\ket{\mathcal{C}_1})/\sqrt2$. The projection of that state on the $xy$ plane would be the mixed state $(\ket{\mathcal{C}_0}\bra{\mathcal{C}_0}+\ket{\mathcal{C}_1}\bra{\mathcal{C}_1})/2$.

\subsection{Comparison with numerics}\label{subs:Benchmark}

\begin{figure*}[t!]
\centering
\includegraphics[width=1.\linewidth]{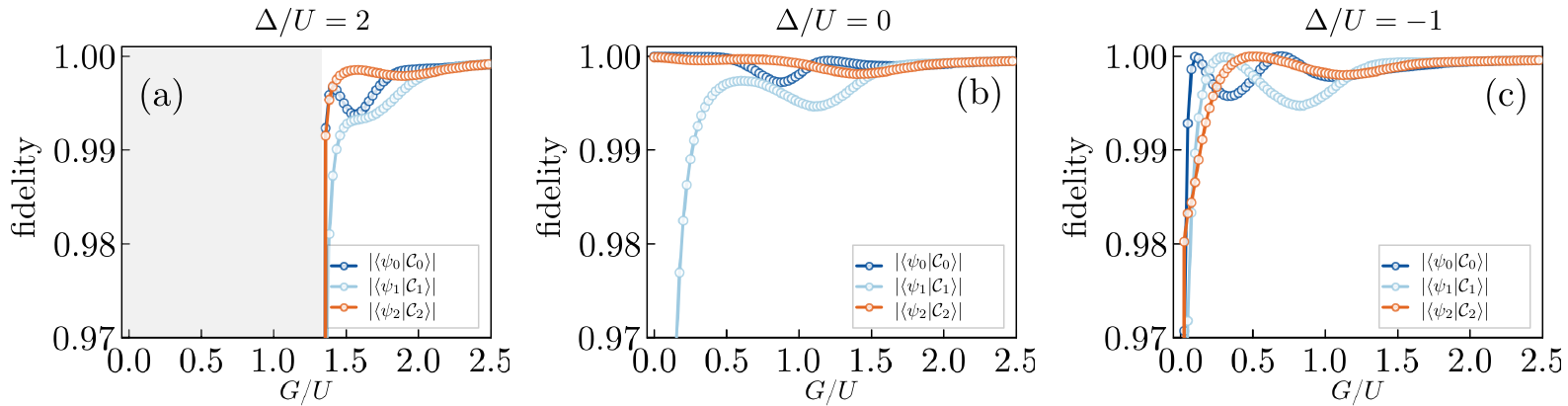}
\caption{
Fidelity $\vert \langle \psi_k\ket{\mathcal{C}_k} \vert$ between the lowest energy state  in the $k$-th symmetry sector, $\ket{\psi_k}$, obtained via numerical diagonalization of Eq.~\eqref{eq:H3KPR}, and the approximate expression of the corresponding three-legged squeezed cat state $\ket{\mathcal{C}_k}$, see Eqs.~\eqref{eq:Cerb0}-\eqref{eq:Cerb2}, for $k=0,1,2$ and for: (a) $\Delta/U=2$, (b) $\Delta/U=0$ and (c) $\Delta/U=-1$. The gray area in panel (a)  marks the region $G<G_\mathrm{thr}$ below the threshold for multi-stability.
}
\label{f:Fidelity}
\end{figure*}

We now benchmark the expressions of the three-legged squeezed cat states Eqs.~\eqref{eq:Cerb0}-\eqref{eq:Cerb2} obtained within our approximate treatment against the actual ground state of the 3KPO Hamiltonian Eq.~\eqref{eq:H3KPR}. Results for three representative cases with positive, zero and negative detuning are shown in Fig.~\ref{f:Fidelity}(a)-(c). We plot the fidelity $\vert \langle \psi_k \ket{\mathcal{C}_k} \vert$ between the lowest energy state in the
$k$-th symmetry sector, obtained numerically, and the corresponding target three-legged squeezed cat state, for $k=0,1,2$. 
For all values of the detuning, unit fidelity is approached for moderately large values of the pump strength. This confirms that, for large enough displacements, the three-legged squeezed cat states are excellent approximations of the corresponding ground states of the 3KPO. This holds true also in the special case $\Delta=G^2/U$ discussed in Sec.~\ref{s:Exact}. 
By decreasing the pump strength, the quasi-degeneracy of the exact spectrum is progressively lifted [cf.  Fig.~\ref{f:Spectrum} (b)], generically leading to sub-unit fidelity at intermediate pump strengths. More dramatically, all three panels of Fig.~\ref{f:Fidelity} also display abrupt drops in the fidelity. These drops occur for different reasons, as we explain below.

For $\Delta>0$, see Fig.~\ref{f:Fidelity}(a), all three fidelities drop in proximity of the threshold value $G_\mathrm{thr}$.
This behaviour can be understood by inspecting the squeezing parameter Eq.~\eqref{eq:SqParameter}. The squeezing parameter $r$ diverges when approaching the threshold value $G_\mathrm{thr}$ from above. To leading order, we find $r\approx -\tfrac18 \log(G-G_\mathrm{thr})$. However, this divergence is an artifact of the approximate treatment restricted to quadratic fluctuations and does not occur in the full model~\eqref{eq:H3KPR}, which instead approaches the threshold with finite squeezing, and hence finite energy. The inadequacy of the approximate description in proximity of  the threshold is therefore responsible for the observed drop in the fidelity. A similar scenario occurs for $\Delta<0$ when approaching $G\rightarrow 0$, see Fig.~\ref{f:Fidelity}(c). In this limit, the approximate treatment  predicts an unphysical  divergence of the squeezing parameter as $r\approx -\tfrac18 \log(G-G_\mathrm{thr})$. 

Different is the case of  $\Delta=0$, shown in Fig.~\ref{f:Fidelity}(b), for which only the fidelity $\vert \langle \psi_1 \ket{\mathcal{C}_1} \vert$ drops to zero for $G\rightarrow 0$. For $\Delta=0$ the squeezing parameter Eq.~\eqref{eq:SqParameter} takes the constant value $r=\tfrac12 \mathrm{arctanh}\left(\frac12\right)$. In the approximate treatment, this implies that for $G\rightarrow0$ the three-legged squeezed cat states reduce to the following  superpositions 
\begin{align}\label{eq:LimitCerb}
\ket{\mathcal{C}_k}&=\frac{\mathcal{N}_k}{\sqrt{\cosh r}}\sum_{n=0}^{+\infty}c_{n,k} \ket{2n},
\end{align}
with $c_{n,k}=\left[1+2\cos\left(\frac{4\pi}{3}(n+k) \right)\right](\tanh r)^n\frac{\sqrt{(2n)!}}{2^n n!}$. Substituting the value of the squeezing parameter, we find the following: for $k=0,2$  Eq.~\eqref{eq:LimitCerb} is well approximated by the Fock state $\ket{\mathcal{C}_0}\approx\ket{0}$ and $\ket{\mathcal{C}_2}\approx\ket{2}$, respectively, while for $k=1$ we instead have $\ket{\mathcal{C}_1}\approx\ket{4}$.
On the other hand, it is clear that in the limit of vanishing pump strength, the actual ground state reduces to $\ket{\psi_k}\rightarrow\ket{n=k}$, with $k=0,1,2$. This explains why Fig.~\ref{f:Fidelity}(b) shows near-unit fidelity for $k=0,2$, but a vanishing fidelity for $k=1$.

\section{Effect of single-photon processes}\label{s:SinglePhoton}

The 3KPO is unavoidably subject to  external perturbations. To examine how noise channels affect the 3KPO, we consider the effect of coupling to a dissipative bath. The creation and annihilation operators act on the squeezed coherent states as follows 
\begin{align}
\hat a\ket{\zeta_k}&=\alpha_k \ket{\zeta_k}-e^{i\theta_k}\sinh r \ket{\zeta_k,1}\,, \label{eq:aOnSqCoh}\\
\hat a^\dagger\ket{\zeta_k}&=\alpha_k^* \ket{\zeta_k}+\cosh r \ket{\zeta_k,1}\,, \label{eq:adagOnSqCoh}
\end{align}
From the orthogonality relation  $\bra{\zeta_k,n}\zeta_k,m\rangle=\delta_{nm}$, it follows that $\bra{\zeta_k}\hat a \ket{\zeta_k}=\alpha_k$, $\bra{\zeta_k}\hat a^\dagger \ket{\zeta_k}=\alpha_k^*$. 

Equivalently, we can look at the action of single-photon processes on the three-legged squeezed cats, which is given by 
\begin{align}
\hat a\ket{\mathcal{C}_k}&= \vert\alpha\vert\frac{\mathcal{N}_k}{\mathcal{N}_{k-1}} \ket{\mathcal{C}_{k-1}} + \sinh r \frac{\mathcal{N}_k}{\mathcal{N}_{k+1}^{(\mathrm{e},1)}}\ket{\psi_{k-1}^{(\mathrm{e},1)}}\,, \label{eq:aOnCerb} \\ 
\hat a^\dagger\ket{\mathcal{C}_k}&= \vert\alpha\vert\frac{\mathcal{N}_k}{\mathcal{N}_{k+1}} \ket{\mathcal{C}_{k+1}} + \cosh r \frac{\mathcal{N}_k}{\mathcal{N}_{k}^{(\mathrm{e},1)}}\ket{\psi_{k+1}^{(\mathrm{e},1)}}\,, \label{eq:adagOnCerb}
\end{align}
where the second terms on the right-hand side are superpositions states in the first excited state manifold 
\begin{equation}\label{eq:SupExcited}
\ket{\psi_{\ell}^{(\mathrm{e},1)}} = \mathcal{N}_{\ell-1}^{(\mathrm{e},1)}\left( \ket{\zeta_0,1}+\omega^{-\ell+1}\ket{\zeta_1,1}+\omega^{\ell-1}\ket{\zeta_2,1}\right), 
\end{equation}
with $\mathcal{N}_{\ell-1}^{(\mathrm{e},1)}$ the normalization factor. Like the three-legged squeezed cat states, these excited states satisfy $\hat Z \ket{\psi_{k}^{(\mathrm{e},1)}}=\omega^{k} \ket{\psi_{k}^{(\mathrm{e},1)}}$, i.e., they  are simultaneous  eigenstates  of $\hat H$ and $\hat Z$. Analogously to Eq.~\eqref{eq:ProjectedState}, they are obtained by projecting a squeezed and displaced single-photon Fock state onto the distinct $\mathbb{Z}_3$ symmetry sectors according to  $\ket{\psi_{k}^{(\mathrm{e},1)}}=3\mathcal{N}_{k-1}^{(\mathrm{e},1)} \hat \Pi_k \ket{\zeta_0,1}$. One can readily extend this construction to  obtain $\mathbb{Z}_3$ symmetric eigenstates of the entire 3KPO spectrum Eq.~\eqref{eq:DispSqNum}, but here we will focus on the first excited state manifold.

The right-hand side of both  Eqs.~\eqref{eq:aOnSqCoh}, \eqref{eq:adagOnSqCoh} and Eqs.~\eqref{eq:aOnCerb}, \eqref{eq:adagOnCerb}   features a contribution from the vacuum manifold and a term that describes leakage to the excited state manifold. The amplitude for leaking outside the vacuum manifold is characterized by different squeezing prefactors. In the limit of vanishing squeezing $r \rightarrow 0$, we see that only photon creation induces leakage outside of the code space, as expected for bosonic codes based on coherent states~\cite{puri2019}. 
Given that, for any finite value of the pump strength, the squeezing does not vanish except for $\Delta=-9G^2/U$, a general feature characterizing the 3KPO is a non-zero leakage outside the code space induced by single photon annihilation~\cite{Zhang2019}. 
As a reference case, for $\Delta>0$ and large three-photon pump, where the squeezing parameter saturates to $r\rightarrow \frac{\log3}{4}$, the two squeezing amplitudes approach the values 
$\sinh{r}\rightarrow \sqrt{3^{-1/2}-2^{-1}}\approx 0.28$ and $\cosh{r}\rightarrow \sqrt{3^{-1/2}+2^{-1}}\approx 1$, respectively. 

Armed with the above expressions, we can evaluate the transition amplitudes between two different logical states due to single-photon perturbations, which read

\begin{widetext}
\begin{align}
\bra{\mathcal{C}_l}\hat a \ket{\mathcal{C}_k}&= \delta_{l,k-1}\left[ \vert \alpha\vert \frac{\mathcal{N}_{k}}{\mathcal{N}_{k-1}} + 6 \sinh r \mathcal{N}_{k} \mathcal{N}_{k-1} \mathrm{Re}\left(\omega^{-(k+1)}\bra{\zeta_0}\zeta_{1},1\rangle\right) \right] \,, \label{eq:aTransCerb}\\
\bra{\mathcal{C}_l}\hat a^\dagger \ket{\mathcal{C}_k}&= \delta_{l,k+1}\left[  \vert \alpha\vert \frac{\mathcal{N}_{k}}{\mathcal{N}_{k+1}} + 6 \cosh r \mathcal{N}_{k} \mathcal{N}_{k+1} \mathrm{Re}\left(\omega^{-k}\bra{\zeta_0}\zeta_{1},1\rangle \right) \right]  \,. \label{eq:adagTransCerb}
\end{align}
\end{widetext}
Importantly, the excited state manifold contributes to the transition amplitude between three-legged squeezed cat states, since the overlap $\bra{\zeta_k,n}\zeta_{\ell\neq k},m\rangle\neq\delta_{nm}$ between displaced squeezed number states with different amplitudes is non-zero. 
The analytical expressions for the overlaps between two distinct squeezed displaced number states are given in Appendix~\ref{app:SqDispNumOverlap}. Although not obvious from  Eqs.~\eqref{eq:aTransCerb}, ~\eqref{eq:adagTransCerb}, one can use the relations provided in Appendix~\ref{app:SqDispNumOverlap} to verify that  $\bra{\mathcal{C}_l}\hat a \ket{\mathcal{C}_k}=\bra{\mathcal{C}_k}\hat a^\dagger \ket{\mathcal{C}_l}^*$. 

We discuss the main features of Eqs.~\eqref{eq:aTransCerb}, ~\eqref{eq:adagTransCerb}. The first terms on the right-hand side describe the contributions to the transition amplitudes from within the vacuum manifold, which are proportional to the coherent amplitude $\vert\alpha\vert$ and controlled by the ratio of normalization factors. These ratios approach unity up to an exponentially small correction 
proportional to the overlap  Eq.~\eqref{eq:OverlapA}, 
namely
\begin{equation}\label{eq:Nratioapprox}
\frac{\mathcal{N}_{k+1}}{\mathcal{N}_{k}}\approx1-\sqrt3\, A \sin\left(\Theta-\frac{(2k+1)\pi}{3}\right)\,,
\end{equation}
with $k=0,1,2$.

The second terms on the right-hand side of Eqs.~\eqref{eq:aTransCerb}, ~\eqref{eq:adagTransCerb} are instead the contributions from the excited state manifold, which are controlled by: (i) the squeezing amplitude, whose behavior we discussed above; (ii) the product of normalization factors, which, similarly to their ratio, can be approximated as $\mathcal{N}_{k} \mathcal{N}_{k\pm1}\approx 1/3$ up to an additive correction proportional to $A$; (iii) the  overlap $\bra{\zeta_0}\zeta_{1},1\rangle$, whose modulus is given by 
\begin{align}
\vert\bra{\zeta_0}\zeta_{1},1\rangle\vert&=2 \sqrt{\frac{2 \cosh 2r-\sinh 2r}{5/3+\cosh 4r}}\vert\alpha\vert A\,.\label{eq:OverlapExc}
\end{align}

Putting everything together, we conclude that for large enough amplitudes, the contributions causing leakage can be safely neglected, as they are exponentially suppressed in $\vert \alpha\vert^2$ due to the presence of $A$ in Eq.~\eqref{eq:OverlapExc}. 
In the large-amplitude regime and upon restricting to the vacuum manifold, the single-photon processes act like 
\begin{align}
\hat a/\vert \alpha\vert & \approx\sum_k \ket{\mathcal{C}_{k}} \bra{\mathcal{C}_{k+1}}\equiv \hat X \,, \label{eq:a} \\
\hat a^\dagger/\vert \alpha\vert &\approx  \sum_k\ \ket{\mathcal{C}_{k+1}} \bra{\mathcal{C}_{k}}\equiv\hat X^\dagger \,. \label{eq:adag}
\end{align}

\begin{figure}[t!]
\centering
\includegraphics[width=1\columnwidth]{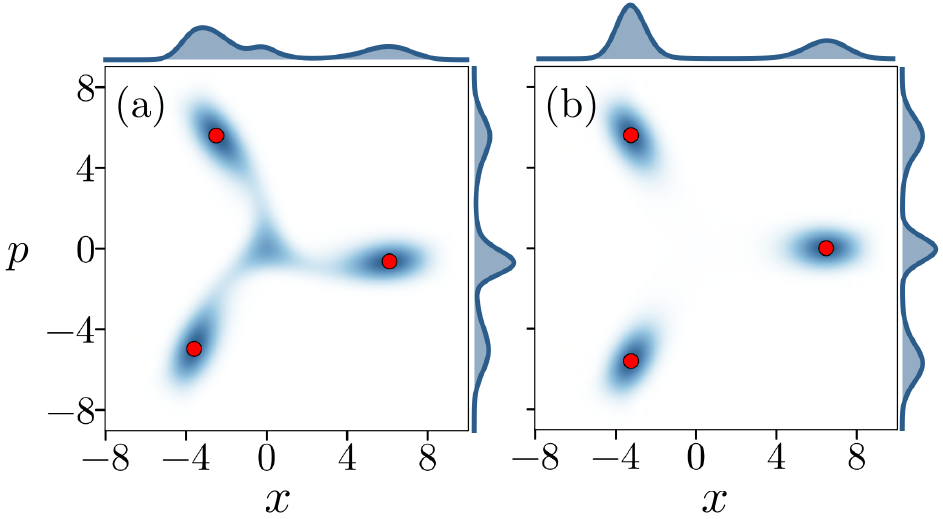}
\caption{Steady-state Wigner function of the 3KPO subject to photon loss at rate $\kappa$ for (a) $\kappa=25 U$ and (b) $\kappa=0.01 U$. Other parameters are $\Delta=2U$ and $G=3.2U$. The red dots correspond to the analytical expressions for the amplitudes given by Eq.~\eqref{eq:Nss} and Eq.~\eqref{eq:Phiss}.
}
\label{f:WignerSteadyState}
\end{figure}

\begin{figure*}[t]
\centering
\includegraphics[width=1.\textwidth]{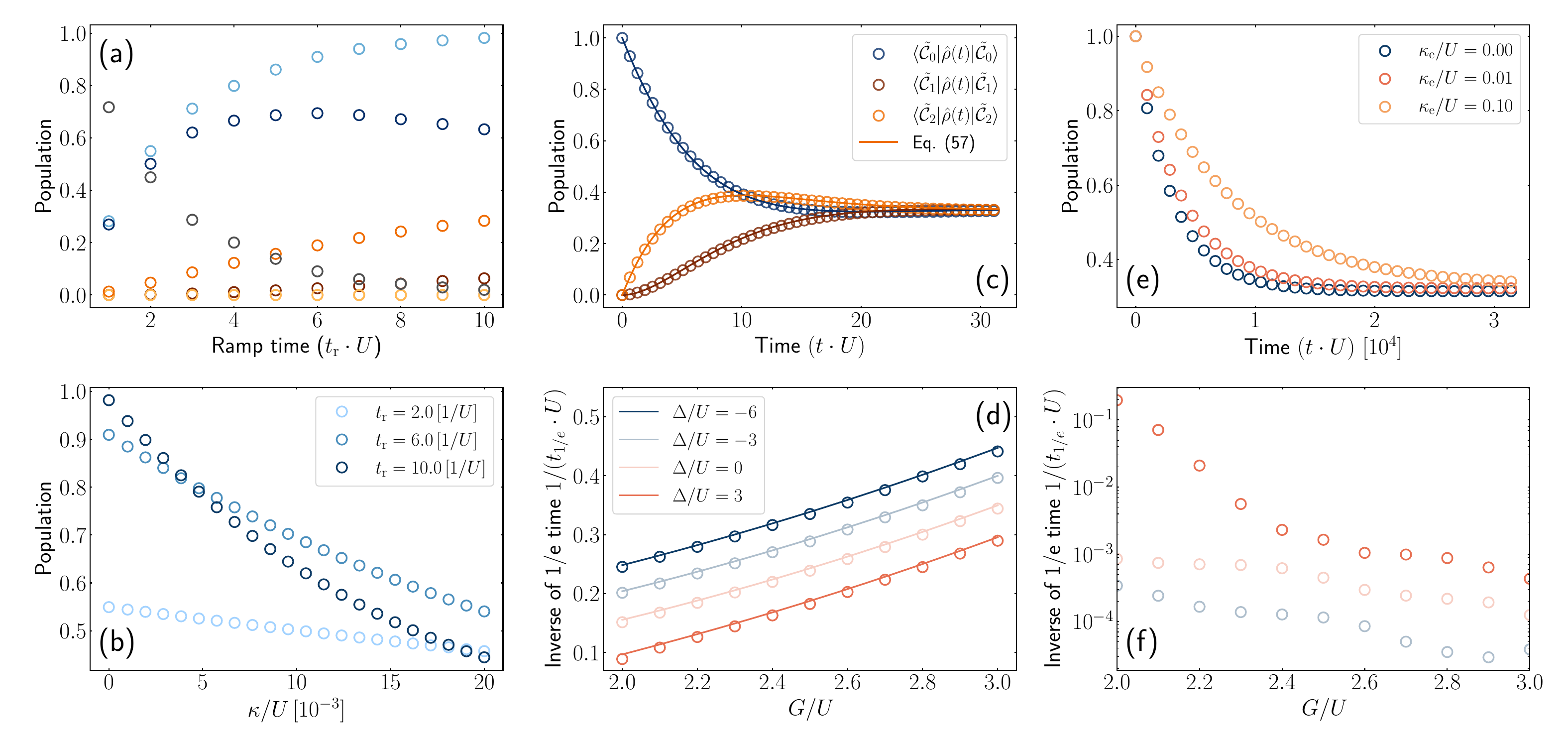}
\caption{\label{f:Cerbspd}
State initialization and evolution under single-photon loss.
        (a) Populations of the three-legged squeezed cat states $|\tilde{\mathcal{C}}_0\rangle$ (blue), $|\tilde{\mathcal{C}}_1\rangle$ (brown), and $|\tilde{\mathcal{C}}_2\rangle$ (orange) versus ramp time $t_\mathrm{r}$, following adiabatic initialization of the $|\tilde{\mathcal{C}}_0\rangle$ state from vacuum. Lighter colors: lossless case ($\kappa = 0$). Darker colors: with single-photon loss rate ($\kappa = 0.01U$). 
        The grey trace indicates leakage to higher excited states for both cases.
        (b) Population remaining in the $|\tilde{\mathcal{C}}_0\rangle$ state for the same initialization protocol as panel (a) as a function of the loss rate $\kappa/U$ for three different ramp times $t_\mathrm{r}$.
        (c) Time evolution under single-photon loss ($\kappa = 0.01U$) of populations of the three-legged squeezed cat states ($|\tilde{\mathcal{C}}_0\rangle, |\tilde{\mathcal{C}}_1\rangle, |\tilde{\mathcal{C}}_2\rangle$), when initializing in the $|\tilde{\mathcal{C}}_0\rangle$ state, for $G/U=2$ and $\Delta/U=-3$. Solid lines are theoretical curves from Eq. \eqref{eq:cerbrho}.
        (d)  Effective decay rates found as inverse of $1/e$ time for the evolution of ($|\tilde{\mathcal{C}}_0\rangle, |\tilde{\mathcal{C}}_1\rangle, |\tilde{\mathcal{C}}_2\rangle$) as a function of $G$ for four different detunings $\Delta/U$. Theoretical estimates from Eq. \eqref{eq:cerbrho} are overlaid as solid lines ($\kappa_d^t=3/2\,\kappa|\alpha|^2$).
        (e) Population of the superposition state $|\tilde{\zeta}_0\rangle$ evolving under single-photon loss. Additional curves illustrate the impact of engineered dissipation for two different rates $\kappa_\mathrm{e}$. 
        (f) Effective decay rates found as inverse of $1/e$ time for the superposition state $|\tilde{\zeta}_0\rangle$ versus $G/U$ for three different detunings $\Delta/U$, using the same color scheme as panel (d).
}
\end{figure*}

\section{Initialization and evolution with single-photon loss}\label{s:PhotonLoss}

\subsection{Steady state}
We now consider the case where the  transitions between three-legged squeezed cat states 
discussed in the previous section are induced by photon loss. As a result of the competition between this dissipative process and the unitary dynamics generated by Eq.~\eqref{eq:H3KPR}, the 3KPO reaches a steady state, which we aim to characterize. Some aspects of the dissipative dynamics of 3KPO under photon loss were addressed in~\cite{Zhang2019, Lorch2019}. 
The system is described by 
the master equation
\begin{equation}\label{eq:MasterEq}
\frac{\mathrm{d}\hat \varrho}{\mathrm{d}t} = -i[\hat H,\hat \varrho]+\kappa \mathcal{D}\left[\hat a \right]\hat \varrho\, ,
\end{equation}
where $\mathcal{D}\left[\hat a \right]\hat \varrho= \hat a \hat \varrho \hat a^\dagger -\frac 12 \{\hat a^\dagger \hat a, \hat \varrho \}$ is the standard dissipator and $\kappa$ is the photon loss rate.
Eq.~\eqref{eq:MasterEq} can be recast as the evolution due to a non-Hermitian Hamiltonian $\hat H_\mathrm{eff}=\hat H - i\frac{\kappa}{2} \hat a^\dagger \hat a$, interrupted by quantum jumps~\cite{Landi2024_tutorial}. A pure state is a steady state of Eq.~\eqref{eq:MasterEq} if and only if it is an eigenstate of both the non-Hermitian Hamiltonian and the  jump operator $\hat a$~\cite{kraus2008}. By applying a displacement and a squeezing transformation to the non-Hermitian Hamiltonian $\hat H_\mathrm{eff}$ and repeating the same analysis of Sec.~\ref{s:Diagonalization}, one can verify that the squeezed coherent states $\ket{\zeta_k}$ are still (approximate) degenerate eigenstates of $\hat H_\mathrm{eff}$, with a modified displacement amplitude  $\alpha_{\mathrm{ss},k}=\sqrt{n_\mathrm{ss}}e^{i \phi_{\mathrm{ss},k}}$,
where
\begin{equation}\label{eq:Nss}
n_{\mathrm{ss}}=\frac{9 G^2 -4\Delta U + \sqrt{9 G^2(9 G^2 -8\Delta U) - 4U^2\kappa^2}}{8 U^2}\,,
\end{equation}
and the phase reads
\begin{equation}\label{eq:Phiss}
\phi_{\mathrm{ss},k}=-\frac{1}{3} \arctan\left( \frac{\kappa/2}{\Delta+2U n_\mathrm{ss} }\right) + \frac{2\pi}{3} k\,,
\end{equation}
with $k=0,1,2$. By substituting these expressions in Eq.~\eqref{eq:SqParameter}, one gets the modified  squeezing parameter, which we do not write here explicitly.

The amplitudes $\alpha_{\mathrm{ss},k}$ coincide with the stationary solutions of 
the mean-field equation obtained from Eq.~\eqref{eq:MasterEq}, namely
\begin{equation}\label{eq:MeanFieldEOM}
\frac{\mathrm{d} \alpha}{\mathrm{d}t} =\left(i\Delta + i 2U \vert\alpha\vert^2 -\frac{\kappa}{2}\right)\alpha -i 3 G\alpha^{*2}\,,
\end{equation} 
whose bifurcation diagram was recently investigated in Ref.~\cite{Xiao2024}.
We notice that the plus sign in front of the square root in Eq.~\eqref{eq:Nss} selects the stable branch of the stationary solution, and the expression reduces to that of $\vert\alpha_+\vert^2$ in Eq.~\eqref{eq:AlphaPlusMin} in the limit $\kappa\rightarrow0$.
For $\Delta>0$, a non-zero stationary amplitude requires the three-photon pump to exceed the modified threshold $G>G_\mathrm{thr,ss}$, with
\begin{equation}\label{GCrit}
G_\mathrm{thr,ss}=\frac{2}{3}\sqrt{\Delta U} \sqrt{1+\sqrt{1+ \left(\frac{\kappa}{2\Delta}\right)^2}}\,,
\end{equation}
which also reduces to the previous  expression for the pump threshold when $\kappa\rightarrow0$.

Crucially, even though the modified squeezed coherent states are eigenstates of $\hat H_\mathrm{eff}$, they are not eigenstates of the jump operator $\hat a$. This is evident from Eq.~\eqref{eq:aOnSqCoh}, which shows that the action of the annihilation operator is always accompanied by a leakage to the excited states. It follows that the steady state cannot be written as a convex combination of squeezed coherent states. This is illustrated by the steady-state Wigner function in Fig.~\ref{f:WignerSteadyState}~(a), which displays a fourth peak in phase space at the origin. The appearance of this additional peak is consistent with the mean field prediction in Eq.~\eqref{eq:MeanFieldEOM}, according to which the zero amplitude solution is always a stationary solution. 
On the other hand, when one enters the regime $\Omega>\kappa\vert\alpha_\mathrm{ss}\vert^2$, the energy gap suppresses leakage to the  excited states and the dynamics is approximately confined to vacuum manifold. Deep enough into this regime, the leakage contribution 
can be neglected  
and the steady state is well described by a mixture of three-legged squeezed cat states with equal weights. We show one such instance in Fig.~\ref{f:WignerSteadyState}~(b), for the regime $\Delta>0$, $G/U \gg1$ and $\kappa/U\ll1$. Note that this ratio between loss rate and Kerr nonlinearity can easily be reached in current experiments~\cite{grimm2020, Frattini2024Arrhenius, ding_quantum_2025, hajr_high-coherence_2024, albornoz_oscillatory_2024, kwon_realisation_2026, Kwon_2022_qec}. 

Once again, we highlight a  key difference with respect to the case of 2KPO. For the 2KPO at zero detuning in the presence of photon loss, the coherent states $\ket{\pm\alpha}$ are both (approximate) eigenstates of the non-Hermitian Hamiltonian and (exact) eigenstates of the jump operator, and hence steady states~\cite{Puri2017Kerr}.  Furthermore, unlike the 2KPO, where exact expressions of the steady state 
can be obtained analytically~\cite{Drummond1980, Minganti2016SciRep}, even in closed form~\cite{Roberts2020}, this is not the case for the 3KPO. For instance,  it was shown that the extended coherent quantum-absorber method cannot be used to obtain the steady state of Eq.~\eqref{eq:MasterEq}~\cite{Roberts2020}.

\subsection{State initialization in the presence of single-photon loss}
We now address the initialization of three-legged squeezed cat states under the action of single-photon loss.
In the following we denote with $|\tilde{\mathcal{C}}_k\rangle$ the exact ground states of $\hat H$ with symmetry $ \hat Z|\tilde{\mathcal{C}}_k\rangle=\omega^k|\tilde{\mathcal{C}}_k\rangle$ and with $|\tilde\zeta_k\rangle=(1/\sqrt{3})\sum_j\omega^{jk}|\tilde{\mathcal{C}}_j\rangle$  their inverse Fourier transform; we use a similar notation for excited states as well.
The three lowest-energy eigenstates $\ket{\tilde{\mathcal{C}}_0}$, $\ket{\tilde{\mathcal{C}}_1}$, $\ket{\tilde{\mathcal{C}}_2}$ can be initialized by ramping on the detuning and pump terms of the Hamiltonian by starting from the initial states $\ket{0}$, $\ket{1}$, $\ket{2}$ respectively. 
An initial state $|\psi(0)\rangle = c_0 |0\rangle + c_1 |1\rangle + c_2 |2\rangle$, under an adiabatic ramp of the parameters $(G,\Delta)$ evolves into the corresponding superposition in the ground state manifold
\begin{equation}
|\psi(t \gg t_r)\rangle \approx 
c_0 |\mathcal{C}_0\rangle + c_1 |\mathcal{C}_1\rangle + c_2 |\mathcal{C}_2\rangle ,
\end{equation}
provided non-adiabatic transitions remain negligible. To achieve adiabatic state preparation, the three-photon drive strength is ramped according to the function:
\begin{equation}
\begin{cases}
G(t)= G\left(3(\frac{t}{t_\mathrm{r}})^2-2(\frac{t}{t_\mathrm{r}})^3\right) \quad \text{for  $t<t_\mathrm{r}$} \\
G(t)=G \qquad \qquad \! \, \text{for  $t\geq t_\mathrm{r}$}
\end{cases}
\end{equation}
where $t_\mathrm{r}$ is the total ramp time. This polynomial ramp ensures a smooth turn-on of the drive, with $G(0)=0$, $G(t_\mathrm{r})=G$, as well as vanishing first derivatives $dG/dt=0$ at $t=0$ and $t=t_\mathrm{r}$.
By imposing $\dot{G}=0 $ at the boundaries, we remove the leading-order nonadiabatic terms and suppress leakage across the 
energy gap
separating the ground state manifold from the first excited states at $G=0$. We additionally ramp on the detuning $\Delta$ with the same function in case the final value $\Delta/U\neq 0$.

We simulate the evolution of the initial state $\ket{0}$ under the action of the time-dependent Hamiltonian and extract the populations 
$\bra{\tilde{\mathcal{C}}_{k}}\hat{\rho}(t)\ket{\tilde{\mathcal{C}}_{k}}$.
Fig.~\ref{f:Cerbspd} (a) illustrates these populations for different ramping times $t_\mathrm{r}$ and fixed Hamiltonian parameters $G/U=2$, $\Delta/U=-3$. In the ideal lossless case, the population in the target state $\ket{\tilde{\mathcal{C}}_{0}}$ converges towards unity for sufficiently long ramp times,
the primary source of imperfection being leakage to the higher excited state $\ket{\psi_{0}^{(\mathrm{e},1)}}$.

The introduction of single-photon loss alters this dynamics and a trade-off emerges: longer ramp times improve adiabaticity, reducing non-adiabatic leakage, but simultaneously increase the duration over which photon loss events can occur. Consequently, the population in $\ket{\tilde{\mathcal{C}}_{0}}$ initially increases with $t_\mathrm{r}$ until reaching a maximum, and then decreases for longer times as photon loss effects accumulate. 
Specifically, substantial population is lost to the state $\ket{\tilde{\mathcal{C}}_{2}}$, since $\hat{a}\ket{\tilde{\mathcal{C}}_{0}}\approx |\alpha|\ket{\tilde{\mathcal{C}}_{2}}$. A smaller population in $\ket{\tilde{\mathcal{C}}_{1}}$ is also present, due to subsequent photon loss events $\hat{a}\ket{\tilde{\mathcal{C}}_{2}}\approx |\alpha|\ket{\tilde{\mathcal{C}}_{1}}$. The rate of non-adiabatic leakage into excited states remains comparable to the lossless case. 

Fig.~\ref{f:Cerbspd} (b) explores this trade-off further by plotting the final population in $\ket{\tilde{\mathcal{C}}_{0}}$ as a function of the loss rate for three different ramp times. In the absence of loss, longer ramp times yield higher fidelity due to increased adiabaticity. However, as $\kappa$ increases, the detrimental effect of loss over longer durations becomes more pronounced. This leads to crossings of the population curves; for example, the $t_\mathrm{r}=10/U$ ramp becomes less effective than the $t_\mathrm{r}=6/U$ beyond $\kappa\approx0.005U$. These crossings highlight that for a given loss rate, an optimal ramp time exists that maximizes the target state fidelity.

\subsection{Evolution of three-legged squeezed cat states and their superpositions under single-photon loss}
Next, we focus on the evolution of the states $\ket{\mathcal{C}_k}$
under the influence of single-photon loss. 
Restricting to the ground state manifold, we obtain from Eq.~\eqref{eq:MasterEq} the following coupled equations for the elements of the density matrix $\rho_{k\ell}(t)=\bra{\mathcal{C}_k}\hat{\rho}(t)\ket{\mathcal{C}_\ell}$
\begin{align}\label{eq:liouveqs}
\frac{\mathrm{d}\rho_{k\ell}}{\mathrm{d}t} &=
\kappa\left[g_k g_\ell \rho_{k+1,\ell+1}-\frac{1}{2}(g^2_{k-1}+g^2_{\ell-1})\rho_{k\ell}\right]\,,
\end{align}
where $g_k=|\alpha|\frac{\mathcal{N}_{k+1}}{\mathcal{N}_k}$.
By taking the large amplitude limit, 
the coefficients $g_k$ reduce to $|\alpha|$ and  the projected annihilation operator  to Eq.~\eqref{eq:a}. Within the manifold, a single photon loss event therefore acts as $\hat X$, that is, imparts a shift error, at the rate $\Gamma_X\equiv\kappa|\alpha|^2$. Since $\hat X$ permutes the three-legged squeezed cat states, the equations for the populations close, giving
\begin{equation}\label{eq:popev}
\frac{\mathrm{d}\rho_{kk}}{\mathrm{d}t}
=\Gamma_X\left(\rho_{k+1,k+1}-\rho_{kk}\right)\,.
\end{equation}
For an initial state $\hat{\rho}(0) = \ket{\mathcal{C}_0}\bra{\mathcal{C}_0}$, the solution is given by
\begin{equation}\label{eq:cerbrho}
\rho_{kk}(t) = \frac{1}{3}+\frac{2}{3}e^{-\frac{3}{2}\Gamma_Xt}\cos{\left(\frac{\sqrt{3}}{2}\Gamma_Xt+\frac{2\pi k}{3}\right)},
\end{equation}
for $k=0,1,2$.
This solution describes an evolution towards a mixed state with equal weights.
The evolution is cyclical, with envelope decay rate $\frac32\Gamma_X$ and oscillation frequency $\frac{\sqrt3}{2}\Gamma_X$. These are the real and imaginary parts of $\Gamma_X(\omega^{\pm1}-1)$, the nonzero eigenvalues of the population equation.
Setting a different initial condition ($\hat{\rho}(0) = \ket{\mathcal{C}_1}\bra{\mathcal{C}_1}$ or $\hat{\rho}(0) = \ket{\mathcal{C}_2}\bra{\mathcal{C}_2}$) yields the same result with different cosine offsets in Eq.~\eqref{eq:cerbrho}. The onset of this cyclical evolution was recently observed experimentally~\cite{kwon_realisation_2026}.
Fig.~\ref{f:Cerbspd}(c) displays these analytical solutions (solid lines) for the populations, showing agreement with the numerical simulations (dots). 
The agreement persists over a broad parameter space, as demonstrated in Fig.~\ref{f:Cerbspd}(d), where we compare the analytical prediction of the effective exponential decay rate in Eq.~\eqref{eq:cerbrho} to the inverse of an $1/e$ time extracted from the simulated curve, see Appendix~\ref{app:oneoveretime} for more details.

We then simulate the evolution of superpositions  of three-legged squeezed cat states Eq.~\eqref{eq:ZetakApprox}. 
From numerical simulations we find that they too relax towards the same equal-weight mixture of three-legged squeezed cat states, albeit on a much slower timescale, as shown by the dark blue points in Fig.~\ref{f:Cerbspd} (e).
While the decay of the three-legged squeezed cat states is generally dominated by single-photon loss rather than leakage, this is not case for the evolution of their superpositions Eq.~\eqref{eq:ZetakApprox}. For these states, 
the leakage rate to excited manifolds can be comparable to or even exceed the evolution rates within the ground state manifold, which makes 
Eq.~\eqref{eq:liouveqs} inadequate.

We then investigate numerically how decoherence of these superposition states depends on the system parameters $G/U$ and $\Delta/U$. 
Fig.~\ref{f:Cerbspd} (f) presents the effective decay rate extracted by inverting the $1/e$ decay time for the population of the superposition state $\ket{\tilde{\zeta}_0}$.  For a fixed $\Delta$, increasing $G$ strongly suppresses this decay rate. 
We attribute this improvement to an increase in the amplitude $|\alpha|$ and a larger phase-space separation, see Eq. \eqref{eq:AlphaPlusMin}. 
Similarly, at fixed $G$ the decay rate is suppressed for increasingly negative detuning. This is consistent with the corresponding increase in $|\alpha|$ and with the accompanying reduction of the squeezing, which sets the amplitude for leakage in Eq.~\eqref{eq:aOnCerb}.
While the datapoints for zero and negative detuning show an approximate exponential suppression of the decay rate with $G/U$, the simulation for positive detuning ($\Delta/U=3$) deviates noticeably from this trend at low values of $G/U$. 

Leakage can be counteracted via state-selective dissipation as has recently been experimentally shown for a Kerr-cat qubit encoded into a 2KPO~\cite{adinolfi_enhancing_2025}. The effective operator of such a state-selective dissipation is:
\begin{equation}\label{eq:StateSelective}
\sqrt{\kappa_\mathrm{e}}\hat{a}_\mathcal{EG}=\sqrt{\kappa_\mathrm{e}}\hat{\Pi}_{\mathcal{G}}\hat{a}\hat{\Pi}_{\mathcal{E}},
\end{equation}
with $\kappa_\mathrm{e}$ the engineered loss rate, $\hat{\Pi}_{\mathcal{G}}$ the projector on the ground-state manifold and $\hat{\Pi}_{\mathcal{E}}$ the projector on the excited-state manifold. This operator is designed to selectively transfer the population from the first excited state manifold back to the ground state manifold. We show the effect of such an engineered dissipation on the state decay in Fig.~\ref{f:Cerbspd} (e). We observe a clear improvement in the decay time for of approximately a factor 2 when applying an experimentally feasible engineered dissipation rate of $\kappa_\mathrm{e}=0.1U$~\cite{adinolfi_enhancing_2025}. 

\section{Tunable Noise Bias}\label{s:NoiseBias} 

\begin{figure}[t]
\centering
\includegraphics[width=0.9\columnwidth]{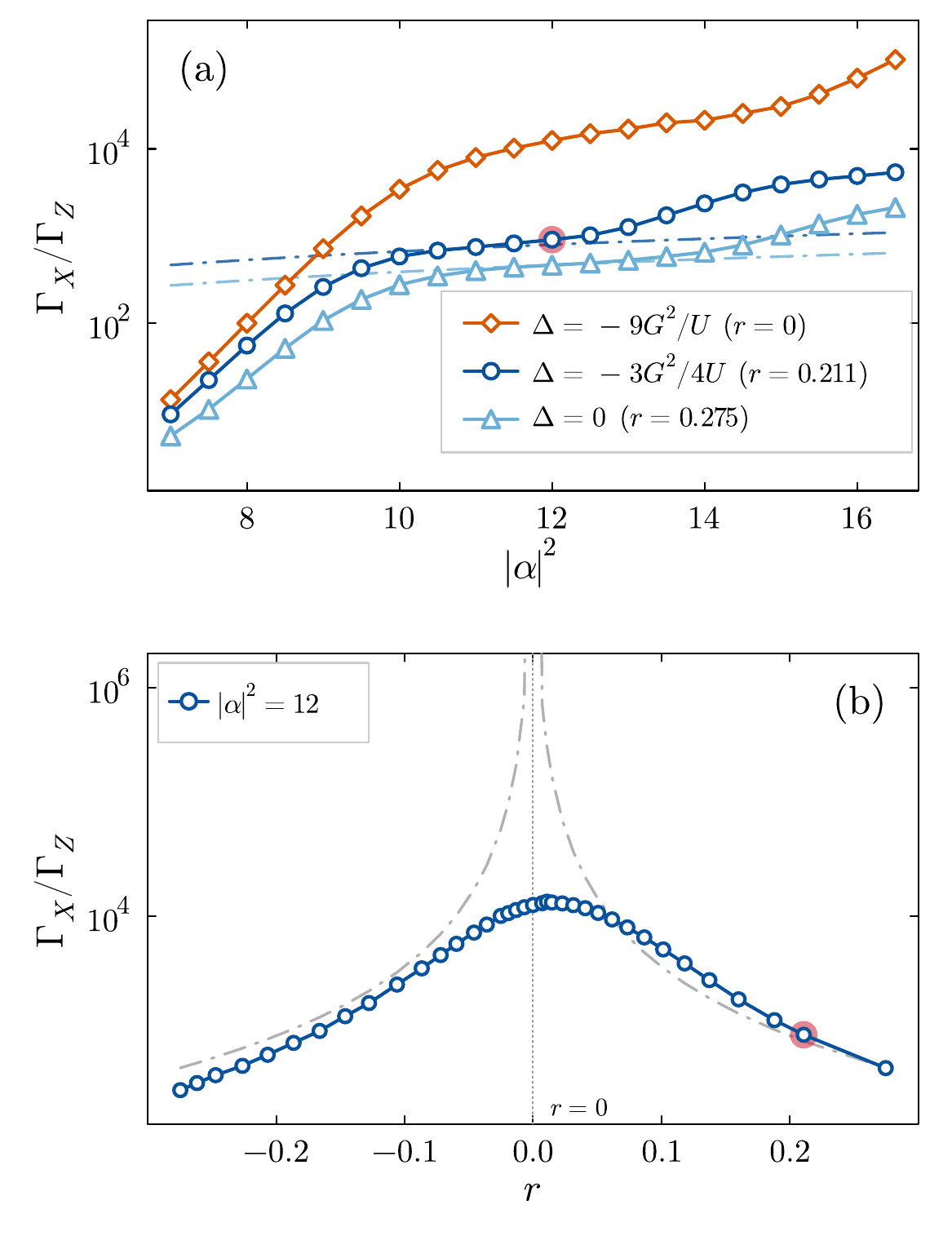}
\caption{\label{f:NoiseBias}
Noise bias $\eta=\Gamma_X/\Gamma_Z$ of the idle qutrit at $\kappa=U/250$; all
rates are exact decay rates of the Liouvillian of Eq.~\eqref{eq:MasterEq}.
(a) $\eta$ versus cat size along three rays of constant squeezing. The
dash-dotted lines are $3|\alpha|^2/\sinh^2 r$, onto which $\eta$ settles once the leakage term of
Eq.~\eqref{eq:Gamma_Z} dominates; for $\Delta=-9G^2/U$ the squeezing vanishes
and there is no such limit. (b) $\eta$ versus the squeezing parameter at fixed
$|\alpha|^2=12$, obtained by varying $G$ and $\Delta$ together along the line
of constant amplitude; the bias is maximal where the squeezing vanishes. The
grey dash-dotted line is  $3|\alpha|^2/\sinh^2 r$. The red dot marks the same point in both panels, $G=2\,U$ and $\Delta=-3\,U$.}
\end{figure}

Fig.~\ref{f:Cerbspd}(c) and Fig.~\ref{f:Cerbspd}(e) show that single-photon loss drives both the codewords $\ket{\mathcal{C}_k}$ and their superpositions $\ket{\zeta_k}$ to the same equal-weight mixture, but the latter decay on a much slower timescale than the former. 
At the logical level, this observation corresponds to the rate of shift errors $\Gamma_X$, i.e., loss-induced cyclic transitions between codewords, dominating over the rate of clock errors $\Gamma_Z$, i.e., transitions between their superpositions $\ket{\zeta_j}$. We now make this separation of timescales quantitative by means of the noise bias.

According to Eq.~\eqref{eq:popev}, photon loss (for large amplitudes) induces shift errors at a rate $\Gamma_X=\kappa|\alpha|^2$ but no clock errors at all.
To account for clock errors and explain the decay of squeezed coherent states  in Fig.~\ref{f:Cerbspd}(e), we then need to depart from this idealized limit. 
We identify three  mechanisms responsible for clock errors. 

The first  is inter-well tunneling.
Contrary to the approximate expressions of the three-legged squeezed cat states Eqs.~\eqref{eq:Cerb0}-\eqref{eq:Cerb2}, tunneling
between the three wells of the metapotential splits the  eigenenergies $\tilde E_k$ of the 3KPO by an exponentially small amount 
$\delta E=\max_k\tilde E_k-\min_k\tilde E_k$, see Sec.~\ref{s:Diagonalization}. 
Denoting the projected Hamiltonian by
$\hat H_{\tilde P}=\hat{\tilde P}\hat H\hat{\tilde P}=\sum_k\tilde E_k\ket{\tilde{\mathcal{C}}_k}
\bra{\tilde{\mathcal{C}}_k}$, the amplitude for hopping between two squeezed coherent states  is given by
\begin{equation}
 h=\left|\langle\tilde\zeta_{j+1}|H|\tilde\zeta_j\rangle\right|
 =\frac13\left|\sum_k \tilde E_k\omega^k\right|,
 \label{eq:h}
\end{equation}
and is proportional to the splitting $\delta E$. Our argument in the following solely relies on the fact that $h\sim\delta E$ is exponentially small in
$|\alpha|^2$. One can also readily check that $h$ vanishes for degenerate energy levels, since $\sum_k\omega^k=0$.
 While a closed form for $h$ 
can in principle be obtained from a WKB treatment, we proceed to evaluate it numerically.  
By itself, the inter-well tunnelling in Eq.~\eqref{eq:h} is a deterministic slow drift and hence removable, but in combination with photon loss it induces an irreversible clock error. 
Indeed, although the populations $\bra{\tilde\zeta_k}\hat\rho\ket{\tilde\zeta_{k}}$ do not decay directly, the coherences
$\bra{\tilde\zeta_k}\hat\rho\ket{\tilde\zeta_{\ell\neq k}}$ built by $h$
decay at a rate $\Gamma_X$ and feed into the evolution of the  populations, inducing clock errors indirectly. 
Since $\Gamma_X\gg h$, the
coherences relax much faster than the populations and eliminating them we obtain the following rate for clock errors 
(see Appendix~\ref{app:ClockError} for
details) 
\begin{equation}
\Gamma_Z^{\rm tun}=\frac{h^2}{\Gamma_X}\,.
\label{eq:Gamma_Ztun}
\end{equation}

The second mechanism is a loss induced round trip out of the code manifold and back. According to Eq.~\eqref{eq:aOnCerb}, photon loss (for large amplitudes) induces leakage at a rate $\Gamma_{\rm leak}=\kappa\sinh^2 r$.
After leaving the qutrit manifold, population can come back  due to a subsequent photon loss event. 
Taking the adjoint of
Eq.~\eqref{eq:adagOnSqCoh}, we obtain the amplitude for returning to the qutrit manifold $\bra{\zeta_k}\hat a\ket{\zeta_k,1}=\cosh r$, which corresponds to a return rate $\kappa|\bra{\zeta_k}\hat a\ket{\zeta_k,1}|^2=\kappa\cosh^2 r$; contributions $\bra{\zeta_k}\hat a\ket{\zeta_{\ell\neq k},1}$ are of order $O(A)$ and can be safely neglected. 
By this mechanism alone, photon loss would return the leaked population to the same squeezed coherent state it came from, without causing any clock error. 
However, since the exact eigenstates of the first excited state manifold are not completely degenerate, the superposition state $\ket{\tilde\zeta_k,1}$ is not a stationary state and therefore starts rotating at a frequency set by the energy splitting
 $\delta E^{(\mathrm{e},1)}$; 
this is the same
mechanism as the one discussed above but for the excited state manifold. 
As a consequence, the which-well information about the label $k$ gets scrambled.  
Whenever the condition
\begin{equation}
\delta E^{(\mathrm{e},1)}\;\gg\;\kappa\cosh^2 r \,,
\label{eq:ConditionScramble}
\end{equation}
holds, 
the well in which a photon was lost is effectively forgotten before the population returns. Returning to the qutrit manifold therefore produces an equal mixture of the three logical states, giving
\begin{equation}
\Gamma_Z^{\rm leak}=\frac13\,\Gamma_{\rm leak} \,.
\label{eq:Gamma_Zleak}
\end{equation}
The expression above is reached deep inside the regime~\eqref{eq:ConditionScramble}, otherwise the coefficient deviates from $1/3$. For  the representative point $G/U=2$, $\Delta/U=-3$, the scrambling condition is well satisfied, with $\delta E^{(\mathrm{e},1)}/(\kappa\cosh^2 r)\simeq16$ and the coefficient correspondingly close to $1/3$, with a value of $0.30$.

A third source of clock errors comes from the breakdown of the large amplitude approximation used above. In this case the
evolution inside the ground state manifold obeys the full expression Eq.~\eqref{eq:liouveqs} instead of Eq.~\eqref{eq:popev}.
This can also cause the decay of superpositions $\ket{\zeta_k}$, as confirmed by inserting the initial condition $\rho_{kj}(0)=1/3$, $k,j=0,1,2$, in Eq.~\eqref{eq:liouveqs}. However, we find that this contribution is typically negligible compared to the other two mechanisms. For instance, at $G/U=2$, $\Delta/U=-3$ 
we estimate a decay time of $t_{1/e} \,U\approx 2.2 \times 10^6$, which is about three orders of magnitude larger than the decay time shown in Fig.~\ref{f:Cerbspd} (e). A similar hierarchy holds true for a wide range of parameters, so this contribution to the clock error budget can be safely neglected. 

Combining the two relevant contributions gives
\begin{equation}
\Gamma_Z=
\frac{h^2}{\kappa|\alpha|^2}+\frac13\kappa\sinh^2 r\,,
\label{eq:Gamma_Z}
\end{equation}
with $\Gamma_Z\ll \Gamma_X$. The noise bias combines the two error rates into a single figure of merit, given by the ratio of the dominant error rate to the suppressed one, $\eta \equiv \frac{\Gamma_X}{\Gamma_Z}$. 
Substituting Eq.~\eqref{eq:Gamma_Z} we find the following expression for the noise bias
\begin{equation}
\eta=\frac{\kappa^2|\alpha|^2}
{h^2/|\alpha|^2+\tfrac13\kappa^2\sinh^2 r}\,.
\label{eq:NoiseBias}
\end{equation}

The two terms of Eq.~\eqref{eq:Gamma_Z} depend inversely on the loss rate, implying an optimal value of $\kappa$. 
The noise bias~\eqref{eq:NoiseBias}, however, does not inherit this structure, because $\kappa$ also appears in
$\Gamma_X$. 
The two terms scale also differently with the amplitude:
the first is exponentially suppressed with $|\alpha|^2$ due to $h$, while the second depends indirectly on $\vert\alpha\vert$ through the squeezing parameter Eq.~\eqref{eq:SqParameter}. 
To  investigate the scaling of $\eta$, we separate the dependence on the amplitude from that on the squeezing.
Since $r$ depends only on the ratio
$\Delta/\Delta_{\rm thr}$, see Eq.~\eqref{eq:SqParameter}, we first change $|\alpha|$ keeping the squeezing constant by sweeping the detuning along the parametric curves $\Delta\propto G^2$. 
Fig.~\ref{f:NoiseBias}(a) shows $\eta$ along three such constant
squeezing curves, 
obtained from exact diagonalization of the Lindbladian associated to the master equation Eq.~\eqref{eq:MasterEq}. 
For small amplitudes, $\Gamma_Z$ is dominated by the first term in Eq.~\eqref{eq:Gamma_Z}, resulting in an exponential noise bias. For larger amplitudes, the first term is no longer dominant, Eq.~\eqref{eq:Gamma_Z} plateaus at the constant value $\kappa\sinh^2 r/3$ and the noise bias becomes $\eta\approx 3|\alpha|^2/\sinh^2 r$, which increases only linearly in $|\alpha|^2$. 
 The exponential scaling of the noise bias is therefore restricted
to three-legged squeezed cats of relatively small size. 
By further increasing $|\alpha|^2$, the noise bias departs from linear scaling and increases again. 
We attribute the departure of the curves to the fact that Eq.~\eqref{eq:ConditionScramble} is no longer satisfied,
so that the coefficient $1/3$ of Eq.~\eqref{eq:Gamma_Zleak} is not attained.
At larger $|\alpha|^2$ the condition analogous to Eq.~\eqref{eq:ConditionScramble} 
 also breaks down for higher excited manifolds, leading to a stepwise increase in the noise bias (not shown), similar to the staircase observed in Ref.~\cite{Frattini2024Arrhenius} for the lifetime of the coherent states of the 2KPO.
 
In Fig.~\ref{f:NoiseBias}(b) we show the noise bias as the squeezing parameter is swept across the squeezing-to-anti-squeezing transition at fixed $|\alpha|^2$. This is achieved by varying both $G$ and $\Delta$ so that the amplitude Eq.~\eqref{eq:AlphaPlusMin} is held constant while the ratio $\Delta/\Delta_{\rm thr}$, which alone sets the squeezing, varies. 
As expected, the noise bias is maximal where
the squeezing vanishes, and falls on both sides. 
The maximum is finite rather than divergent
because the leakage does not vanish exactly at $r=0$. The exact states localized in each well are
not perfectly Gaussian, which causes leakage. At $(G,\Delta)=(2,-3)U$ this residual is more than an order of magnitude below
$\kappa\sinh^2 r$ and, unlike it, decreases as the cat size increases.

Of the two mechanisms featuring in Eq.~\eqref{eq:Gamma_Z}, only the first is
exponentially small in the cat size. For any sufficiently large cat the
clock-error rate is therefore set by the leakage round trip alone, and with it
the noise bias; at $(G,\Delta)=(2,-3)U$ the tunneling term contributes $0.5\%$
of $\Gamma_Z$.
This is where the noise bias connects with our earlier result on squeezing. The
squeezing that the detuning can enhance or suppress is
what sets the leakage $\Gamma_{\rm leak}\simeq\kappa\sinh^2r$. The clock-error
rate approaches $\tfrac13\kappa\sinh^2 r$ in the leakage-round-trip approximation, so the noise bias approaches
$3|\alpha|^2/\sinh^2 r$. The detuning is thus a single
knob on both the shape of the squeezed cat states and the noise bias of the
encoding.
This limit, however,  is not fundamental:
returning the leaked population to the code space before it scrambles removes
it. This is precisely the role of the state-selective dissipation
$\hat a_{\mathcal{EG}}$ introduced in Eq.~\eqref{eq:StateSelective}, whose effect on
the decay of superpositions of codewords  is already visible in Fig.~\ref{f:Cerbspd}(e), i.e., it reduces the leakage.

\section{Quantum gates on the Kerr-cat qutrit}
\label{s:Gates}

\begin{figure*}[t]
\centering
\includegraphics[width=1.0\textwidth]{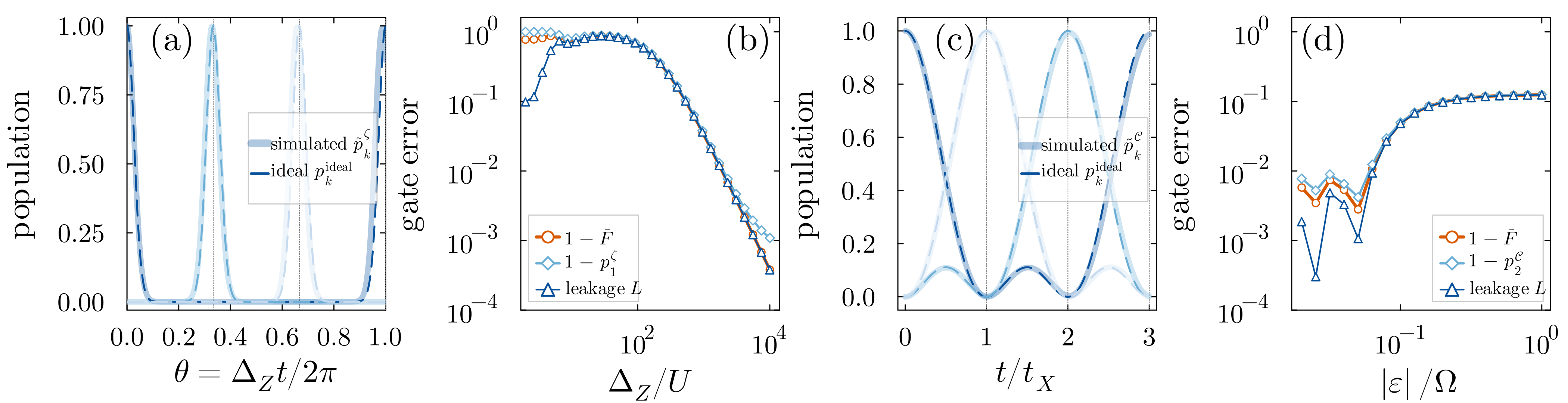}
\caption{\label{fig:gates}
Validation of the clock gate $\hat Z$ [(a),(b)] and of the continuous
$X$-diagonal gates [(c),(d)], obtained by solving the master equation
Eq.~\eqref{eq:MasterEq} for $G/U=2$, $\Delta/U=-3$ and single-photon loss
$\kappa/U=10^{-3}$.
(a) Population dynamics over one full $2\pi$ rotation starting from
$|\tilde\zeta_0\rangle$, at $\Delta_Z/U=1500$. Thick solid curves: simulated
$\tilde p_k^{\zeta}(t)=\langle\tilde\zeta_k|\hat\rho(t)|\tilde\zeta_k\rangle$;
dashed curves: ideal action $p_k^{\mathrm{ideal}}(t)=|\langle\zeta_k|e^{i\Delta_Z\hat n t}|\zeta_0\rangle|^2$
[free rotation, Eq.~\eqref{eq:Z_pk}], same color per lobe.
(b) Gate error at $t_Z=2\pi/(3\Delta_Z)$ vs.\ pulse strength $\Delta_Z/U$:
gate infidelity $1-\bar F$ (circles), analytic population error
$1-p_1^{\zeta}$ (diamonds), leakage $L(t_Z)$ (triangles).
(c) Codeword populations starting from $|\tilde{\mathcal{C}}_0\rangle$, at the
sweep minimum $|\epsilon|/\Omega\simeq0.05$. Thick solid curves: simulated
$\tilde p_k^{\mathcal C}(t)=\langle\tilde{\mathcal{C}}_k|\hat\rho(t)|\tilde{\mathcal{C}}_k\rangle$;
dashed curves: ideal action $p_k^{\mathrm{ideal}}(t)$
[Eq.~\eqref{eq:3k-num-xl-Pk}], same color per codeword.
(d) Gate error at $t_X$ vs.\ drive strength $|\epsilon|/\Omega$, showing the
same three quantities as in (b), with the population error now
$1-p_2^{\mathcal C}$.}
\end{figure*}

In this Section, we provide a universal gate set for quantum computation with Kerr cat qutrits. For each gate, we validate the ideal action, obtained by restricting the relevant generator to the qutrit manifold, against a realistic simulation of the gate in 
the 3KPO Hamiltonian, supplemented by additional drives and including single-photon loss.
We numerically solve the corresponding Lindblad master equation Eq.~\eqref{eq:MasterEq} and characterize the gate of interest via two sets of quantities.
First, the populations $\tilde p_k^{\mathcal C}(t)=\langle\tilde{\mathcal{C}}_k|\hat\rho(t)|\tilde{\mathcal{C}}_k\rangle$
and
$\tilde p_k^{\zeta}(t)=\langle\tilde\zeta_k|\hat\rho(t)|\tilde\zeta_k\rangle$, from which we  obtain the leakage $L=1-\mathrm{Tr}\,\hat{\tilde{P}}\hat\rho(t)$, i.e., 
the fraction of probability that escapes the qutrit manifold, with $\hat{\tilde{P}}=\sum_k|\tilde{\mathcal{C}}_k\rangle\langle\tilde{\mathcal{C}}_k|=\sum_k|\tilde{\zeta_k}\rangle\langle\tilde{\zeta}_k|$. We also denote by $p_k^{\mathcal C}$ and $p_k^{\zeta}$ their analytic counterparts based on the approximate expressions Eqs.~\eqref{eq:Cerb0}-\eqref{eq:Cerb2} and Eq.~\eqref{eq:ZetakApprox}, respectively. Second, since populations  are blind to the relative phase between the
codewords, we certify the gate by means of the average gate fidelity~\cite{Nielsen2002}.  
If we denote by $\hat V$ the target logical unitary describing the gate and by  $|\psi\rangle$ a pure input state, the gate fidelity is given by $F(\psi)=\langle\psi|\hat V^\dagger\,\hat\rho(t_{\rm gate})\,\hat V|\psi\rangle$, where $\hat\rho(t_{\rm gate})$ represents the input state evolved under the full master equation up until time $t_{\rm gate}$. The average gate fidelity $\bar F$ is then obtained by averaging $F(\psi)$ over the four mutually unbiased bases of the code space. Besides  $\{|\tilde{\mathcal{C}}_k\rangle\}$, the other three sets of basis states are the equal-weight superpositions
$\tfrac{1}{\sqrt3}\sum_j\omega^{\,bj^2+mj}|\tilde{\mathcal{C}}_j\rangle$, $b,m=0,1,2$~\cite{WoottersFields1989}. Averaging $F(\psi)$ over these twelve states (projective 2-design) reproduces exactly the Haar average over all pure input states. This is the qutrit analogue of the more familiar qubit case, where the average is taken over the three pairs of Pauli eigenstates along the three Cartesian axes of the Bloch sphere~\cite{Bowdrey2002}.

\subsection{Clock gate $\hat Z$}
\label{s:ZGate}

To implement $\hat Z$, whose action is given in Eq.~\eqref{eq: Z_Lcatzeta}, 
we take the 3KPO Hamiltonian and quickly bring it far off-resonance with respect to the three-photon pump, so that it approximately evolves following the free evolution.
We model this by an instantaneous quench of the detuning,  $\Delta \rightarrow \Delta_Z$. 
The ideal evolution of the populations, starting from $|\zeta_0\rangle$, is given by 
\begin{equation}
p_k^{\mathrm{ideal}}(t)=\bigl|\langle\zeta_k|e^{i\Delta_Z\hat n t}|\zeta_0\rangle\bigr|^2
=\bigl|\langle\zeta_0|e^{i(\Delta_Z t-2\pi k/3)\hat n}|\zeta_0\rangle\bigr|^2,
\label{eq:Z_pk}
\end{equation}
where we used $|\zeta_k\rangle=\hat Z^k|\zeta_0\rangle$. 
In Fig.~\ref{fig:gates}(a) we compare this ideal evolution with the simulated populations $\tilde p_k^{\zeta}$, starting from the state $|\tilde\zeta_0\rangle$. After $\Delta_Z t_Z=2\pi/3$ the population is almost entirely 
transferred to $|\tilde\zeta_1\rangle$, with vanishing $\tilde p_0^{\zeta}$ and $\tilde p_2^{\zeta}$; the mismatch
$1-\tilde p_1^{\zeta}\approx 1.5\times10^{-2}$ thus coincides with the leakage $L(t_Z)$. At intermediate times, the state leaves the code space and all three populations drop to zero.
We also see that the simulated curves progressively advance relative
to the ideal ones, which is due to the Kerr nonlinearity (in the frame rotating at $\Delta_Z$ the
fast pump oscillations at $3\Delta_Z$ average out) and the mismatch is suppressed as $\Delta_Z$ is increased. 
This Kerr-induced shift is removable by referencing the gate time to the actual shift rather than the bare $\Delta_Z$.
Fig.~\ref{fig:gates}(b) shows the  gate infidelity $1-\bar F$ and the leakage $L(t_Z)$ when sweeping the quench detuning $\Delta_Z/U$ at the gate time $t_Z=2\pi/(3\Delta_Z)$. For a  fast enough  gate,  photon loss is negligible so that the two curves coincide. For $\Delta_Z\gg\Omega$, both curves fall as
$(\Delta_Z/U)^{-2}$, while the analytic population error $1-p_1^{\zeta}$ displays the same behaviour until it saturates at a floor value given by $1-|\langle\tilde\zeta_1|\zeta_1\rangle|^2\approx7.0\times10^{-4}$.
For $\Delta_Z\lesssim\Omega$, the quench induces transitions to the  excited-state manifold and the leakage is of order one, peaking at around $\Delta_Z/U\simeq30$.

Finally, we note that the clock gate does not always require a physical implementation. As shown
at the end of Sec.~\ref{s:Xdiag}, it can instead be absorbed into the phases of the drives that
generate the $X$-diagonal gates.

\subsection{Continuous $X$-diagonal gates}
\label{s:Xdiag}

To implement this class of gates, we supplement the 3KPO with a single-photon drive
$\hat H_\mathrm{1ph}=\epsilon\hat a^\dagger+\epsilon^*\hat a$. Once projected onto the qutrit
manifold, $\hat H_{\mathrm{1ph},P} = \hat P \hat H_\mathrm{1ph} \hat P$, the drive Hamiltonian reads
\begin{equation}
\hat H_{\mathrm{1ph},P}=\mathcal{E}\,\hat X+\mathcal{E}^*\hat X^\dagger\,,
\label{eq:H_1ph}
\end{equation}
with $\mathcal{E}=\epsilon^*|\alpha|$ and shift operator $\hat X$ given by Eq.~\eqref{eq:a}. We notice that, since $\hat X^\dagger=\hat X^{2}$, a two-photon drive
$\hat H_\mathrm{2ph}=\epsilon\hat a^{\dagger\,2}+\epsilon^*\hat a^2$ would result in the same 
generator, only with $\mathcal{E}\propto|\alpha|^2$, so we focus on the case of single-photon drive in the following. The generator is diagonal in the
squeezed coherent states, $\hat H_{\mathrm{1ph},P}\ket{\zeta_j}=\mu_j\ket{\zeta_j}$, with eigenvalues
\begin{equation}
\mu_j=2|\mathcal{E}|\cos\!\left(\arg \mathcal{E}+\frac{2\pi j}{3}\right)\,,
\label{eq:eigenvalue-single-photon}
\end{equation}
with $j=0,1,2$. The drive then generates the continuous $X$-diagonal gate $\hat U_X(t)=e^{-i\hat H_{\mathrm{1ph},P}\,t}$ given by
\begin{equation}
\hat U_X(t)=\mathrm{diag}_X\!\big(e^{-i\mu_0 t},\,e^{-i\mu_1 t},\,e^{-i\mu_2 t}\big)\,,
\label{eq:U_X}
\end{equation}
which depends only on the complex parameter $\mathcal E t$.
The gates $\hat U_X(t)$ commute with the dominant shift error $\hat X$ and are therefore bias-preserving at the logical level.
Sweeping both  $\arg \mathcal{E}$ and $|\mathcal{E}|t$ yields a two-parameter family of $X$-diagonal gates, whose image is a
maximal torus of $SU(3)$, we will come back to this aspect later in the Sec.~\ref{s:Universality}. This is the
qutrit counterpart of the continuous rotation $\exp(-i\theta/2 \hat X_2)$ about Pauli $\hat X_2$ found in a Kerr-cat qubit, which is also generated by a single-photon drive~\cite{Puri2017Kerr,grimm2020}. The difference is that in our case
the shift operator $\hat X$ is not Hermitian, so the generator Eq.~\eqref{eq:H_1ph} 
combines the two independent terms $\hat X+\hat X^\dagger$ and $i(\hat X-\hat X^\dagger)$, weighted by the real and
imaginary parts of $\mathcal{E}$ respectively, while for the qubit case ($\hat X_2^\dagger=\hat X_2$, with $\hat X_2$ Pauli-$X$) only a single real parameter survives. 
The leakage out of the
manifold is controlled by keeping the drive weak relative to the spectral gap, i.e., $|\epsilon|\ll\Omega$, which implies
\begin{equation}
t_{\rm gate}\gg\frac{\pi}{9\,|\alpha|^3U}.
\label{eq:t_gate}
\end{equation}

Out of this continuous family, we now single out three relevant  $X$-diagonal qutrit gates, obtained by specific choices of the drive phase and interaction time: the shift gate $\hat X$, the Clifford phase gate $\hat S_X$ and the non-Clifford gate $\hat T_X$.

\emph{Shift gate $\hat X$.}
For a drive along the phase quadrature, i.e., $\arg \mathcal{E}=\pi/2$, Eq.~\eqref{eq:eigenvalue-single-photon} becomes
$\mu=|\mathcal{E}|(0,-\sqrt3,+\sqrt3)$ and the associated unitary realizes  the shift gate
\begin{equation}\label{eq:ShiftGateX}
\hat U(t_X)=\mathrm{diag}_X(1,\omega,\omega^2)\equiv\hat X\,,
\end{equation}
evaluated at the gate time $t_X=\frac{2\pi}{3\sqrt3\,|\epsilon|\,|\alpha|}$.
We can verify the action of the gate on the codewords Eq.~\eqref{eq: X_Lcatzeta} by taking
$|\mathcal{C}_0\rangle\approx(|\zeta_0\rangle+|\zeta_1\rangle+|\zeta_2\rangle)/\sqrt{3}$ and computing $p_k^{\mathrm{ideal}}(t) = |\langle \mathcal{C}_k| e^{-i\hat H_{\mathrm{1ph},P} t}|\mathcal{C}_0\rangle|^2$, which for the ideal unitary evolution gives 
\begin{equation}
p_k^{\mathrm{ideal}}(t)=\frac{1}{9}\bigl|1+\omega^k e^{i\sqrt{3}|\mathcal{E}|t}
           +\omega^{2k}e^{-i\sqrt{3}|\mathcal{E}|t}\bigr|^2,
\label{eq:3k-num-xl-Pk}
\end{equation}
showing cyclic transfer $|\mathcal{C}_0\rangle\to|\mathcal{C}_2\rangle$, etc. at multiples of time $t_X$ as expected.
In Fig.~\ref{fig:gates}(c) we simulate the full Hamiltonian including the single-photon drive, starting from  $|\tilde{\mathcal{C}}_0\rangle$, and 
compare it with the ideal evolution  of Eq.~\eqref{eq:3k-num-xl-Pk}. The latter never leaves the qutrit manifold while, at $t=t_X$, the transfer $|\tilde{\mathcal{C}}_0\rangle\to|\tilde{\mathcal{C}}_2\rangle$ reaches $\tilde p_2^{\mathcal C}=0.997$.
In Fig.~\ref{fig:gates}(d) we plot the average gate infidelity $1-\bar F$, the leakage $L(t_X)$, and $1-p_2^{\mathcal C}(t_X)$ (which saturates at  $1-|\langle\tilde{\mathcal{C}}_2|\mathcal{C}_2\rangle|^2=7.0\times10^{-4}$) by sweeping the drive strength $|\epsilon|/\Omega$ at fixed $G$ and $\Delta$.
For a weak drive, the gate error shows oscillations due to coherent exchange of a small population of order $(|\epsilon|/\Omega)^2$ between the code space and the excited manifold, which are controlled by the fraction not yet returned at time $t_X$. For intermediate drive strengths, $|\epsilon|/\Omega\simeq0.05$, the gate error has a minimum while for stronger drives population leaks irreversibly out of the code manifold: the error is dominated by leakage and the three curves merge. The minimum can be further decreased by increasing the size of the cat state and the gap $\Omega$, either by choosing a stronger pump or a more negative value of detuning. 

\emph{Phase gate $\hat S_X$ and $T$ gate $\hat T_X$.}
For implementing a phase gate we require that two eigenvalues in Eq.~\eqref{eq:U_X} coincide. From
Eq.~\eqref{eq:eigenvalue-single-photon}, this happens whenever $\arg \mathcal{E}$ is a multiple of $\pi/3$. In particular, for 
$\arg \mathcal{E}=-\frac{\pi}{3}$ we have $\mu_j=|\mathcal{E}|\,(1,1,-2)$, which corresponds to a drive with amplitude $\epsilon=|\epsilon|e^{i\pi/3}$. The resulting gate reads, up to a global phase,
\begin{equation}
\hat U_X(\lambda)\equiv\mathrm{diag}_X\big(1,1,e^{i2\pi\lambda/3}\big)\,,
\label{eq:PhaseGatesU_X}
\end{equation}
with $\lambda=9|\mathcal{E}|t/(2\pi)$. This choice of the drive amplitude therefore selects a one-parameter family of $X$-diagonal phase gates, which contains both the Clifford phase gate $\hat S_X=\mathrm{diag}_X(1,1,\omega)$, realized for $\lambda=1$, and the non-Clifford $T$ gate
$\hat T_X=\mathrm{diag}_X(1,1,e^{2\pi i/9})$, realized for $\lambda=\tfrac13$. Indeed, one can readily check that, while $\hat S_X\,\hat Z\,\hat S_X^\dagger=\omega^{2}\,\hat X\hat Z$, i.e., $\hat S_X$ maps generalized Paulis to generalized Paulis, the same test for the choice $\lambda=\tfrac13$ sends $\hat Z$ to a non-Pauli string.
Since both gates $\hat S_X$ and $\hat T_X$ come from the same single-photon drive as the shift gate
$\hat X$, the only difference being the drive phase and the gate time, given by  $t_{S_X}=2\pi/(9|\mathcal{E}|)$ and $t_{T}=2\pi/(27|\mathcal{E}|)$ instead of
$t_X=2\pi/(3\sqrt3\,|\mathcal{E}|)$, they therefore inherit the same qualitative error budget as $\hat X$. At fixed cat parameters, changing only the drive phase and pulse duration changes the logical operation while retaining the same competition between photon loss  and drive-induced leakage.

We conclude by discussing the relation between clock gate and $X$-diagonal gates. 
For a single-photon drive $\hat H_\mathrm{1ph}(\epsilon)$ we have 
$\hat Z^\dagger\hat H_\mathrm{1ph}(\epsilon)\hat Z=\hat H_\mathrm{1ph}(\omega^{2}\epsilon)$, from which it follows
\begin{equation}
\hat U_X(\epsilon,t)\,\hat Z=\hat Z\,\hat U_X(\omega^{2}\epsilon,t)\,.
\label{eq:Zconj}
\end{equation}
A clock gate can thus be commuted through any subsequent
$X$-diagonal gate by shifting the phase of the drive accordingly. Since
the drive phase is a freely adjustable parameter, every $\hat Z$ appearing in a circuit can be moved to the end of it, provided that 
 the outcomes of a measurement in the remaining three mutually unbiased bases are relabelled to account for the total number of clock gates. This is similar to the cat qubit case, where 
commuting the photon-number parity through the single-photon drive amounts to a change of sign of
the drive amplitude~\cite{Puri2020}.

\subsection{Discrete phase gate  $\hat S_Z$}
\label{s:KerrGate}

The Kerr Hamiltonian $\hat H_K = -U\hat a^{\dagger2}\hat a^2$ at the time
$t_K=2\pi/(3U)$ generates the  gate $\hat U_K$ that acts as the identity for $k=0,1,$ and as $\hat U_K|\mathcal{C}_k\rangle=\omega^{2}|\mathcal{C}_k\rangle$ for $k=2$. 
Therefore, $\hat U_K$ coincides with (the inverse of) the standard
qutrit phase gate $\hat S_Z:=\mathrm{diag}(1,1,\omega)$ in the clock basis, namely
\begin{equation}
\hat U_K=\hat S_Z^\dagger=\hat S_Z^{\,2}.
\label{eq:U_K}
\end{equation}
The action of  $\hat U_K$ on the squeezed coherent states reads 
\begin{equation}
  \hat U_K|\zeta_j\rangle
  =  \tfrac{1}{\sqrt3}\bigl(|\mathcal{C}_0\rangle+\omega^{j}|\mathcal{C}_1\rangle
      +\omega^{2j+2}|\mathcal{C}_2\rangle\bigr)\,,
  \label{eq:U_KonZeta}
\end{equation}
with $j=0,1,2$.
Since $|\langle\mathcal{C}_j|\hat U_K|\zeta_k\rangle|^2=|\langle\zeta_j|\hat U_K|\zeta_k\rangle|^2=1/3$ for all $j,k$, the set $\{\hat U_K|\zeta_k\rangle\}$ is a third mutually
unbiased basis of the qutrit manifold, distinct from and unbiased with
respect to both clock and shift bases.
The gate $\hat U_K$ is a logical Clifford operation that commutes with the clock while rotating the shift as $\hat U_K\,\hat X\,\hat U_K^\dagger=\omega^{2}\,\hat X\hat Z$.

Physically, $\hat U_K$ is realized by switching off the pump and detuning and letting the 
Kerr nonlinearity act for the fixed time $t_K$. Unlike the gates implemented via a
single-photon drive, the Kerr gate takes the state out of the code space, returning it only at $t=t_K$. 
The reason is that the Kerr imparts a number-dependent phase to each Fock state component and only at $t=t_K$ does the phase 
depend solely on $n\bmod3$. This mechanism is reversible at $\kappa=0$, where the state returns with unit probability,
so it is not itself an error, in contrast to the leakage of the gates implemented via a
single-photon drive. The only error is photon loss: throughout the evolution the state is spread
over $\sim|\alpha|^2$ photons and exposed to decay, so that $1-\bar F$ is set entirely by
$\kappa$. The infidelity grows linearly with the loss rate, following $\kappa|\alpha|^2 t_K$, and
displays no coherent floor. At $\kappa/U=10^{-3}$ we find $1-\bar F=0.024$.

\subsection{Arbitrary single-qutrit operations}
\label{s:Universality}

We now show that the gates introduced above generate every single-qutrit unitary.
From Eq.~\eqref{eq:U_X} we know that the single-photon drive
produces, by sweeping the drive phase and strength, every $X$-diagonal gate. 
These gates form a two-parameter subgroup
of $SU(3)$, a maximal torus, see Appendix~\ref{app:Universality}.
The clock gate $\hat Z$ permutes the squeezed coherent states as per Eq.~\eqref{eq: Z_Lcatzeta}.
Equivalently, clock gates can be commuted through the continuous drive by
shifting its phase as in Eq.~\eqref{eq:Zconj}. Therefore sequences of $X$-diagonal gates and clock gates
alone remain confined to this two-dimensional set.

The element that allows us to leave the torus is the Kerr gate. Indeed, although $\hat U_K$ is
diagonal in the number basis, just like $\hat Z$, it is not a permutation but a superposition of
clock powers, $\hat U_K=g_0\mathds{1}+g_1\hat Z+g_2\hat Z^2$. Consequently,
$\hat U_K\hat X\hat U_K^\dagger=\omega^2\hat X\hat Z$ is not diagonal in
$\{\ket{\zeta_j}\}$. Conjugating the $X$-diagonal gates by $\hat U_K$ therefore produces a second,
distinct maximal torus, diagonal in the rotated basis $\{\hat U_K\ket{\zeta_j}\}$, which is
mutually unbiased with respect to the original basis, see Eq.~\eqref{eq:U_KonZeta}.

There are two complementary ways to formulate the resulting universality. First, with continuous control of the single-photon drive, 
the two tori generate the full Lie algebra $\mathfrak{su}(3)$. The two drive generators, $\hat X+\hat X^\dagger$ and
$i(\hat X-\hat X^\dagger)$, span the diagonal part of the algebra in the basis ${\ket{\zeta_j}}$, 
 their conjugates by $\hat U_K$ reach all three off-diagonal sectors, 
and, for a generic drive phase, repeated commutators produce both directions within each sector, giving the eight 
dimensions of $\mathfrak{su}(3)$, see Appendix~\ref{app:Universality}.
The continuous drive together with the fixed Kerr gate therefore has full $SU(3)$ as its connected reachable group. Thus any single-qutrit unitary can be synthesized exactly, up to a global phase, by a finite sequence of available continuous-drive and Kerr evolutions. 

Second, one can formulate universality in terms of a fixed discrete gate set. A Clifford phase gate
($\hat S_X$, or equivalently $\hat S_Z=\hat U_K^\dagger$), the qutrit Hadamard $\hat H_3$, and the
non-Clifford gate $\hat T_X$ are together universal for a single qutrit~\cite{Howard2012_qudit_pi8,Campbell2012_magicstate}. We show in Appendix~\ref{app:Hadamard} that 
$\hat H_3$ can be synthesized from the gates discussed above, taking the form
\begin{equation}
\hat H_3=e^{-i\pi/6}\hat X\hat S_X\hat U_K^2\hat S_X .
\label{eq:app-hadamard-main}
\end{equation}
The fixed discrete set generates a dense subgroup of $SU(3)$ and can therefore approximate arbitrary single-qutrit unitaries to finite accuracy~\cite{DawsonNielsen2006_solovaykitaev}.

\subsection{Entangling gate}
\label{sec:3k-entangling}

We conclude by discussing a two-qutrit entangling gate, which is the last ingredient required for universal quantum computation. We consider a hopping Hamiltonian between two 3KPOs,
$\hat H_{\rm int}=J\,\hat a^\dagger\hat b+J^*\,\hat a\hat b^\dagger$.
After projecting onto the two-qutrit code space, we get
\begin{equation}
\hat H_{{\rm int},P}=\mathcal{A}\,\hat X_a^\dagger\hat X_b+\mathcal{A}^*\,\hat X_a\hat X_b^\dagger,
\label{eq:ent-HP}
\end{equation}
with amplitude $\mathcal{A}=J|\alpha|^2$. 
Analogously to the single-photon drive, the interaction
is diagonal in the joint basis $\ket{\zeta_j}_a\ket{\zeta_k}_b$, with eigenvalues
$\mu_d=2|\mathcal{A}|\cos(\arg\mathcal{A}+2\pi d/3)$, where
$d=(k-j)\bmod3$.

Whether the  corresponding gate
$\hat U_{XX}=e^{-i\hat H_{{\rm int},P}t}$ acting on $\ket{\zeta_j}_a\ket{\zeta_k}_b$ entangles or not depends on how the phase varies with $d$. If the accumulated phase satisfies  $-\mu_d t=\phi_0+s d$ with a real offset $\phi_0$ and  $s$ a
multiple of $2\pi/3$, then the gate factorizes into a product of single-mode gates and is  non-entangling.
As an example, for  $\arg\mathcal{A}=\pi/2$, 
we get
$-\mu_d t=2\pi d/3$ and
$\hat U_{XX}=\hat X_a^\dagger\otimes\hat X_b$ at the gate time
$t=2\pi/(3\sqrt3\,|\mathcal{A}|)$. 
When the dependence on $d$  
cannot be cast in this form  
the gate does not factorize and is entangling. For $\arg\mathcal{A}=-\pi/3$, we obtain the one-parameter family
\begin{equation}
\hat U_{XX}(\lambda)=e^{-i2\pi\lambda/9}\,\mathrm{diag}_d\big(1,1,\omega^{\lambda}\big)\,,
\label{eq:ent-family}
\end{equation}
at  $t_{\rm gate}(\lambda)=2\pi\lambda/(9|\mathcal{A}|)$,
where $\mathrm{diag}_d$ denotes the diagonal entries for $d=0,1,2$. For $\lambda=1$, the nontrivial phase can be written explicitly as
\begin{equation}
\mathrm{diag}_d(1,1,\omega)=\big(\hat S_X\hat Z^\dagger\hat S_X^2\hat Z\big)_a\otimes\big(\hat S_X\big)_b\,
\widehat{CZ}_X,
\label{eq:ent-CZ}
\end{equation}
where $\widehat{CZ}_X=\sum_{jk}\omega^{jk}\ket{\zeta_j\zeta_k}\bra{\zeta_j\zeta_k}$ is the two-qutrit controlled-phase gate, diagonal in the squeezed coherent states of both modes. Thus $\lambda=1$ is locally equivalent to a controlled-phase gate and is an entangling Clifford operation. The choice $\lambda=1/3$ yields instead a non-Clifford entangling gate, extending the single-qutrit construction of Sec.~\ref{s:Xdiag} to two qutrits.

The exchange interaction is bias-preserving. The logical Hamiltonian in Eq.~\eqref{eq:ent-HP} is a function of $\hat X_a$ and $\hat X_b$, so it commutes with the dominant shift errors on both modes. Moreover, the physical exchange coupling conserves the total photon number, $[\hat H_{\rm int},\hat n_a+\hat n_b]=0$, and therefore also the total $\mathbb Z_3$ charge $\hat Z_a\hat Z_b$. Thus the interaction preserves the separation between the dominant shift-error sector and the complementary clock sector.

For $\lambda=1$, a full two-mode simulation with $\kappa=0$ confirms the entangling action. Starting from the product state $\ket{\tilde{\mathcal C}_0}_a\ket{\tilde{\mathcal C}_0}_b$, 
the entanglement entropy of each mode reaches its maximum $S=\ln3$ at $t_{\rm gate}$. 
The simulation shows that the error budget of the entangling gate is qualitatively the same as that of the single-qutrit $X$-diagonal gates, with the coupling strength in place of the drive strength. 
Together with the universal single-qutrit operations of Sec.~\ref{s:Universality}, this gate provides the final ingredient for universal quantum computation.

\section{Discussion and Conclusion}\label{s:Conclusions}

In this work we have studied the quantum properties of a nonlinear Kerr oscillator driven by a detuned three-photon pump (3KPO). We identified an exact spectral degeneracy at which the interacting ground state admits a nonperturbative analytical solution described asymptotically by a threefold-degenerate manifold. More generally, we showed that the resulting threefold-degenerate states are well described by three-legged squeezed cat states throughout the regime of semiclassical tristability. The detuning provides a direct control of the displacement and squeezing of the quantum fluctuations, allowing the system to be tuned continuously from squeezed to anti-squeezed states. In the large-amplitude regime, these states define a Kerr-cat qutrit with a tunable noise bias: single-photon loss predominantly produces cyclic shift errors, while clock errors are strongly suppressed.

Building on this encoding, we have  constructed a set for arbitrary single-qutrit operations from physical controls of the 3KPO, together with an entangling gate between two Kerr-cat qutrits, and validated their logical action through simulations of the full oscillator dynamics including leakage and photon loss. In this way, we provided a universal gate set for quantum computation with Kerr-cat qutrits. Our results establish the 3KPO as a platform combining analytically tractable macroscopic quantum states with the possibility of encoding and manipulating qutrit quantum information in a bosonic system.

Several directions follow naturally from these results. On the quantum-information side, an important next step is to investigate quantum error correction tailored to the Kerr-cat qutrit. The error structure identified here provides a natural starting point: the error channel is strongly biased, with a bias that is itself tunable and, for sufficiently large cat states, with the clock-error rate eventually dominated by residual leakage rather than by the cat size. It would be interesting to determine whether this bias can be exploited by suitable qutrit error-correction codes, while retaining the bias under the available logical gates. 

Second, while in this work we focused on the case of Hamiltonian stabilization of a qutrit manifold in the 3KPO, it would be interesting to extend this idea to dissipative protocols. This is in the same spirit as cat qubits, which can be realized by means of both the 2KPO Hamiltonian~\cite{Puri2017Kerr, grimm2020} and two-photon driven dissipation~\cite{leghtas2015, touzard2018, Lescanne2020}. Specifically, the condition in Eq.~\eqref{eq:Dark} can be readily interpreted as a dark state condition for the dissipative stabilization of three-legged squeezed cat states. This corresponds to a purely dissipative evolution  $\nobreak{\partial_t\hat \rho=\gamma \mathcal{D}[\hat a^2-g\,\hat a^\dagger]\hat \rho}$, which stabilizes a steady-state qutrit  manifold generated by the dark states $\ket{\mathcal{C}_k}$, $k=0,1,2$. Exploring deviations from this fine tuned working point and more generally combining higher order parametric pump terms with engineered nonlinear dissipation is a promising direction for future research.

Finally, the 3KPO could find applications in analog quantum simulation of chemical dynamics. Recent experiments leveraged a 2KPO realized in a SNAIL transmon to simulate quantum dynamics of elementary chemical reactions~\cite{DeAlbornoz2026_chemicalactivation}, mapped to an asymmetric double well potential, see also~\cite{Cabral2024_quchemistry}. 
The 3KPO  extends this paradigm to genuinely two-dimensional energy landscapes, enabling the exploration of more complex
free energy surfaces beyond the one-dimensional double-well framework. 
A natural target class are pseudorotational dynamics in 
three-fold symmetric
molecules, e.g. Berry pseudorotation in phosphorus pentafluoride $\mathrm{PF}_5$~\cite{Berry1960, Raynaud2006}. These systems are characterized by a two-dimensional surface spanned by a degenerate pair of vibrational modes, with three equivalent pathways connecting symmetry-related configurations, whose local connectivity matches the topology generated by the 3KPO.

Overall, our work  offers new insight into a model of nonlinear quantum optics relevant for theoretical~\cite{Reynoso_2025, Minganti2023nphoton} as well as experimental~\cite{kwon_realisation_2026} investigations, and more generally into the quantum properties of nonlinear Kerr oscillators driven by higher-order parametric drives. At the same time, it charts a concrete path toward hardware-efficient bosonic quantum computation with qutrits featuring built-in noise bias.

\begin{acknowledgments}
We acknowledge useful discussions with Francesco Adinolfi. M.B. and P.P. ~acknowledge funding from the Swiss National Science Foundation under grant No.~PCEFP2\_194268. M.B. acknowledges funding from the European Research Council (ERC) under the European Union’s Horizon 2020 research and innovation program (Grant agreement No. 101002955 – CONQUER). A.B. acknowledges funding from the Swiss Nanoscience Institute Fellowship Grant No. P2101.
\end{acknowledgments}

Note: At the time of writing this manuscript, we became aware of a recent pre-print discussing experimental results on a qutrit encoded in a three-photon Kerr parametric oscillator~\cite{kwon_realisation_2026}.

\appendix

\section{Exact ground state solution via a wave function approach}\label{app:WaveFunction}

\begin{figure}[t!]
\centering
\includegraphics[width=\columnwidth]{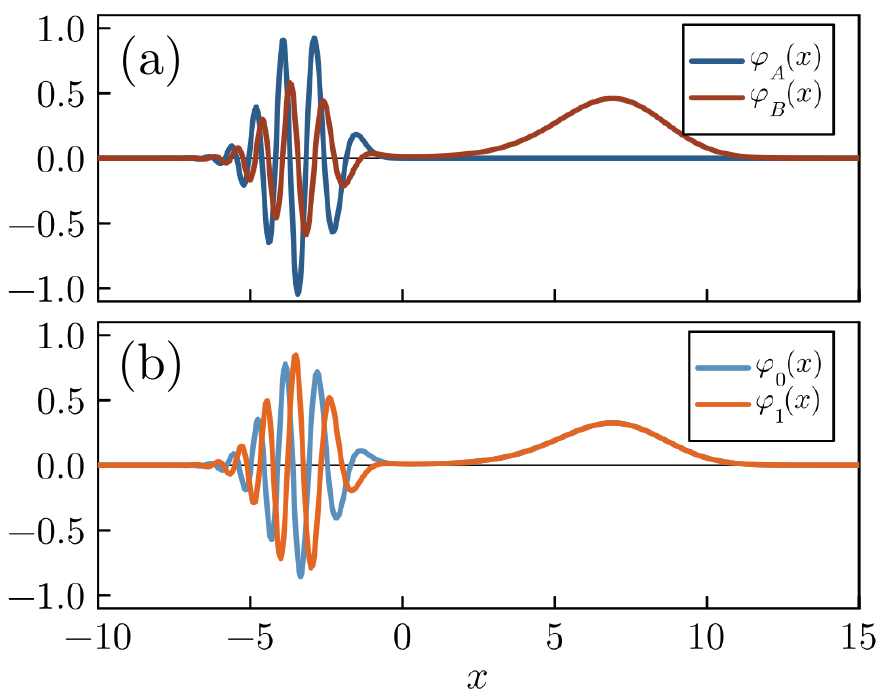}
\caption{\label{f:CerberusWaveFunction}  (a)-(b)  Exact wavefunctions  Eqs.~\eqref{eq:PhiA}, ~\eqref{eq:PhiB} describing  the vacuum manifold spanned by the two highest quasi-energy states of the 3KPO Hamiltonian~\eqref{eq:H3KPR} for $\Delta=G^2/U$ and (d) their symmetric and antisymmetric superposition, as per Eq.~\eqref{eq:Phi01Sup}. The pump strength is $G=\sqrt2 \times 3.5 U\approx 4.95 U$.}
\end{figure}

An alternative way to analyze Eq.~\eqref{eq:Dark} is by means of a wave function approach. To do that, we simply project Eq.~\eqref{eq:Dark} onto the continuous position basis $\{\ket{x}\}$, obtaining a differential equation for the wave function $\varphi(x)\equiv\bra{x}\varphi\rangle$ associated to the stationary Schrodinger's equation $\hat H\varphi(x)=E\varphi(x)$~\cite{Brunelli2018, Brunelli2019}. Since the canonical position and momentum operators $\hat{x}=\left(\hat a+\hat a^\dagger\right)/\sqrt2$ and $\hat{p}=-i\left(\hat a-\hat a^\dagger\right)/\sqrt2$ act on the wavefunction as multiplication by $x$ and  $-i\partial_x$, respectively, Eq.~\eqref{eq:Dark}  becomes 
\begin{equation}\label{eq:RecurrencePos}
   \partial_x^2\varphi(x)+ 2(x+\widetilde{g}) \partial_x\varphi(x)+(x^2-2\widetilde{g} x+1) \varphi(x)=0 \, ,
\end{equation}
where for convenience we set $\widetilde{g}=g/\sqrt2$. This is a second-order differential equation (with non-constant coefficients) 
and therefore admits two independent solutions, each corresponding to a pure state. Which linear combination of the two is selected depends upon the choice of the boundary terms. The two independent solutions, $\varphi_A(x)$ and $\varphi_B(x)$, have the following elegant expressions 
\begin{align}
	\varphi_A(x)&=\mathcal{N}_A\, e^{-\frac12(x+\widetilde{g})^2}\mathrm{Ai}\left[(4\widetilde{g})^\frac{1}{3}\left(x+\widetilde{g}/4\right)\right]\, , \label{eq:PhiA}\\
	\varphi_B(x)&=\mathcal{N}_B \, e^{-\frac12(x+\widetilde{g})^2}\mathrm{Bi}\left[(4\widetilde{g})^\frac{1}{3}\left(x+\widetilde{g}/4\right)\right]\, , \label{eq:PhiB}
\end{align}
where $\mathrm{Ai}(x)$, $\mathrm{Bi}(x)$ are the Airy function of the first and second kind and $\mathcal{N}_{A,B}$ normalization factors.  Both  wave functions are the product of a Gaussian function times an Airy function. In particular, we notice that the function $\mathrm{Bi}(x)$ is not square integrable on the real axis, as it diverges exponentially for positive $x$, but here the multiplication by a Gaussian ensures convergence of the square modulus. The Airy functions confer to the expressions~\eqref{eq:PhiA}, \eqref{eq:PhiB} peculiar properties, as can be seen in Fig~\ref{f:CerberusWaveFunction}(a).  In particular, both develop the shape of an oscillating wave-packet for negative $x$, while $\varphi_B(x)$ gets also delocalized over two spatially separated regions. 

The states $\ket{\varphi_{A,B}}$ associated to Eqs.~\eqref{eq:PhiA}, \eqref{eq:PhiB} are related to the states $\ket{\varphi_{0,1}}$ in Eqs.~\eqref{eq:Phi0}, \eqref{eq:Phi1} in the following way:
\begin{equation}\label{eq:Phi01Sup}
\ket{\varphi_0}=\mathcal{N_+}(\ket{\varphi_A}+\ket{\varphi_B}),\quad \ket{\varphi_1}=\mathcal{N_-}(-\ket{\varphi_A}+\ket{\varphi_B})
\end{equation}
with $\mathcal{N_\pm}=[2(1\pm\langle\varphi_A\ket{\varphi_B})]^{-1/2}$. The corresponding wave functions are shown in Fig.~\ref{f:CerberusWaveFunction}(b).
We can then think of the two states $\ket{\varphi_{0,1}}$ in direct analogy to Schr\"{o}dinger's cat states.
Like cat states, these states are superpositions of non-orthogonal states and  eigenstates of the (three-photon) parity. Unlike cat states, however, they are  superpositions of nonlinear coherent states $\ket{\varphi_{A,B}}$, since Eq~\eqref{eq:Dark} cannot be factorized as the product of two linear transformations.

\section{Explicit expression of $\omega_\pm$ and $\lambda_\pm$}\label{app:Explicit}
We provide the explicit expression of Eq~\eqref{eq:OmegaPM} 
and Eq~\eqref{eq:LambdaPM}, which we report below for convenience  
\begin{align}
\omega&=U\left(4\vert\alpha\vert^2 +\frac{\Delta}{U} \right)\,, \\
\lambda&=U\vert\alpha\vert \left(\vert\alpha\vert -3\frac{G}{U}\right)\,,
\end{align}
in terms of the system parameters, for the two possible choices of displacement $\vert\alpha_{\pm}\vert$, see Eq.~\eqref{eq:AlphaPlusMin}. These are given by  
\begin{align}
\omega_\pm&=-\Delta + \frac{3G(3G\pm\sqrt{9G^2-8U\Delta})}{2U},\,\\
\lambda_\pm&=  -\frac{\Delta}{2}-\frac{3G(3G\pm\sqrt{9G^2-8U\Delta})}{8U}\,.
\end{align}

\section{Overlap between squeezed coherent states}\label{app:SqCohOverlap}
We evaluate the overlap between two coherent squeezed states $\langle \zeta_k \ket{\zeta_{k+1}}=A e^{i\Theta}$ given in 
Eqs.~\eqref{eq:OverlapA},~\eqref{eq:OverlapTheta}. We first consider two general squeezed coherent states $\ket{\alpha,\xi}=\hat D(\alpha)\hat S(\xi)\ket{0}$ and compute the overlap  $\langle \alpha',\xi'\vert \alpha,\xi\rangle$.  It is convenient to move the position representation 
\begin{equation}\label{overlap_cohsq}
\langle \alpha',\xi'\vert \alpha,\xi\rangle=\int_\mathbb{R}dx \,\psi_{\alpha',\xi'}^*(x)\psi_{\alpha,\xi}(x)\,,
\end{equation}
where the wavefunction  $\psi_{\alpha,\xi}(x)=\langle x\ket{\alpha,\xi}$ of a coherent squeezed state is given by
\begin{align}
\psi_{\alpha,\xi}(x)&=\frac{1}{\pi^{1/4}\sqrt{\cosh r-e^{i\theta}\sinh r}}e^{-\frac{i}{2}x_\alpha p_\alpha} \nonumber\\
&\times e^{i p_\alpha x}e^{-\frac{1}{2}\left(\frac{\cosh r+e^{i\theta}\sinh r}{\cosh r-e^{i\theta}\sinh r}\right)(x-x_\alpha)^2}    
\end{align}
and we set $\alpha=\vert\alpha\vert e^{i\phi}$, $\xi=r e^{i\theta}$, $x_\alpha=\sqrt2 \vert\alpha\vert \cos\phi$, $p_\alpha=\sqrt2 \vert\alpha\vert \sin\phi$.
The Gaussian integral is easily evaluated. We then get the following expression 
\begin{widetext}
\begin{align}\label{overlap_cohsq_general}
\langle \alpha',\xi'\vert \alpha,\xi\rangle&=\frac{e^{-\frac{i}{2}(x_\alpha p_\alpha-x_{\alpha'} p_{\alpha'})} }{\sqrt{(\cosh r'-e^{-i\theta'}\sinh r')(\cosh r-e^{i\theta}\sinh r)}} \nonumber \\
&\times \exp\left(-\frac{1}{2} \frac{A(\xi)A(\xi')^*(x_{\alpha}-x_{\alpha'})^2+(p_{\alpha}-p_{\alpha'})^2-2i(A(\xi)x_\alpha+A(\xi')^*x_{\alpha'}) (p_{\alpha}-p_{\alpha'})}{A(\xi)+A(\xi')^*}\right)\,.
\end{align}
\end{widetext}
where we set $A(\xi)=(\cosh r+e^{i\theta}\sinh r)/(\cosh r-e^{i\theta}\sinh r)$.
One can easily check that in the limit of vanishing squeezing $\xi,\,\xi' \rightarrow 0$, the familiar expression of the coherent state overlap $\langle \alpha'\vert \alpha\rangle=e^{-\frac{|\alpha|^2}{2}}e^{-\frac{|\alpha'|^2}{2}}e^{\alpha'^* \alpha}$ is recovered.

We can now particularize the general expression~\cref{overlap_cohsq_general} to the case of our interest. A first simplification comes from the fact that for us, any two coherent squeezed states differ only for the phases of displacement and squeezing, their magnitudes  staying constant. We then set $|\alpha'|=|\alpha|$ and $r'=r$. After some simplifications we get the expression \begin{align}\label{overlap_phase}
\langle \alpha',\xi'\vert \alpha,\xi\rangle&=
\frac{e^{-\frac{i}{2}\vert \alpha\vert^2(\sin 2\phi-\sin 2\phi' )}}{\sqrt{\cosh^2 r-e^{i\theta_-}\sinh^2 r}} e^{-C\vert \alpha\vert^2 } 
\end{align}

\begin{equation}
C= \frac{M \cosh^2 r + M^* e^{i\theta_-}\sinh^2 r -2 N e^{i\frac{\theta_-}{2}}\sinh 2r  }{\cosh^2 r-e^{i\theta_-}\sinh^2 r}    
\end{equation}
where we introduced the angles $\phi_\pm =\phi\pm \phi'$, $\theta_\pm =\theta\pm \theta'$ and the quantities 
\begin{align}
M&=1-\cos\phi_- -2 i\cos^2\left(\frac{\phi_+}{2} \right) \sin\phi_- \\
N&= \cos\left(\phi_+ -\frac{\theta_+}{2} \right) \sin^2\left(\frac{\phi_-}{2} \right)\,.
\end{align}
These expressions evaluated at the pairs of angles in Eqs.~\eqref{eq:DS0}-\eqref{eq:DS2} yield the expression of the overlap provided in Eq.~\eqref{eq:OverlapA} and Eq.~\eqref{eq:OverlapTheta}.

\section{Overlap between squeezed displaced number states}
\label{app:SqDispNumOverlap}
Following Ref.~\cite{Moller1996_dispsqueeznum}, we rewrite the overlap between squeezed coherent states, see Eqs.~\eqref{eq:OverlapA} and~\eqref{eq:OverlapTheta}, in the following equivalent form
\begin{equation}\label{eq:OverlapSqCoh}
\bra{\zeta_k}\zeta_{l}\rangle = \frac{e^{i\vert\alpha\vert^2 \sin(\phi_l-\phi_k)}e^{\frac{\eta_{lk}\eta_{kl}^*}{2\sigma_{lk}}}}{\sqrt{\sigma_{lk}}}\,,
\end{equation}
where we introduced the quantities
\begin{align}
\sigma_{kl}&=\cosh^2 r - e^{i(\theta_{k}-\theta_{l})}\sinh^2 r\,, \label{eq:OverlapSigma} \\
\eta_{kl}&= \vert\alpha\vert \cosh r\left(e^{i\phi_{k}}-e^{i\phi_{l}} \right) \nonumber \\
&- \vert\alpha\vert e^{i\theta_{k}}\sinh r\left(e^{-i\phi_{k}}-e^{-i\phi_{l}} \right) \,. \label{eq:OverlapEta}
\end{align}
This rewriting allows us  to express the overlap between squeezed displaced number states with $n=1$ in the compact form~\cite{Moller1996_dispsqueeznum}
\begin{align}
\bra{\zeta_k,1}\zeta_{l}\rangle &= \bra{\zeta_k}\zeta_{l}\rangle \frac{\eta_{lk}}{\sigma_{lk}} \,, \\
\bra{\zeta_k}\zeta_{l},1\rangle &= \bra{\zeta_k}\zeta_{l}\rangle \frac{\eta_{kl}^*}{\sigma_{lk}} \,, \\
\bra{\zeta_k,1}\zeta_{l},1\rangle &= \bra{\zeta_k}\zeta_{l}\rangle \left(\frac{1}{\sigma_{lk}}+\frac{\eta_{lk}\eta_{kl}^*}{\sigma_{lk}^2}\right) \,.
\end{align}
From the relation above one also gets the useful identities
\begin{equation}
\bra{\zeta_k,1}\zeta_{l}\rangle = \bra{\zeta_k}\zeta_{l},1\rangle  \frac{\eta_{lk}}{\eta_{kl}^*}
\end{equation}
Unlike for the squeezed coherent states overlap, the overlap involving number states (squeezed displaced single photon states) depends on the individual phases, not only on the difference between the phases.
One can also verify that the following relations hold
\begin{align}
\bra{\zeta_0}\zeta_{1},1\rangle &= \bra{\zeta_0}\zeta_{2},1\rangle^*\,, \\
\bra{\zeta_1}\zeta_{2},1\rangle &= \bra{\zeta_2}\zeta_{1},1\rangle^* \,, \\
\bra{\zeta_1}\zeta_{0},1\rangle &= \bra{\zeta_2}\zeta_{0},1\rangle^* \,.
\end{align}
These overlaps allow us to evaluate the expression in Eq.~\eqref{eq:aTransCerb} and Eq.~\eqref{eq:adagTransCerb}.

\section{$1/e$ time}\label{app:oneoveretime}
To compare the analytical prediction of the decay rate with numerical simulations, we extract an effective $1/e$ time $t_{1/e}$ from the population dynamics of the three-legged squeezed cat states.
In our case, the populations exhibit damped oscillations of the form
\begin{equation}
\rho_{00}(t) = \frac{1}{3} + \frac{2}{3} e^{-\Gamma t} \cos\left(\frac{\Gamma t}{\sqrt{3}}\right),
\end{equation}
with $\Gamma = \frac{3}{2}\kappa |\alpha|^2$.

We define $t_{1/e}$ operationally as the first time at which the excess population above the steady-state value has decayed by a factor $1/e$, i.e.
\begin{equation}
\rho_{00}(t_{1/e}) - \frac{1}{3} = \frac{1}{e}\left(\rho_{00}(0)-\frac{1}{3}\right),
\end{equation}
which yields
\begin{equation}
e^{-x}\cos\!\left(\frac{x}{\sqrt{3}}\right)=e^{-1}.
\end{equation}
Solving this numerically yields $x \approx 0.85$, implying that the actual decay time extracted from simulations is related to the envelope rate by $t_{1/e} \approx 0.85/\Gamma$.

\section{Derivation of the clock-error rate Eq.~\eqref{eq:Gamma_Z}}
\label{app:ClockError}

This appendix contains the derivation of the two contributions Eqs.~\eqref{eq:Gamma_Ztun} and
\eqref{eq:Gamma_Zleak} to the clock-error rate Eq.~\eqref{eq:Gamma_Z} provided in Sec.~\ref{s:NoiseBias}. We use the same notation as in the main text, the analytic three-legged squeezed cat states
and squeezed coherent states are denoted as $\ket{\mathcal{C}_k}$ and $\ket{\zeta_j}$ respectively, while 
their numerically exact counterparts carry a tilde.

Since a clock error implies a transfer of population between squeezed coherent states, we start from the matrix elements $\rho^{\zeta}_{jj'}=\langle\zeta_j|\hat\rho|\zeta_{j'}\rangle$ of the density matrix restricted to the qutrit manifold. Within
the description of Sec.~\ref{s:Approximate} and assuming large amplitudes,   
$\ket{\zeta_j}$ are degenerate and photon loss acts as the cyclic shift, of which the $\ket{\zeta_j}$ are eigenstates, see Eqs.~\eqref{eq:a} and \eqref{eq: X_Lcatzeta}.
It follows that, up to corrections of the order of the overlap $A$, the evolution of the matrix elements $\rho^{\zeta}_{jj'}$ obtained from the master equation
Eq.~\eqref{eq:MasterEq} is given by
\begin{equation}
\partial_t\,\rho^{\zeta}_{jj'}=-\Lambda_{j-j'}\,\rho^{\zeta}_{jj'} \,,
\label{eq:eomZeta}
\end{equation}
where $\Lambda_{j-j'}=\Gamma_X\big(1-\omega^{j-j'}\big)$ and $\Gamma_X=\kappa | \alpha |^2$ the shift-error rate.

The off-diagonal terms of Eq.~\eqref{eq:eomZeta} give
$\Lambda_{\pm1}=\tfrac32\Gamma_X\mp i\tfrac{\sqrt3}{2}\Gamma_X$, i.e., the coherences
between squeezed coherent states decay and rotate. We notice that the codeword populations in Eq.~\eqref{eq:cerbrho} are the discrete Fourier transform of the matrix elements $\rho^{\zeta}_{jj'}$, and therefore decay at these same complex rates.
The diagonal element, on the other hand, gives $\Lambda_0=0$: clock errors are identically zero.
The non-zero clock-error rate observed from the numerical simulations in Sec.~\ref{s:PhotonLoss} must thus come from corrections to
Eq.~\eqref{eq:eomZeta}. We identify the two following mechanisms.

\subsection{Inter-well tunneling}

Unlike the approximate eigenstates of Sec.~\ref{s:Approximate}, the exact eigenstates of the 3KPO are not degenerate, their energies $\tilde E_k$ differing by an exponentially
small amount $\delta E=\max_k\tilde E_k-\min_k\tilde E_k$ due to inter-well tunneling, so that the projected
Hamiltonian
$\hat{\tilde P}\hat H\hat{\tilde P}=\sum_k\tilde E_k\ket{\tilde{\mathcal{C}}_k}
\bra{\tilde{\mathcal{C}}_k}$ is no longer proportional to the identity. 
The amplitude for inter-well tunneling $\langle\tilde\zeta_j|\hat H|\tilde\zeta_{j'}\rangle
=\tfrac13\sum_k\tilde E_k\,\omega^{\,k(j'-j)}\equiv\varepsilon_{j'-j}$ is then non-zero. In particular, the diagonal element is
$\varepsilon_0=\tfrac13\sum_k\tilde E_k \equiv \bar E$, while the off-diagonal elements
have modulus $h=|\varepsilon_{\pm1}|=\tfrac13\big|\sum_k\tilde E_k\,\omega^{k}\big|$, which is the hopping amplitude of Eq.~\eqref{eq:h}.
Inverting the expression above we find
$\tilde E_k=\bar E+2h\cos(\varphi-2\pi k/3)$, with $\varphi$ the phase of
$\varepsilon_1$;  $h$ vanishes when the three energies coincide, and
otherwise measures their spread, $\delta E/2\sqrt3\le h\le\delta E/3$.

Projecting the master equation Eq.~\eqref{eq:MasterEq} onto the $\ket{\tilde\zeta_j}$ basis, and using the fact that the dissipator acts as in Eq.~\eqref{eq:eomZeta}, since $\hat X\ket{\tilde\zeta_j}=\omega^{\,j}\ket{\tilde\zeta_j}$, we find
the following coupled equations
\begin{align}
\partial_t\,\tilde\rho^{\zeta}_{jj}
&=2\,\mathrm{Im}\sum_{l\neq j}\varepsilon_{l-j}\,\tilde\rho^{\zeta}_{lj} \,,
\label{eq:eomZetaTilde1}\\
\partial_t\,\tilde\rho^{\zeta}_{jj'}
&=-i\,\varepsilon_{j'-j}\big(\tilde\rho^{\zeta}_{j'j'}-\tilde\rho^{\zeta}_{jj}\big)
-\Lambda_{j-j'}\,\tilde\rho^{\zeta}_{jj'} \nonumber \\
&-i\sum_{l\neq j,j'}\Big(\varepsilon_{l-j}\,\tilde\rho^{\zeta}_{lj'}
-\varepsilon_{j'-l}\,\tilde\rho^{\zeta}_{jl}\Big)\,,
\label{eq:eomZetaTilde2}
\end{align}
from which we see that the populations decay only indirectly: the populations couple to the coherences, see Eq.~\eqref{eq:eomZetaTilde1}, which in turn decay and are driven by the population difference, see Eq.~\eqref{eq:eomZetaTilde2}.
Since $h\ll|\Lambda_{\pm1}|=\sqrt3\,\Gamma_X$, the coherences adiabatically follow the populations. Setting $\partial_t\tilde\rho^{\zeta}_{jj'}=0$ in
Eq.~\eqref{eq:eomZetaTilde2} and discarding the third term, which is a factor $h/|\Lambda_{\pm1}|$ smaller than the second one, gives
$\tilde\rho^{\zeta}_{jj'}=-i\,\varepsilon_{j'-j}
(\tilde\rho^{\zeta}_{j'j'}-\tilde\rho^{\zeta}_{jj})/\Lambda_{j-j'}$.  
Substituting back into
Eq.~\eqref{eq:eomZetaTilde1} 
and using $\mathrm{Re}\,\Lambda_{\pm1}^{-1}
=\mathrm{Re}\,\Lambda_{\pm1}/|\Lambda_{\pm1}|^2=1/(2\Gamma_X)$, we finally find 
\begin{equation}
\partial_t\,\tilde\rho^{\zeta}_{jj}=\Gamma_Z^{\rm tun}\sum_{l\neq j}
\big(\tilde\rho^{\zeta}_{ll}-\tilde\rho^{\zeta}_{jj}\big) ,
\label{eq:eomZetaTilde_final}
\end{equation}
where the rate of transfer between a given pair of states is $\Gamma_Z^{\rm tun}=h^2/\Gamma_X$, recovering Eq.~\eqref{eq:Gamma_Ztun}, which is inversely proportional to $\Gamma_X$, i.e., transfer of population between wells is suppressed for large dissipation, akin to the quantum Zeno effect.
Eq.~\eqref{eq:eomZetaTilde_final} is closed and can be solved. Using
$\sum_l\tilde\rho^{\zeta}_{ll}=1$ we get
\begin{equation}
\tilde\rho^{\zeta}_{jj}(t)=\frac13
+\Big[\tilde\rho^{\zeta}_{jj}(0)-\frac13\Big]
e^{-3\Gamma_Z^{\rm tun}t} .
\label{eq:new-tunsol}
\end{equation}
The squeezed-coherent-state populations relax to the uniform mixture at $3\Gamma_Z^{\rm tun}$, without oscillation. This is in contrast with the codeword populations of
Eq.~\eqref{eq:popev}, whose eigenvalues $\Gamma_X(\omega^{\pm1}-1)$ are
complex. The shift operator transfers population around a directed cycle, whereas tunneling connects each pair of states in both directions at the same rate, which makes the rate matrix in Eq.~\eqref{eq:eomZetaTilde_final} symmetric and its eigenvalues real.

\subsection{Leakage out of the manifold and back}
\label{app:new-leak}

The second correction to Eq.~\eqref{eq:eomZeta} is due to the fact that the action of photon loss is not confined inside the qutrit
manifold. According to Eq.~\eqref{eq:aOnSqCoh}, a photon loss event acting on $\ket{\zeta_k}$ is associated with a leakage rate
$\Gamma_{\rm leak}=\kappa\sinh^2 r$. 
From the relation
\begin{equation}
\hat a\ket{\zeta_k,1}=\cosh r\ket{\zeta_k}+\alpha_k\ket{\zeta_k,1}-\sqrt2 e^{i\theta_k}\sinh r \ket{\zeta_k,2}\,,
\label{eq:aOnZeta1}
\end{equation}
we see that a portion of the leaked population can return to the qutrit manifold due to a
subsequent loss event, at a rate
$\kappa| \bra{\zeta_k}\hat a\ket{\zeta_k,1}|^2 = \kappa \cosh^2 r$.
Notice that, in evaluating the return rate, we neglected transitions to a different well $k'\neq k$ since they are exponentially suppressed due to the overlap~\eqref{eq:OverlapA}.
Within this approximation, the population  returns to the same squeezed coherent state it
left, resulting in no clock error. 

However, just like for the ground-state case discussed above,
inter-well tunneling lifts the degeneracy of
the first excited manifold, at the level of the exact eigenstates
$\ket{\tilde\psi^{(\mathrm{e},1)}_\ell}$ [of which
Eq.~\eqref{eq:SupExcited} is the approximate form], inducing an energy splitting
$\delta E^{(\mathrm{e},1)}$. Their Fourier combinations
\begin{equation}
\ket{\tilde\zeta_k,1}=\frac{1}{\sqrt3}\sum_\ell\omega^{\,\ell k}
\ket{\tilde\psi^{(\mathrm{e},1)}_\ell} ,
\label{eq:new-excexp}
\end{equation}
are then superpositions of three states of different energy,
and hence not stationary. The transition amplitude to another state $\ket{\tilde\zeta_{k'},1}$ localized in the well
$k'$ after a time $t$ is given by
\begin{equation}
\langle\tilde\zeta_{k'},1|e^{-i\hat Ht}|\tilde\zeta_k,1\rangle
=\frac13\sum_\ell\omega^{\,\ell(k-k')}
e^{-i\tilde E^{(\mathrm{e},1)}_\ell t} ,
\label{eq:new-excamp}
\end{equation}
so that the probability for the state to be found in the original well $k$ reads
\begin{equation}
p^{(\mathrm{e},1)}_{kk}(t)=\frac13+\frac29\sum_{\ell<\ell'}
\cos\big[(\tilde E^{(\mathrm{e},1)}_\ell
-\tilde E^{(\mathrm{e},1)}_{\ell'})t\big] \,,
\label{eq:new-pkk}
\end{equation}
which oscillates about $1/3$ at frequencies of order
$\delta E^{(\mathrm{e},1)}$ and equals unity at $t=0$.
The population returns to the qutrit manifold after a random time, exponentially
distributed with mean $\tau=1/(\kappa\cosh^2 r)$.
Provided
$\tau \gg1/\delta E^{(\mathrm{e},1)}$, i.e. Eq.~\eqref{eq:ConditionScramble},
the fast-oscillating terms in
Eq.~\eqref{eq:new-pkk} average away, and we have
$\bar p^{(\mathrm{e},1)}_{kk}=1/3$.
Treating the round trip as instantaneous, i.e. short compared with
$1/\Gamma_{\rm leak}$, is justified by the population found outside the code
manifold at any time, which is a fraction
$\tau\,\Gamma_{\rm leak}=\tanh^2 r$ of that inside  (population leaves at
$\Gamma_{\rm leak}$ and returns at $1/\tau$). As a reference, for $G=2U$ and
$\Delta=-3U$ we have $\tau\,\Gamma_{\rm leak}\approx0.04$.
Population then leaves a given squeezed coherent state at a rate $\Gamma_{\rm leak}$ and comes back to the same state with
probability $\bar p^{(\mathrm{e},1)}_{kk}$, hence with probability
$(1-\bar p^{(\mathrm{e},1)}_{kk})/2$ to each of the other two, giving a rate for clock error 
\begin{equation}
\Gamma_Z^{\rm leak}=\left( \frac{1-\bar p^{(\mathrm{e},1)}_{kk}}{2} \right)\Gamma_{\rm leak} \,.
\label{eq:GammaZ_leakApp}
\end{equation}
The complete scrambling of which-well information therefore gives $\Gamma_Z^{\rm leak}=\tfrac13 \Gamma_{\rm leak}$, which is Eq.~\eqref{eq:Gamma_Zleak} of the main text.

Putting everything together, we obtain the following rate equation for the populations of  squeezed coherent states 
\begin{align}
\partial_t\tilde\rho^{\zeta}_{jj}
&=-\Gamma_{\rm leak}\tilde\rho^{\zeta}_{jj}
+\Gamma_{\rm leak}\Big[\bar p^{(\mathrm{e},1)}_{kk}\,\tilde\rho^{\zeta}_{jj}
+\frac{1-\bar p^{(\mathrm{e},1)}_{kk}}{2}\sum_{l\neq j}\tilde\rho^{\zeta}_{ll}\Big]
\nonumber\\
&=\Gamma_Z^{\rm leak}\sum_{l\neq j}
\big(\tilde\rho^{\zeta}_{ll}-\tilde\rho^{\zeta}_{jj}\big) \,.
\label{eq:eomZetaTilde_leak}
\end{align}
The above equation has the
same form as Eq.~\eqref{eq:eomZetaTilde_final}, and the solution is given by
\begin{equation}
\tilde\rho^{\zeta}_{jj}(t)=\frac13
+\Big[\tilde\rho^{\zeta}_{jj}(0)-\frac13\Big]
e^{-3\Gamma_Z^{\rm leak}t} \,.
\label{eq:LeakSol}
\end{equation}
When the condition Eq.~\eqref{eq:ConditionScramble} is not satisfied, the which-well
information is only partially lost, which results in $\bar p^{(\mathrm{e},1)}_{kk}>\tfrac13$, and the
ratio $\Gamma_Z^{\rm leak}/\Gamma_{\rm leak}$ falls below $\tfrac13$ accordingly.

\subsection{The two mechanisms combined}
\label{app:new-budget}

\begin{figure}[t]
\centering
\includegraphics[width=1\columnwidth]{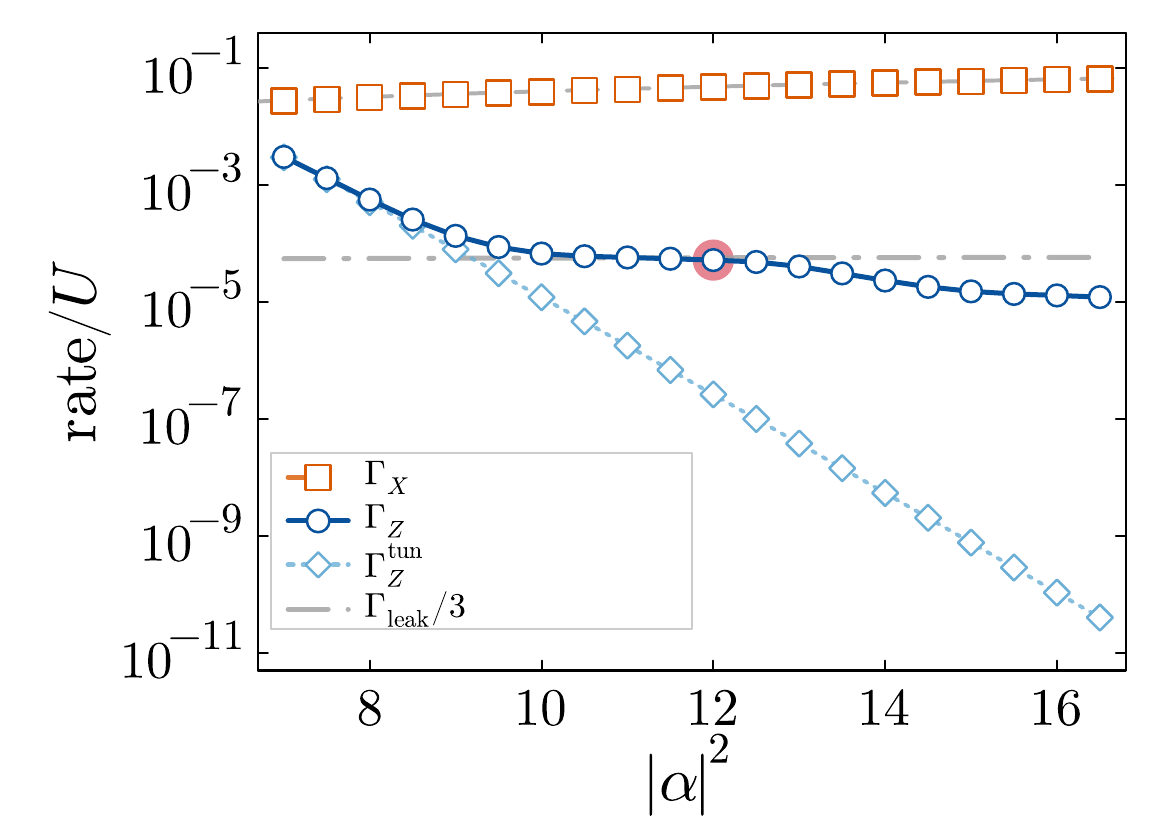}
\caption{\label{fig:Gamma_Zrates}
Error rates along the curve $\Delta=-3G^2/4U$, on which the squeezing parameter
is constant at $r=0.211$ and only $|\alpha|^2$ varies, for $\kappa=U/250$.
$\Gamma_Z$ and $\Gamma_Z^{\rm tun}$ are eigenvalues of the Liouvillian of
Eq.~\eqref{eq:MasterEq}, $\Gamma_X$ and $\Gamma_{\rm leak}$ matrix elements of
$\hat a$ in the code manifold. The
shift-error rate $\Gamma_X$ (squares) follows $\kappa|\alpha|^2$ (dashed). The
clock-error rate $\Gamma_Z$ (circles) follows the tunneling term
$\Gamma_Z^{\rm tun}$ (diamonds; the dotted line is $h^2/\Gamma_X$, with no free
parameter) while that term dominates, is then held at one third of the leakage rate,
$\Gamma_{\rm leak}/3$ (dash-dotted), over some three units of $|\alpha|^2$, and
falls below it beyond $|\alpha|^2\simeq12.5$, where the leaked population no
longer loses the label of the state it came from. The red dot marks $G=2U$,
$\Delta=-3U$.}
\end{figure}

We combine the two contributions Eqs.~\eqref{eq:eomZetaTilde_final}
and~\eqref{eq:eomZetaTilde_leak} into the total clock-error rate
\begin{equation}
\Gamma_Z=\Gamma_Z^{\rm tun}+\Gamma_Z^{\rm leak}
=\frac{h^2}{\kappa|\alpha|^2}+\frac{\kappa\sinh^2 r}{3} \,,
\label{eq:GammaZ_budget}
\end{equation}
which recovers Eq.~\eqref{eq:Gamma_Z} of Sec.~\ref{s:NoiseBias}. 
In Fig.~\ref{fig:Gamma_Zrates} we show the total rate $\Gamma_Z$ and its tunneling
contribution, together with the shift-error rate $\Gamma_X$ and one third of the
leakage rate, $\Gamma_{\rm leak}/3$, by varying the three-photon pump along the
curve $\Delta=-3G^2/4U$ of constant squeezing. The squeezing parameter is held
fixed at $r=0.211$, so that $\Gamma_{\rm leak}=\kappa\sinh^2 r$ stays constant
and only $|\alpha|^2$ varies.

The rates $\Gamma_Z$ and $\Gamma_Z^{\rm tun}$ are obtained by diagonalising
the Lindbladian associated to Eq.~\eqref{eq:MasterEq}. The  Lindbladian  splits
into three blocks, $q=0$ and $q=\pm1$, with $q=(k-k')\bmod3$, where the labels refer to the operator $\ket{\tilde{\mathcal{C}}_k}\bra{\tilde{\mathcal{C}}_{k'}}$ built from two codewords.
The $q=0$ block holds the steady state plus the two decaying modes of eigenvalue $-\Lambda_{\pm1}$.
The coherences between distinct codewords,
$\bra{\tilde{\mathcal{C}}_k}\hat\rho\ket{\tilde{\mathcal{C}}_{k\pm1}}$, lie instead in the other two blocks, which contain no stationary mode and whose slowest eigenvalue is $-3\Gamma_Z$; the factor of three follows from using $\sum_l\tilde\rho^{\zeta}_{ll}=1$ in Eqs.~\eqref{eq:new-tunsol} and~\eqref{eq:LeakSol}. The squeezed coherent state populations are computed from those coherences as $\tilde\rho^{\zeta}_{jj}=\tfrac13+\tfrac13\sum_{d=\pm1}\omega^{\,jd}
\sum_k \bra{\tilde{\mathcal{C}}_{k}} \hat \rho \ket{\tilde{\mathcal{C}}_{k+d}}$. The deviation from $1/3$, which quantifies clock errors, is therefore entirely contained in the $q=\pm1$ blocks.

We extract the tunneling term $\Gamma_Z^{\rm tun}$
by applying the same procedure to the projected model
$\dot{\hat\rho}=-i[\hat{\tilde P}\hat H\hat{\tilde P},\hat\rho]
+\kappa\mathcal{D}[\hat{\tilde P}\hat a\hat{\tilde P}]\hat\rho$, which retains the
tunneling but has no leakage by construction, and whose $q=\pm1$ eigenvalue is
$-3\Gamma_Z^{\rm tun}$. We obtain the leakage term by taking the difference,
$\Gamma_Z^{\rm leak}=\Gamma_Z-\Gamma_Z^{\rm tun}$. The rates $\Gamma_X$ and
$\Gamma_{\rm leak}$ are computed as
matrix elements of $\hat a$ in the code manifold,
$\kappa|\bra{\tilde{\mathcal{C}}_{k}}\hat a\ket{\tilde{\mathcal{C}}_{k+1}}|^2$ and
$\kappa\Vert(1-\hat{\tilde P})\hat a\ket{\tilde{\mathcal{C}}_k}\Vert^2$,
averaged over $k=0,1,2$. Since $\Gamma_{\rm leak}$ is obtained from a matrix element
and not from the spectrum, the agreement of $\Gamma_Z$ with
$\tfrac13\Gamma_{\rm leak}$ over the plateau tests
the validity of Eq.~\eqref{eq:Gamma_Zleak}.

The two terms of Eq.~\eqref{eq:GammaZ_budget} scale differently with the size of
the three-legged squeezed cat states: the tunneling term is exponentially
suppressed, while the leakage term does not depend on $|\alpha|$ at
fixed squeezing. Similar to Sec.~\ref{s:NoiseBias} we identify three regimes of $\Gamma_Z$, for increasing values of $|\alpha|^2$: exponential decay, plateau at 
$\tfrac13\Gamma_{\rm leak}$, followed by a further decrease not contained in Eq.~\eqref{eq:GammaZ_budget}.
We attribute the latter behaviour to the
failure of Eq.~\eqref{eq:ConditionScramble}. The splitting
$\delta E^{(\mathrm{e},1)}$ decreases with cat size as well, until the
which-well information is no longer completely erased, the leaked population
returns increasingly to the state it left, $\bar p^{(\mathrm{e},1)}_{kk}$ rises
above $\tfrac13$ and the coefficient of $\Gamma_{\rm leak}$ in
Eq.~\eqref{eq:GammaZ_leakApp} falls below it.

\section{Universality of single-qutrit operations}
\label{app:Universality}

In this Appendix we show that the gates discussed in Section~\ref{s:Universality} generate
every single-qutrit unitary. In particular, we show that the family of gates obtained from single-photon drive Eq.~\eqref{eq:H_1ph} together with the Kerr gate Eq.~\eqref{eq:U_K} reach the whole $SU(3)$.
In what follows $\mathrm{diag}(\cdot)$ and $\mathrm{diag}_X(\cdot)$ denote matrices
diagonal in the codewords $\{\ket{\mathcal{C}_k}\}$ and in the squeezed coherent states
$\{\ket{\zeta_j}\}$, respectively.

We first characterize the set of gates generated by the single-photon drive
Eq.~\eqref{eq:H_1ph}.
We separate real and imaginary part of the drive amplitude in the projected Hamiltonian Eq.~\eqref{eq:H_1ph} as
\begin{equation}
\hat H_{\mathrm{1ph},P}=\mathrm{Re}\,\mathcal{E}\,\hat D_1
+\mathrm{Im}\,\mathcal{E}\,\hat D_2 \,,
\label{eq:app-univ-D}
\end{equation}
where we denoted the two generators with $\hat D_1=\hat X+\hat X^\dagger$
and $\hat D_2=i(\hat X-\hat X^\dagger)$. Their representation in terms of squeezed coherent state is
$\hat D_1=\mathrm{diag}_X(2,-1,-1)$ and $\hat D_2=\mathrm{diag}_X(0,-\sqrt3,\sqrt3)$.
The generators are traceless and independent. Therefore, as $\mathcal{E}t$ varies in the complex plane,
the exponent of the Hamiltonian covers the whole two-dimensional space of traceless diagonal Hermitian matrices.
Hence no $\mathrm{diag}_X(\cdot)$ gate of unit determinant is missed, and the reachable set is
\begin{equation}
T=\big\{\,\mathrm{diag}_X(\cdot) \;\; \text{with unit determinant}\,\big\} ,
\label{eq:app-univ-torus}
\end{equation}
namely a maximal torus of $SU(3)$.
Finally, the generators commute, so no sequence of $X$-diagonal gates produces a new generator.
The torus $T$ is two-dimensional, against the eight dimensions of $SU(3)$.

We next show that the clock gate $\hat Z$ cannot provide the missing directions, as its action is given by
\begin{equation}
\hat Z\,\mathrm{diag}_X(\mu_0,\mu_1,\mu_2)\,\hat Z^\dagger
=\mathrm{diag}_X(\mu_2,\mu_0,\mu_1) \,,
\label{eq:app-univ-clock}
\end{equation}
that is, conjugation by $\hat Z$ only shifts the drive phase by $-2\pi/3$. A clock gate inserted in a sequence of $X$-diagonal gates can thus be moved to the end
\begin{equation}
\mathrm{diag}_X\big(e^{-i\mu_j t}\big)\,\hat Z\,\mathrm{diag}_X\big(e^{-i\mu_j' t'}\big)
=\mathrm{diag}_X\big(e^{-i(\mu_j t+\mu_{j-1}' t')}\big)\,\hat Z \,,
\label{eq:app-univ-collapse}
\end{equation}
so that sequences of $X$-diagonal and clock gates produce only the elements of
$T$ and its two copies $\hat Z T$ and $\hat Z^{2} T$, and the reachable set stays
two-dimensional within each connected component.

We then consider the Kerr gate Eq.~\eqref{eq:U_K}. Unlike the clock gate, the Kerr gate does not permute the squeezed coherent states but maps
the shift operator to $\hat U_K\hat X\hat U_K^\dagger=\omega^{2}\hat X\hat Z$, which is not diagonal in
$\{\ket{\zeta_j}\}$. Conjugation by $\hat U_K$ therefore maps $T$ into a second maximal torus, diagonal in the rotated basis
$\{\hat U_K\ket{\zeta_j}\}$ and mutually unbiased with the first, i.e.,
$|\bra{\zeta_j}\hat U_K\ket{\zeta_k}|^2=1/3$.

If we take a sequence made of $X$-diagonal gates and Kerr gates, the available continuous generators
are $\hat D_1$ and $\hat D_2$ together with their conjugates by $\hat U_K$ and by $\hat U_K^{2}$
(we remind that $\hat U_K^{3}=\mathds{1}$). Indeed, a Kerr gate can be used to conjugate any
continuous drive evolution, so that
\begin{equation}
\hat U_K e^{-i\hat D_j t}\hat U_K^\dagger
=e^{-i(\hat U_K\hat D_j \hat U_K^\dagger)t}.
\end{equation}
Thus the dimension of the Lie algebra generated by these operators determines the connected
reachable set. It is enough to consider one conjugate:
\begin{equation}
\dim\mathrm{Lie}\{i\hat D_1,\ i\hat D_2,\ i\,\hat U_K\hat D_1\hat U_K^\dagger\}
=8=\dim\mathfrak{su}(3) \,,
\label{eq:app-univ-dim}
\end{equation}
where $\mathrm{Lie}\{\cdot\}$ is the algebra generated by the listed elements, obtained by
adding to their span every repeated commutator.

To verify Eq.~\eqref{eq:app-univ-dim} we count the number of independent real parameters that the
commutators produce. It is convenient to represent elements of $\mathfrak{su}(3)$ as $i$ times
traceless Hermitian matrices. A generic traceless Hermitian matrix in the shift basis reads
\begin{equation}
\hat A=\mathrm{diag}_X(a_0,a_1,a_2)+\sum_{j<k}\big(x_{jk}\hat S_{jk}+y_{jk}\hat T_{jk}\big) ,
\label{eq:app-univ-decomp}
\end{equation}
with $a_j$, $x_{jk}$ and $y_{jk}$ real coefficients, $\sum_j a_j=0$, and with each off-diagonal sector spanned by the pair of Hermitian operators
\begin{equation}
\hat S_{jk}=\ket{\zeta_j}\bra{\zeta_k}+\ket{\zeta_k}\bra{\zeta_j} \,,
\qquad
\hat T_{jk}=i\big(\ket{\zeta_j}\bra{\zeta_k}-\ket{\zeta_k}\bra{\zeta_j}\big)\,.
\end{equation}
In this basis, these coincide with the Gell-Mann matrices, as one has
$\hat S_{01}=\lambda_1$ and $\hat T_{01}=-\lambda_2$ for the first
sector and likewise for the other two, while $\lambda_3$ and $\lambda_8$ are linear
combinations of $\hat D_1$ and $\hat D_2$. Eq.~\eqref{eq:app-univ-decomp} is the
standard decomposition of $\mathfrak{su}(3)$ into a Cartan subalgebra (diagonal part) and the three root spaces (off-diagonal sectors).
The diagonal part is supplied by $\hat D_1$ and $\hat D_2$, so the commutators have to provide the three sectors.

Taking the commutators with the drive generator Eq.~\eqref{eq:H_1ph} gives
\begin{align}
\big[\mathrm{diag}_X(\mu_0,\mu_1,\mu_2),\hat S_{jk}\big]&=-i(\mu_j-\mu_k)\hat T_{jk} \,, \\
\big[\mathrm{diag}_X(\mu_0,\mu_1,\mu_2),\hat T_{jk}\big]&=i(\mu_j-\mu_k)\hat S_{jk} \,,
\label{eq:app-univ-ST}
\end{align}
so each sector is mapped to itself, $\hat S_{jk}$ and $\hat T_{jk}$ just being exchanged.
Commuting twice returns them unexchanged and both multiplied by the same factor $(\mu_j-\mu_k)^2$.
Iterating and commuting $\hat A$ (minus the diagonal part, since it is in the algebra already) with the drive generator $2n$ times we get the matrix
\begin{equation}
\hat A^{(2n)}=\sum_{j<k}(\mu_j-\mu_k)^{2n}\big(x_{jk}\hat S_{jk}+y_{jk}\hat T_{jk}\big) \, .
\label{eq:app-univ-iter}
\end{equation}
For $n=0,1,2$, Eq.~\eqref{eq:app-univ-iter} gives three matrices, $\hat A^{(0)}$,
$\hat A^{(2)}$ and $\hat A^{(4)}$, each containing all three sector components, the one of
the pair $(j,k)$ weighted by $1$, $(\mu_j-\mu_k)^2$ and $(\mu_j-\mu_k)^4$ respectively.
When the three numbers $(\mu_j-\mu_k)^2$ are distinct, the corresponding Vandermonde matrix is invertible.
Therefore, for each pair $(j,k)$ the component $x_{jk}\hat S_{jk}+y_{jk}\hat T_{jk}$ is a linear combination of those
three matrices and hence belongs to the algebra. Commuting this component once more
with the drive generator exchanges $\hat S_{jk}$ and $\hat T_{jk}$, see
Eq.~\eqref{eq:app-univ-ST}, and gives a second element of the same sector, independent of
the first as long as $\mu_j\neq\mu_k$. The two span the sector, so both its parameters are
obtained.

In summary, when the three numbers
$(\mu_j-\mu_k)^2$ are distinct and nonzero, the commutators of a matrix $\hat A$ give both
elements of every sector in which $\hat A$ has a nonzero component.
The numbers $(\mu_j-\mu_k)^2$ are distinct and nonzero for almost every choice of the drive phase.
By
Eq.~\eqref{eq:eigenvalue-single-photon}, we have
$(\mu_j-\mu_k)^2=12|\mathcal{E}|^2\sin^2(\arg\mathcal{E}+\pi d/3)$ with $d=0,1,2$.
They become degenerate only when $\arg\mathcal{E}$ is a multiple of $\pi/6$; the shift gate
Eq.~\eqref{eq:ShiftGateX} $(\arg\mathcal{E}=\pi/2)$ and
the phase gates Eq.~\eqref{eq:PhaseGatesU_X} $(\arg\mathcal{E}=-\pi/3)$ are among these exceptions.
We may therefore choose any generic drive phase for the Lie-algebra construction.

The last step is to take
a matrix with a nonzero component in every sector, and $\hat U_K\hat D_1\hat U_K^\dagger$ featuring in Eq.~\eqref{eq:app-univ-dim} is
one such matrix, since we have
\begin{equation}
\hat U_K\hat D_1\hat U_K^\dagger
=\omega^{2}\,\hat X\hat Z+\omega\,\hat Z^\dagger\hat X^\dagger ,
\label{eq:app-univ-conj}
\end{equation}
which, by $\hat X\hat Z\ket{\zeta_j}=\omega^{\,j+1}\ket{\zeta_{j+1}}$, has vanishing diagonal
part and nonzero components in each of the three off-diagonal sectors.
Its commutators therefore give the six off-diagonal parameters, and with the two diagonal ones
the count is $2+6=8$, as stated in Eq.~\eqref{eq:app-univ-dim}.

In conclusion, the continuous drive together with the fixed Kerr gate generates a Lie algebra
equal to $\mathfrak{su}(3)$. Since $U_K$ and $U_K^\dagger=U_K^2$ are available, the conjugated
continuous evolutions used above are themselves finite sequences of available gates. The connected
reachable subgroup therefore has Lie algebra $\mathfrak{su}(3)$ and is $SU(3)$ itself. Thus every
single-qutrit unitary can be synthesized exactly, up to a global phase, by a finite sequence of
available continuous-drive and Kerr evolutions.

\section{Qutrit Hadamard gate}
\label{app:Hadamard}

In this Appendix we show that the qutrit  Hadamard gate, the missing element from Sec.~\ref{s:Universality} to build  a standard
qutrit universal gate set, can be synthesised from a
short sequence of gates already discussed in the main text, namely the shift gate Eq.~\eqref{eq:ShiftGateX}, the Kerr gate Eq.~\eqref{eq:U_K} and the phase gate Eq.~\eqref{eq:PhaseGatesU_X}.

The qutrit Hadamard gate $\hat H_3$ maps between clock and shift bases, $\hat H_3\hat Z\hat H_3^\dagger=\hat X$ and
$\hat H_3\hat X\hat H_3^\dagger=\hat Z^\dagger$, and is  defined by action $\hat H_3\ket{\mathcal{C}_k}=\ket{\zeta_k}$ with matrix elements
$(\hat H_3)_{jk}=\omega^{jk}/\sqrt3$, that is the discrete Fourier transform. 
The following identity holds
\begin{equation}
\hat H_3=e^{-i\pi/6}\hat X \hat S_X \hat U_K^{2} \hat S_X ,
\label{eq:app-hadamard}
\end{equation}
which shows that the qutrit Hadamard can be realized starting from available gates in the 3KPO: $\hat X$, $\hat S_X$ and $\hat U_K$. 
Eq.~\eqref{eq:app-hadamard} is the qutrit analogue of the Euler decomposition of the qubit Hadamard,
$\hat H_2\propto e^{-i\pi\hat Z_2/4}e^{-i\pi\hat X_2/4}e^{-i\pi\hat Z_2/4}$ for Pauli
operators $\hat X_2$ and $\hat Z_2$.
The qutrit Fourier gate is the inverse of the Hadamard, $\hat F_3=\hat H_3^\dagger$, and is
produced by the related sequence
\begin{equation}
\hat F_3=e^{i\pi/6}\;\hat Z^{2}\,\hat S_X^{2}\,\hat U_K\,\hat S_X^{2} .
\label{eq:app-fourier-alt}
\end{equation}

Together with either phase gate, $\hat S_X$ or $\hat S_Z=\hat U_K^\dagger$, which are
conjugate to one another by $\hat H_3$, the Hadamard generates the single-qutrit Clifford
group. Adding the non-Clifford gate $\hat T_X$ allows one to obtain any single-qutrit
unitary operation to arbitrary accuracy. By the Solovay--Kitaev theorem any single-qutrit unitary
is approximated, up to a global phase, to accuracy $\varepsilon$ by
$\mathrm{polylog}(1/\varepsilon)$ gates of the set. Adding the two-qutrit entangling gate makes the set universal for computation.

\bibliographystyle{apsrev4-2}
\bibliography{bibliography}

\end{document}